\documentclass[prc,aps,amsfonts,twocolumn,floatfix,superscriptaddress,nofootinbib]{revtex4-1}
\usepackage{graphicx,amssymb,color,amsmath}
\usepackage{natbib}
\usepackage{ulem}
\usepackage{braket}
\usepackage{cases}
\usepackage{color}
\usepackage{mathtools}
\usepackage{caption}
\usepackage{subcaption}
\usepackage{multirow}

\definecolor{orange}{rgb}{1,0.5,0}

\newcommand {\aA} [1] {\hat{\rm a}_{#1}}
\newcommand {\aC} [1] {\hat{\rm a}_{#1}^{\dag}}
\newcommand {\aT} [1] {\widetilde{a}_{#1}}
\newcommand {\elmx}   [3] {\bra{#1}#2\ket{#3}}
\newcommand {\ketbar} [1] {\ket{\overline{#1}}}
\newcommand {\multieq}    [1]   {\begin{gather} #1 \end{gather}}
\newcommand {\multieqal}  [2][] {\begin{align#1} #2 \end{align#1}}

\newcommand {\Pn}          {\Psi_n}
\newcommand {\Pm}          {\Psi_m}

\newcommand {\rhobab}  {\boldsymbol{\rho}_{ji,\lambda\mu}^{n\rightarrow m}}
\newcommand {\rhobabz} {\boldsymbol{\rho}_{ji,J0}^{0\rightarrow n}}
\newcommand {\rhoij}   {\rho_{ij}^{n\rightarrow m}}

\newcommand{\D}{\text{d}}

\DeclareMathAlphabet{\mathpzc}{OT1}{pzc}{m}{it}

\begin{document}

\title{Description of nuclear systems with a self-consistent configuration-mixing approach. II: Application to structure and reactions in even-even $sd$-shell nuclei}

\author{C. Robin}
\email{caroline.robin@wmich.edu}
\affiliation{CEA, DAM, DIF, F-91297 Arpajon, France}
\affiliation{Department of Physics, Western Michigan University, Kalamazoo, MI 49008-5252, USA}

\author{N. Pillet}
\email{nathalie.pillet@cea.fr}
\affiliation{CEA, DAM, DIF, F-91297 Arpajon, France}

\author{M. Dupuis}
\affiliation{CEA, DAM, DIF, F-91297 Arpajon, France}

\author{J. Le Bloas}
\affiliation{CEA, DAM, DIF, F-91297 Arpajon, France}

\author{D. Pe\~na Arteaga}
\affiliation{CEA, DAM, DIF, F-91297 Arpajon, France}

\author{J.-F. Berger}
\affiliation{CEA, DAM, DIF, F-91297 Arpajon, France}


\begin{abstract}
\begin{description}
 \item[Background] The variational multiparticle-multihole configuration mixing approach to nuclei has been proposed about a decade ago. 
 While the first applications followed rapidly, the implementation of the full formalism of this method has only been recently completed and applied in Ref.~\cite{Robin} to $^{12}$C as a test-case. 
 \item[Purpose] The main objective of the present paper is to carry on the study that was initiated in that reference, in order to put the variational multiparticle-multihole configuration mixing method to more stringent tests. To that aim we perform a systematic study of even-even $sd$-shell nuclei.
 \item[Method] The wave function of these nuclei is taken as a configuration mixing built on orbitals of the $sd$-shell, and both the mixing coefficients of the nuclear state and the single-particle wave functions are determined consistently from the same variational principle. As in the previous works, the calculations are done using the D1S Gogny force. 
 \item[Results] Various ground-state properties are analyzed. In particular, the correlation content and composition of the wave function as well as the single-particle orbitals and energies are examined. Binding energies and charge radii are also calculated and compared to experiment. The description of the first excited state is also examined and the corresponding transition densities are used as input for the calculation of reaction processes such as inelastic electron and proton scattering. Special attention is paid to the effect of the optimization of the single-particle states consistently with the correlations of the system.
 \item[Conclusions] The variational multiparticle-multihole configuration mixing approach is systematically applied to the description of even-even $sd$-shell nuclei. Globally, the results are satisfying and encouraging. In particular, charge radii and excitation energies are nicely reproduced. 
However, the chosen valence-space truncation scheme precludes achieving maximum collectivity in the studied nuclei. Further refinement of the method and a better-suited interaction are necessary to remedy this situation.
\end{description}
\end{abstract}


\maketitle


\section{Introduction}

Atomic nuclei are among the most complex and challenging quantum many-body systems to describe. On the one hand they exhibit properties of independent-particle systems, as evidenced, for example, by the existence of magic numbers. On the other hand many phenomena such as superfluidity, deformation, clustering or emergent collective phenomena are proof of the deep correlations that exist between the nucleons. 
Following these observations, historically two types of approaches to the description of nuclei appeared. The first is based on the self-consistent mean-field (SCMF) method~\cite{Ring_Schuck}, which rests on the assumption that in first approximation, the nucleons can be described as evolving in a mean potential which emerges from the underlying effective nuclear interaction. The nucleus is thus described as a system of independent nucleons which are dressed by their averaged interaction with the other particles. 
The second approach, known as interacting Shell-Model (SM)~\cite{SM}, starts from a given set of single-particle states, and directly tackles the correlations between the nucleons in a truncated many-body model space.
Both of these approaches have advantages and drawbacks. For instance, in the SCMF method, accounting for correlations beyond the first order approximation often requires symmetry-breaking and restoration techniques. Also, ground and excited states as well as even-even, odd-even and odd-odd nuclei can usually not be treated on the same footing. On the other hand this approach is applicable to a very large range of masses, and excited states can usually be calculated up to high energies. While the traditional Shell-Model is mainly applicable up to mid-mass nuclei (or near closed-shell) and restricted to low-lying states, the description of ground and excited states of all types of nuclei can be achieved in a unified way, while explicitly preserving important symmetries. 
\\
Several approaches aiming to reconcile SM and SCMF methods have been developed in the past. We can cite for instance the Shell-Model Monte-Carlo (SMMC) \cite{SMMC1,SMMC2,SMMC3,SMMC4,SMMC5,SMMC6} and Monte-Carlo Shell-Model (MCSM) \cite{MCSM1,MCSM2,MCSM3,MCSM4,MCSM5}, that use stochastic sampling of many-body configurations around a deformed mean-field solution and diagonalize the Hamiltonian in the resulting subspace. From the mean-field side the VAMPIR-family methods (see \textit{e.g.} \cite{VAMPIR1,VAMPIR2,VAMPIR3,VAMPIR4,VAMPIR5,VAMPIR6,VAMPIR7,VAMPIR8}) diagonalize a Hamiltonian in the space spanned by a symmetry-projected Hartree-Fock-Bogoliubov ground state and quasiparticle excitations on top of it.  
\\
The present work aims to unify SM and SCMF views in an alternative manner, without making use of symmetry-breaking and projection techniques. The resulting approach is called "variational multiparticle-multihole configuration mixing method", which we will abbreviate by MPMH. More specifically, this method considers a wave function written in the form of configuration interaction methods, as a superposition of many-nucleon configurations. Such an ansatz allows one to preserve the advantages of Shell-Model techniques, such as explicit preservation of rotational and particle-number symmetry, and the ability to calculate energies, densities and transition probabilities from ground or excited states in a single framework. Additionally, full self-consistency is obtained as both correlations and orbitals are calculated via a common variational principle applied to the same energy functional of the system. The single-particle states are therefore not considered as frozen inputs, but are instead determined consistently with the correlations of the system. 
\\
While such kinds of approaches have been widely employed in the past in atomic physics and quantum chemistry \cite{MCHF1,MCHF2,MCSCF1,MCSCF2}, their applications to nuclear systems are more recent. The very first attempts were restricted to analytical models \cite{VariaRPA,Faessler1,Faessler1b}, and were later followed by applications in the intrinsic frame that were however restrained by computational limits \cite{Satpathy,Faessler2,HoKim,Faessler3,Faessler4,Faessler5}. 
Last decade has seen renewed interest and effort in the development and application of the MPMH approach to nuclei in realistic scenarios.
The first studies have partly applied the formalism of the MPMH method: analyses of the spectroscopy of $sd$-shell nuclei were performed using frozen Hartree-Fock orbitals \cite{Pillet2,LeBloas}, and the description of pairing correlations in the ground states of Sn isotopes was studied in Ref.~\cite{Pillet1} by making important approximations in the determination of the single-particle states. Finally the full formalism of the MPMH approach was implemented and applied for the first time to the $^{12}$C nucleus in Ref. \cite{Robin}. In the present study we continue the work initiated in that reference, and perform systematic calculations of ground and excited states of $sd$-shell nuclei.
\\
In section \ref{sec:formalism} we briefly remind the formalism of the MPMH method. In section \ref{sec:ground-state} we investigate various properties of the ground state of even-even $sd$-shell nuclei. We analyze in more detail some aspects of a few benchmark nuclei, such as their correlation content, wave-function composition, single-particle energies and single-particle orbitals. More systematically, we calculate binding energies and charge radii, which we compare to experimental data. In section \ref{sec:excited_states}, we investigate the description of low-lying excited states, and use the calculated transition densities as inputs for reaction calculations such as inelastic electron and proton scattering. Finally we give conclusions and perspectives to this work in section \ref{sec:conclusion}.


\section{Formalism of the MPMH method} \label{sec:formalism}
We briefly remind the formalism of the MPMH method. For more details we refer the reader to Ref.~\cite{Robin}.
\\
In the MPMH approach, the nuclear state $\ket{\Psi}$ is taken as a superposition of Slater determinants $\ket{\phi_\alpha}$ built on a certain single-particle basis $\{i\}$:
\begin{eqnarray}
\ket{\Psi} = \sum_\alpha A_\alpha \ket{\phi_\alpha} \; , \mbox{  with  } \ket{\phi_\alpha} = \prod_{i \in \alpha} a^\dagger_i \ket{0} \; ,
\end{eqnarray}
where $i$ denotes a set of quantum numbers and $\ket{0}$ refers to the true particle vacuum.
\\
In the case of a two-body density-dependent interaction like the D1S Gogny force \cite{D1S} used in this study, the minimization of the energy with respect to the expansion coefficients $A_\alpha$, and to the single-particle orbitals $\{i\}$ leads to the following system of coupled equations:
\begin{numcases}{}
\sum_{\beta} A_{\beta} \braket{\phi_{\alpha}|\hat{\mathpzc{H}}[\rho,\sigma] |\phi_{\beta}} = \lambda A_{\alpha}, \;\; \forall \alpha  \label{e:eq1_Gogny} \\
\left[ \hat{\mathpzc{h}}[\rho,\sigma] , \hat{\rho} \right] = \hat{G}[\sigma] \; , \label{e:eq2_Gogny}
\end{numcases}
where $\rho$ and $\sigma$ denote the one-body density matrix and two-body correlation matrix of the state $\ket{\Psi}$, respectively:
\begin{eqnarray}
 \rho_{ij} &=& \braket{\Psi|a^{\dagger}_j a_i|\Psi} \; , \\
\sigma_{il,jk} &=& \braket{\Psi|a^{\dagger}_i a^{\dagger}_j a_{k} a_{l}|\Psi} - \rho_{li} \rho_{kj} + \rho_{lj} \rho_{ki} \; .
\end{eqnarray}
In Eq. (\ref{e:eq1_Gogny}), the two-body operator $\hat{\mathpzc{H}}[\rho,\sigma]$ is the sum of the kinetic energy\footnote{In this work, only the one-body corrections to the center-of-mass motion are implemented.} $\hat K$, the D1S interaction $\hat{V}^{2N}_{D1S}[\rho]$, and a rearrangement term $\hat{\mathpzc{R}}[\rho,\sigma]$ arising from the density-dependence of the interaction:
\begin{eqnarray}
\hat{\mathpzc{H}}[\rho,\sigma] = \hat K + \hat{V}^{2N}_{D1S}[\rho] + \hat{\mathpzc{R}}[\rho,\sigma] \; ,
\label{e:GognyHamilt}                                                              
\end{eqnarray}
where the rearrangement term can be written as
\begin{eqnarray}
\hat{\mathpzc{R}}[\rho,\sigma] &=& \frac{1}{4} \int \D^3 r \sum_{klmn} \braket{kl| \frac{\delta {V}^{2N}[\rho]}{\delta \rho(\vec{r})}|\widetilde{mn}} \nonumber \\ 
                               && \hspace{0.7cm} \times \left( \rho_{mk} \rho_{nl} - \rho_{ml} \rho_{nk} + \sigma_{km,ln} \right) \hat{\rho}(\vec{r}) \; ,
\end{eqnarray}
with $\ket{\widetilde{mn}} \equiv \ket{mn} - \ket{nm}$.
\\
In Eq. (\ref{e:eq2_Gogny}), $\mathpzc{h}[\rho,\sigma]$ is a general mean-field Hamiltonian:
\begin{eqnarray}
\mathpzc{h}_{ij}[\rho,\sigma]  &=& K_{ij} + \sum_{kl} \braket{ik|\widetilde{V}^{2N}_{D1S}[\rho]|jl} \rho_{lk}+ \mathpzc{R}_{ij}[\rho,\sigma] \; ,
\label{e:mean-field}
\end{eqnarray}
and $G[\sigma]$ is the source term containing the effect of two-body correlations beyond the mean-field $\mathpzc{h}$:
\begin{eqnarray}
G[\sigma]_{ij} &=&  \frac{1}{2} \sum_{klm} \sigma_{ki,lm} \braket{kl|{V}^{2N}_{D1S}[\rho]| \widetilde{jm}} \nonumber \\
               &&  - \frac{1}{2}  \sum_{klm} \braket{ik|{V}^{2N}_{D1S}[\rho]| \widetilde{lm}}  \sigma_{jl,km}  \; .
\label{e:G_Gogny}                        
\end{eqnarray}

Eq. (\ref{e:eq1_Gogny}) is responsible for introducing explicit correlations into the nuclear wave function, as it determines the mixing coefficients $A_\alpha$ via the diagonalization of the matrix $\mathpzc{H}[\rho,\sigma]$, 
while Eq. (\ref{e:eq2_Gogny}) determines the single-particle orbitals that are consistent with these correlations. Practically, they are taken as natural orbitals, \textit{i.e.} eigenstates of the one-body density satisfying Eq. (\ref{e:eq2_Gogny}). A detailed analysis of the role of the orbital equation (\ref{e:eq2_Gogny}) can be found in Ref. \cite{Robin}.


\section{Ground-state description of even-even $sd$-shell nuclei} \label{sec:ground-state} 

In this section we investigate various properties of the ground state of even-even $sd$-shell nuclei, described within the MPMH method.
\\ \\
This systematic study is performed using the following scheme: The single-particle states are expanded on axially deformed harmonic oscillator states at the spherical point ($\beta = 0$), and we choose to use $N_0=9$ major oscillator shells. To select the relevant many-body configurations included in the wave function $\ket{\Psi}$ we use a "Shell-Model" scheme, \textit{i.e.} we allow for all possible excitations of nucleons within the $sd$-shell, on top of a core of $^{16}$O. Since this model space is rotationally invariant, the final correlated state is characterized by a good total angular momentum $J$.
The minimum number of configurations (418, using time-reversal invariance) is obtained in the case of $^{20}$Ne, having only 4 valence nucleons that can lead to excitations from 0$p$-0$h$ up to 4$p$-4$h$. Conversely, the maximum number of configurations (56~937) occurs in the case of $^{28}$Si, for which excitations up to 12$p$-12$h$ are possible within the valence space. 
This type of truncation scheme was already used to test the MPMH approach in $^{12}$C (Ref. \cite{Robin}), where it led to reasonable results using the same Gogny interaction.
\\ \\
In this study, we find it interesting to introduce self-consistency step by step, in order to understand the implications of its full implementation in the MPMH method.
Therefore, we show the theoretical results at three levels of implementation of the MPMH method:
\begin{enumerate}
 \item Without any self-consistency, \textit{i.e.} after one single diagonalization of the many-body matrix $H[\rho_{HF}] = K + V^{D1S}[\rho_{HF}]$, in the $sd$-shell of pure Hartree-Fock (HF) orbitals, when the pure HF density $\rho_{HF}$ is used in the interaction and therefore, no rearrangement terms are present. \label{level1}
 \item With partial self-consistency, \textit{i.e.} after solving the full Eq. (\ref{e:eq1_Gogny}) alone, on HF orbitals, including the correlated density in the interaction and the rearrangement terms. This is achieved by diagonalizing $\mathpzc{H}[\rho,\sigma]=H[\rho]+\mathpzc{R}[\rho,\sigma]$ iteratively, until convergence. In this work, convergence is said to be reached when the difference $|\rho_{ij}^{(N-1)}-\rho_{ij}^{(N)}|$ between any element of the one-body density matrix between two iterations $N-1$ and $N$ is less than $1.0 \times 10^{-5}$. \label{level2}
 \item With full self-consistency, \textit{i.e.} when both equations (\ref{e:eq1_Gogny}) and (\ref{e:eq2_Gogny}) are solved together and consistency between correlations and orbitals is reached. This is achieved using the doubly-iterative procedure described in detail in Ref. \cite{Robin}. The convergence criteria on the density matrix is also set to $1.0 \times 10^{-5}$ for both types of iterations. \label{level3}
\end{enumerate}
This comparison makes it possible to separate the effects of (i) the use of the correlated density in the interaction - which is not justified \textit{a priori} - and (ii) the orbital equation (\ref{e:eq2_Gogny}).
\\ \\
The present study follows the work presented in Ref. \cite{LeBloas} which provided a description of even-even $sd$-shell nuclei at the level 1 of implementation of the MPMH approach stated above. However, the aforementioned study did not include the exact exchange Coulomb term in the HF field, and $N_0=11$ major oscillator shells were used to expand the HF single-particle states. 
\\ \\
In the first part of this section we focus on the analysis of a few noteworthy nuclei. In particular we are interested in their correlation matrices, source terms $G[\sigma]$, wave-function composition and single-particle properties.
In the second part of the section we give a systematic description of ground-state observables such as charge radii, binding and two-nucleon separation energies.

\subsection{Analysis of a few benchmark nuclei}
\begin{figure}[h!]
\centering
\subcaptionbox{$^{28}$Ne\label{28Ne_triax}}
{\includegraphics[width=.49\columnwidth] {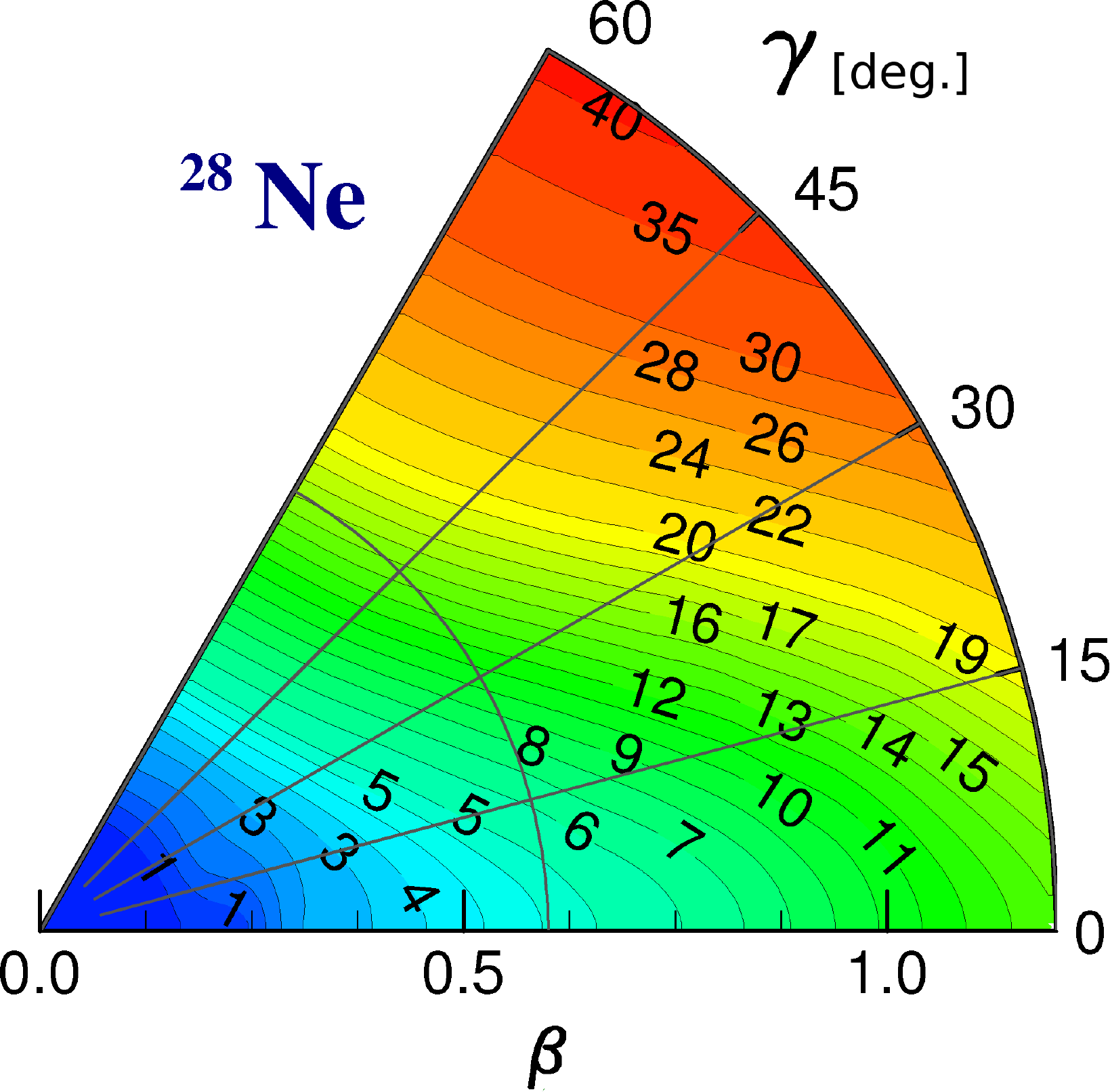}}  \hfill %
\subcaptionbox{$^{26}$Ne\label{26Ne_triax}}%
{\includegraphics[width=.49\columnwidth] {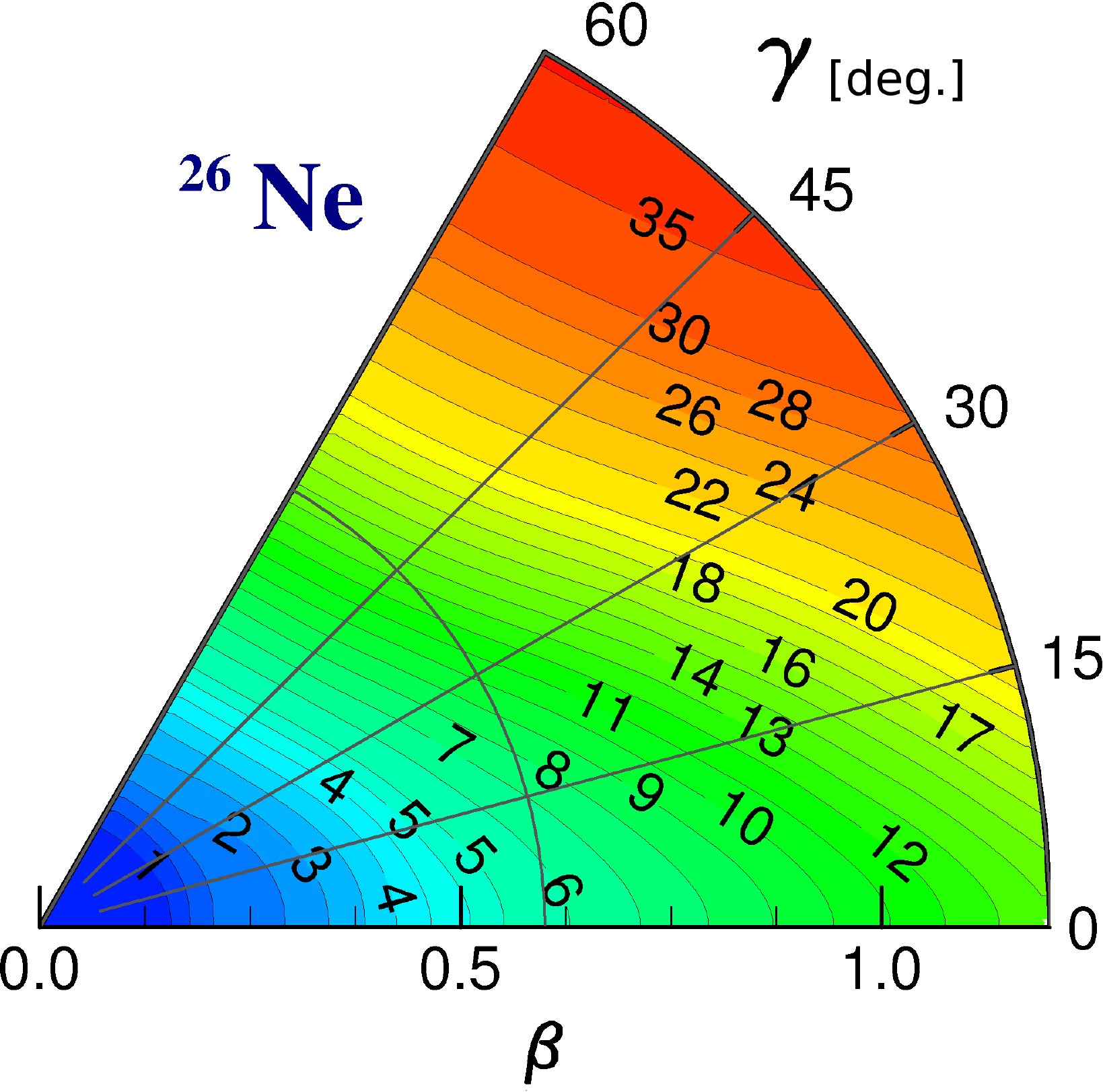}} \hfill %
\subcaptionbox{$^{24}$Ne\label{24Ne_triax}}%
{\includegraphics[width=.49\columnwidth] {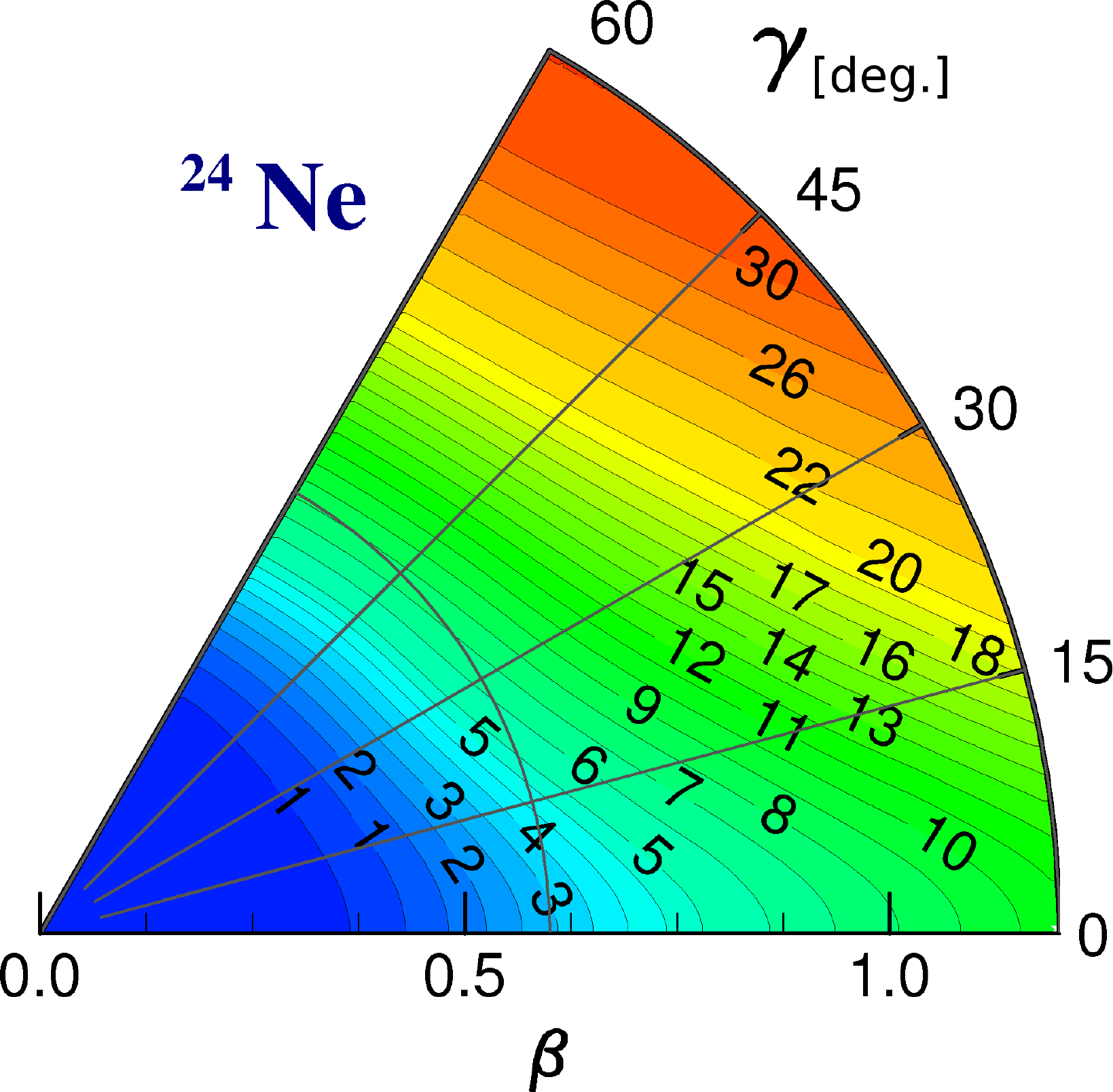}} \hfill %
\subcaptionbox{$^{22}$Ne\label{22Ne_triax}}%
[.49\linewidth]{\includegraphics[width=.49\columnwidth] {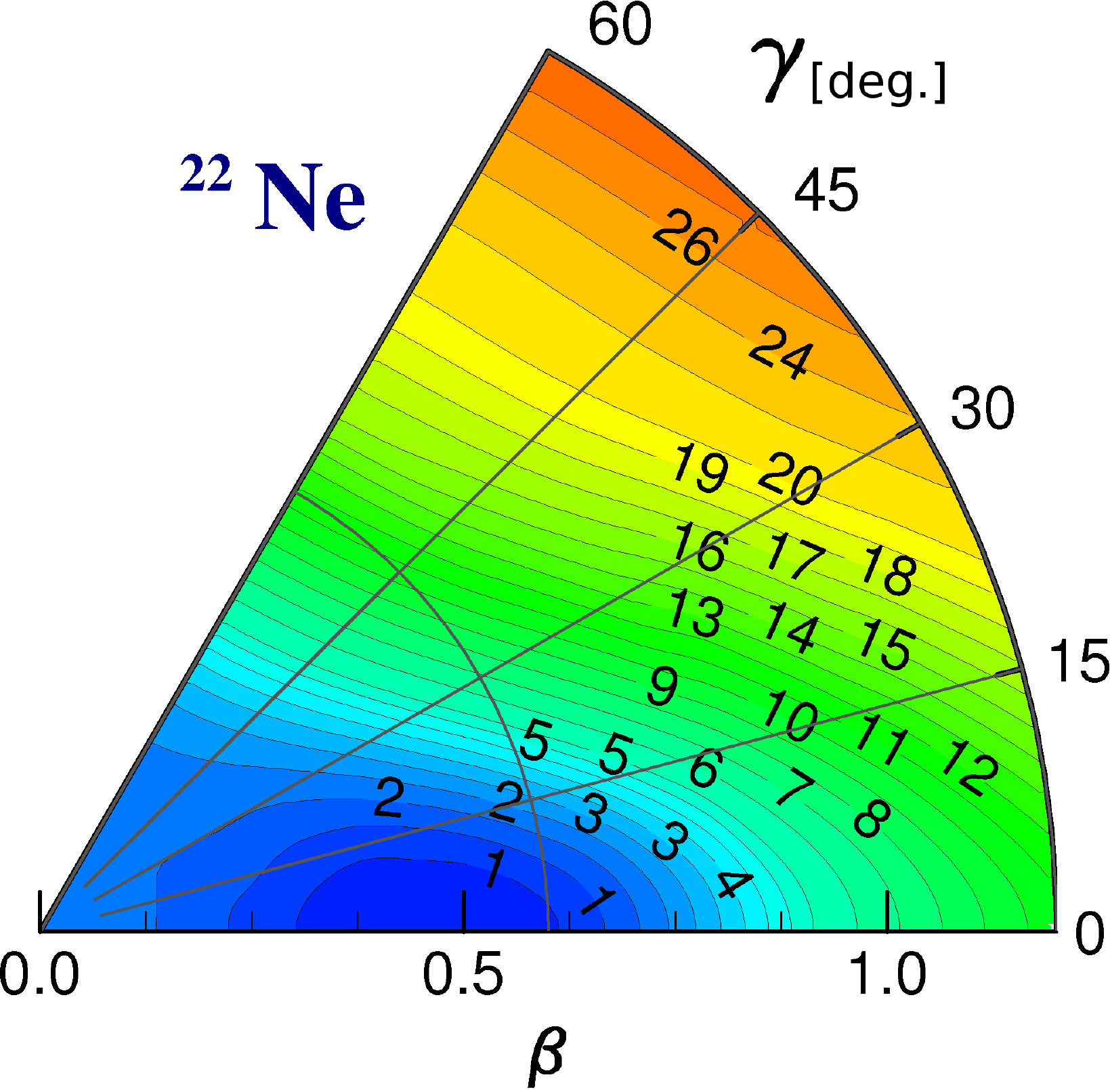}} \hfill %
\subcaptionbox{$^{20}$Ne\label{20Ne_triax}}%
[.49\linewidth]{\includegraphics[width=.49\columnwidth] {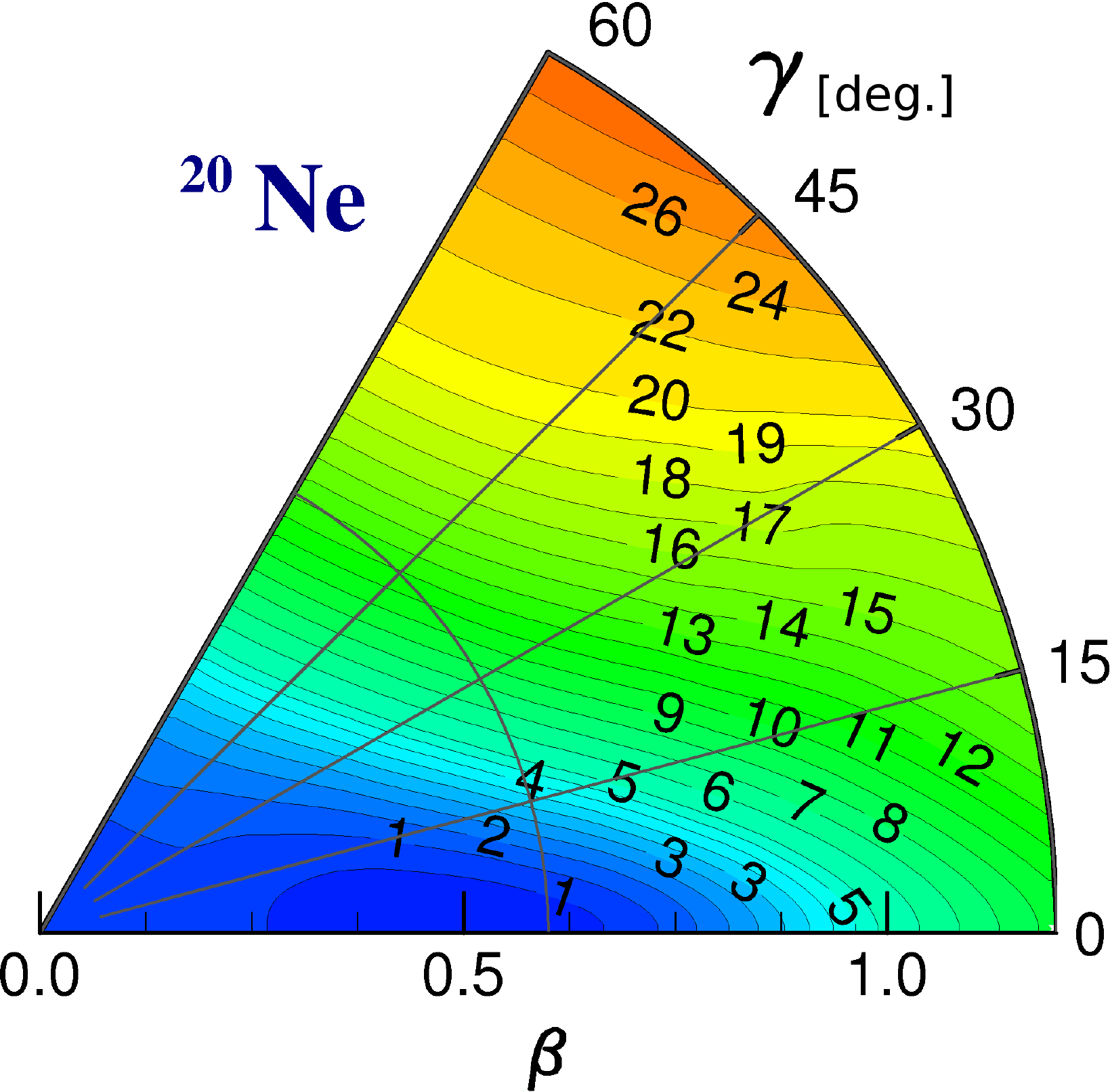}} \hfill %
\caption{(Color Online) HFB Potential Energy Surface (PES) of the Neon isotopes. The red curve includes zero-point energy corrections.}
 \label{Ne_triax} 
\end{figure}

\begin{figure}[h!]
\centering
\subcaptionbox{$^{24}$Mg\label{24Mg_triax}}%
{\includegraphics[width=.49\columnwidth] {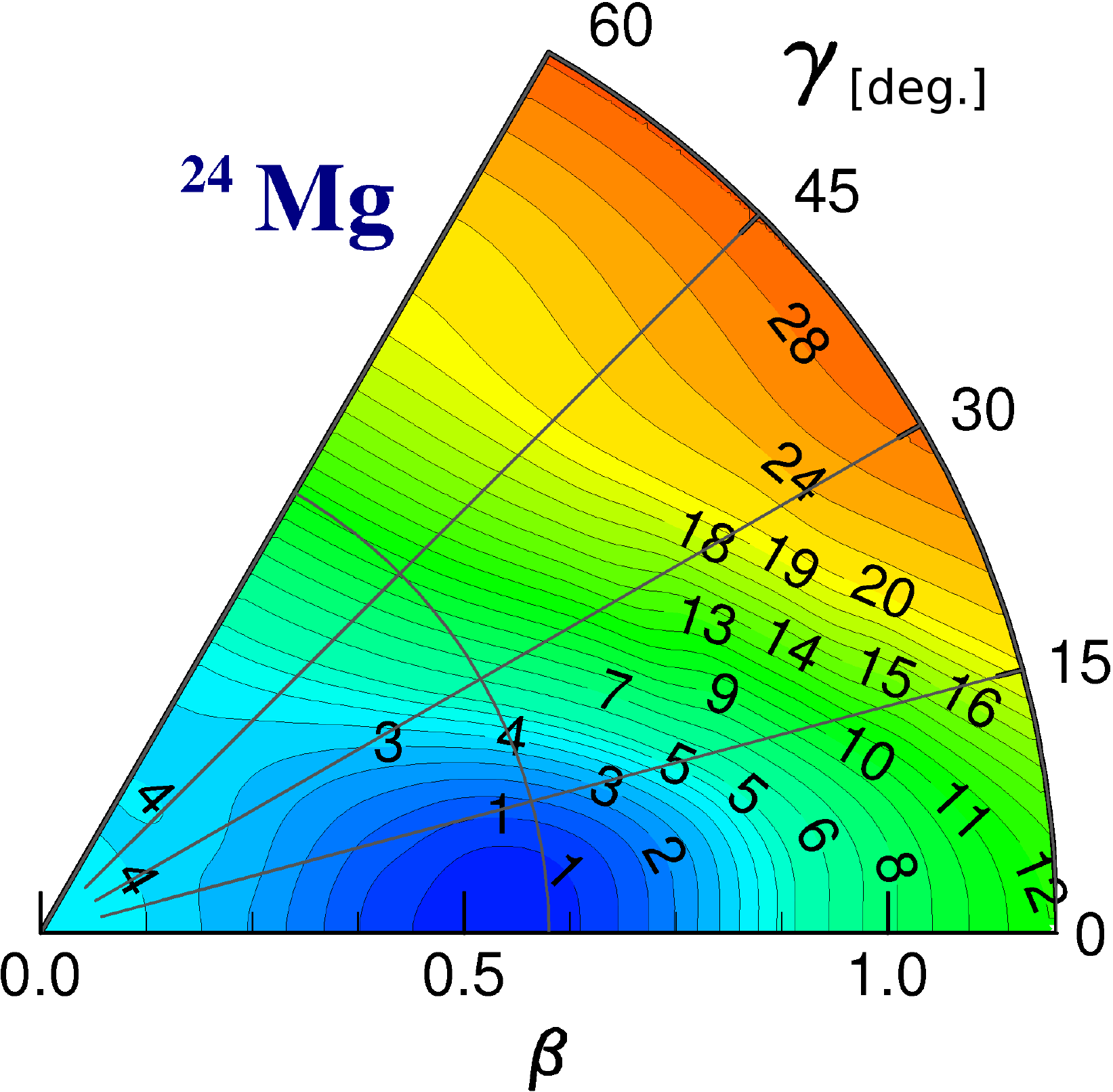}} \hfill %
\subcaptionbox{$^{28}$Si\label{28Si_triax}}%
{\includegraphics[width=.49\columnwidth] {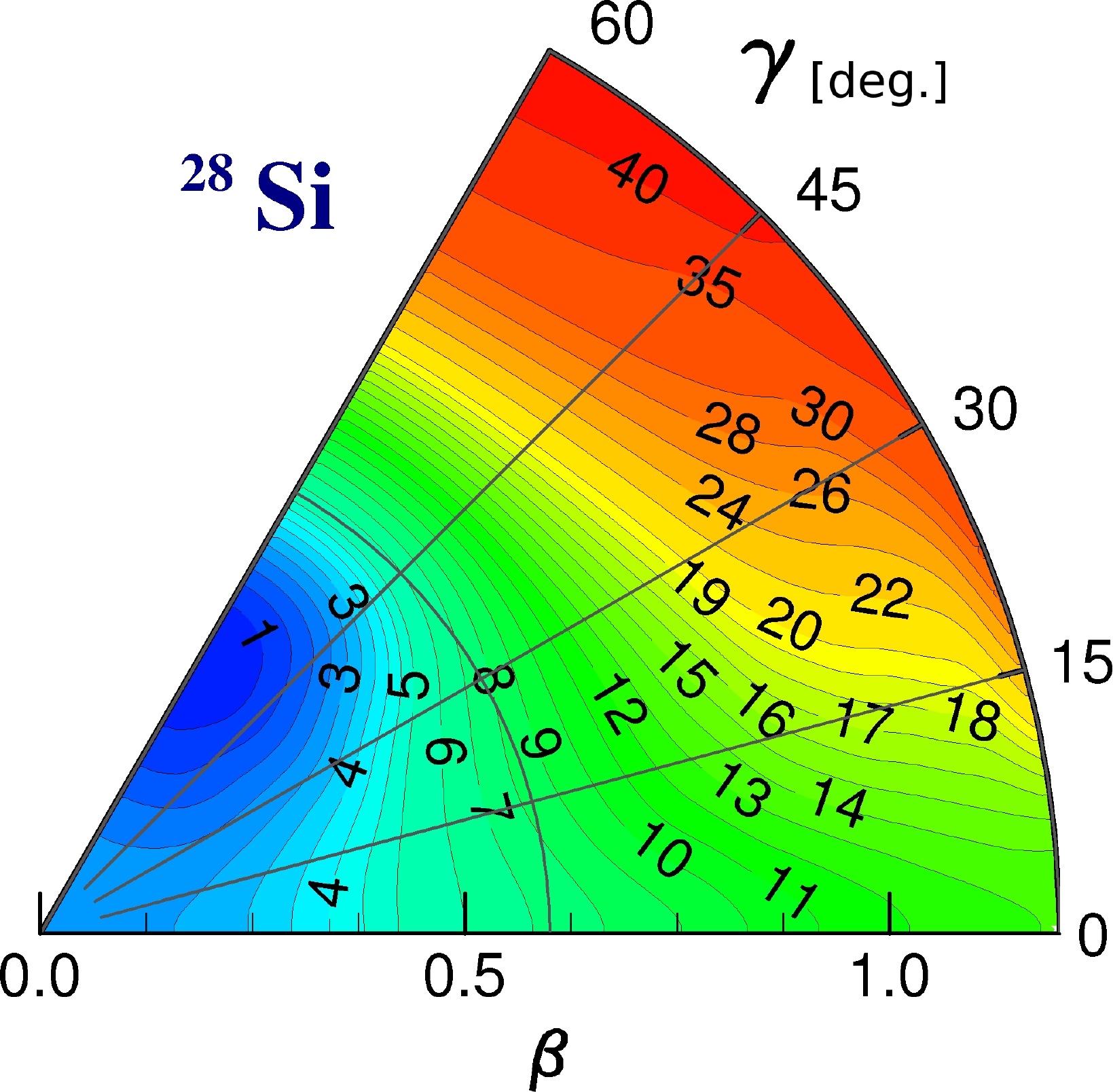}} \hfill %
\subcaptionbox{$^{32}$S\label{32S_triax}}%
{\includegraphics[width=.49\columnwidth] {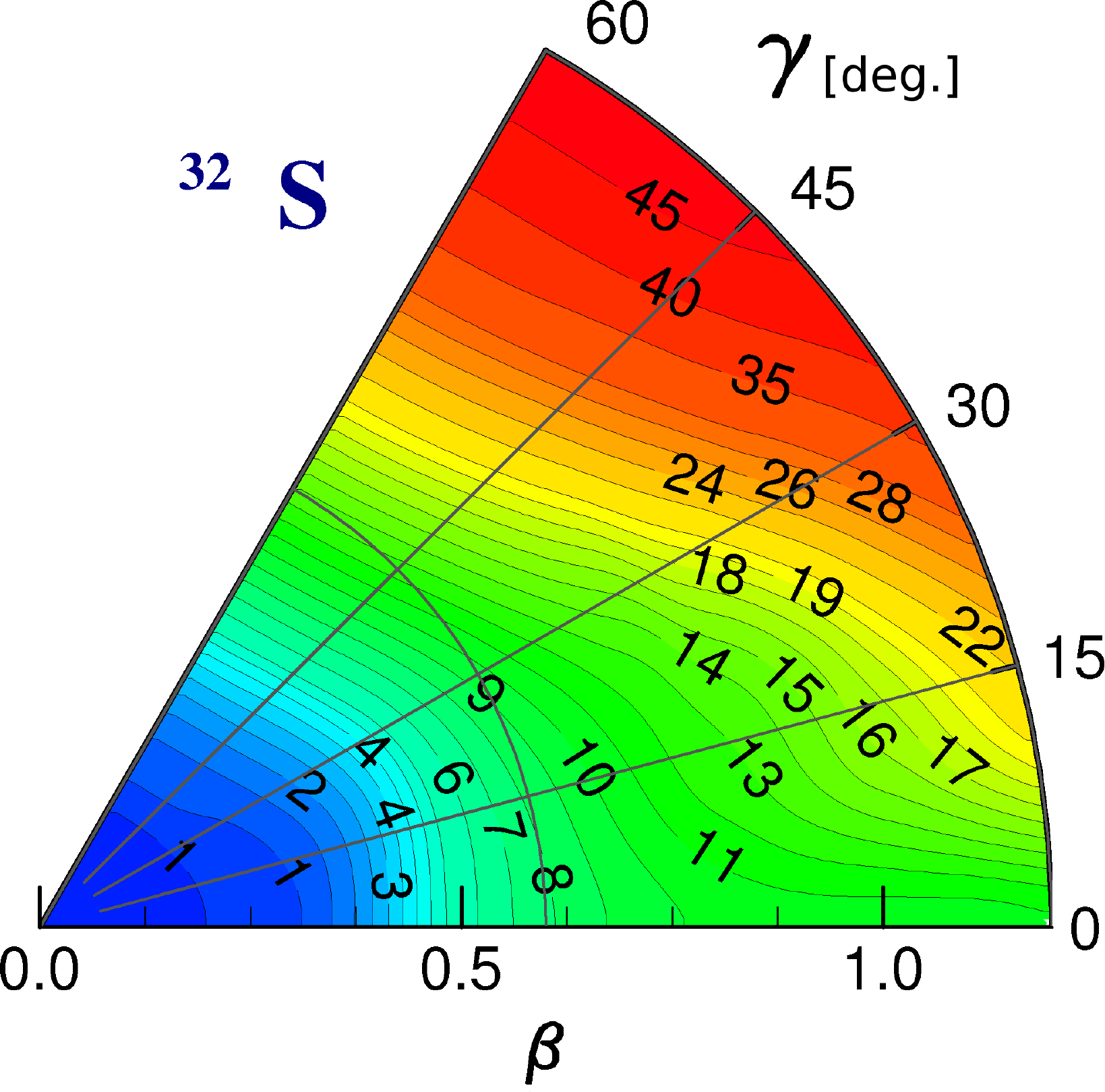}} \hfill %
\caption{(Color Online) HFB PES of $^{24}$Mg (top left), $^{28}$Si (top right) and $^{32}$S (bottom).}
\label{24Mg_28Si_32S_triax} 
\end{figure}

The nuclei of the $sd$-shell exhibit diverse correlation properties. In particular their deformation profile can be very different. As an illustration, we show in Fig. (\ref{Ne_triax}) triaxial potential-energy surfaces (PES) of Neon isotopes, obtained within the Hartree-Fock-Bogoliubov (HFB) approach using the same D1S Gogny interaction. One observes a transition of shape along this isotopic chain, as the number of neutrons $N$ decreases. The heaviest isotopes appear spherical while the lightest ones are predicted oblate ($^{24}$Ne) or prolate ($^{20-22}$Ne). We also display in Fig. (\ref{24Mg_28Si_32S_triax}) the PES of three other nuclei of the $sd$-shell: $^{24}$Mg, $^{28}$Si and $^{32}$S. The $^{24}$Mg and $^{28}$Si nuclei exhibit a large axial deformation characterized by $\beta \sim 0.6$ and $\beta  \sim - 0.4$ respectively. The $^{32}$S nucleus is predicted spherical and soft in its ground state, and exhibits a super-deformed second minimum at $\beta \sim 1.2$. 

\subsubsection{Correlation matrices and source terms $G[\sigma]$}
\begin{figure}
 \centering
 \includegraphics[width=\columnwidth] {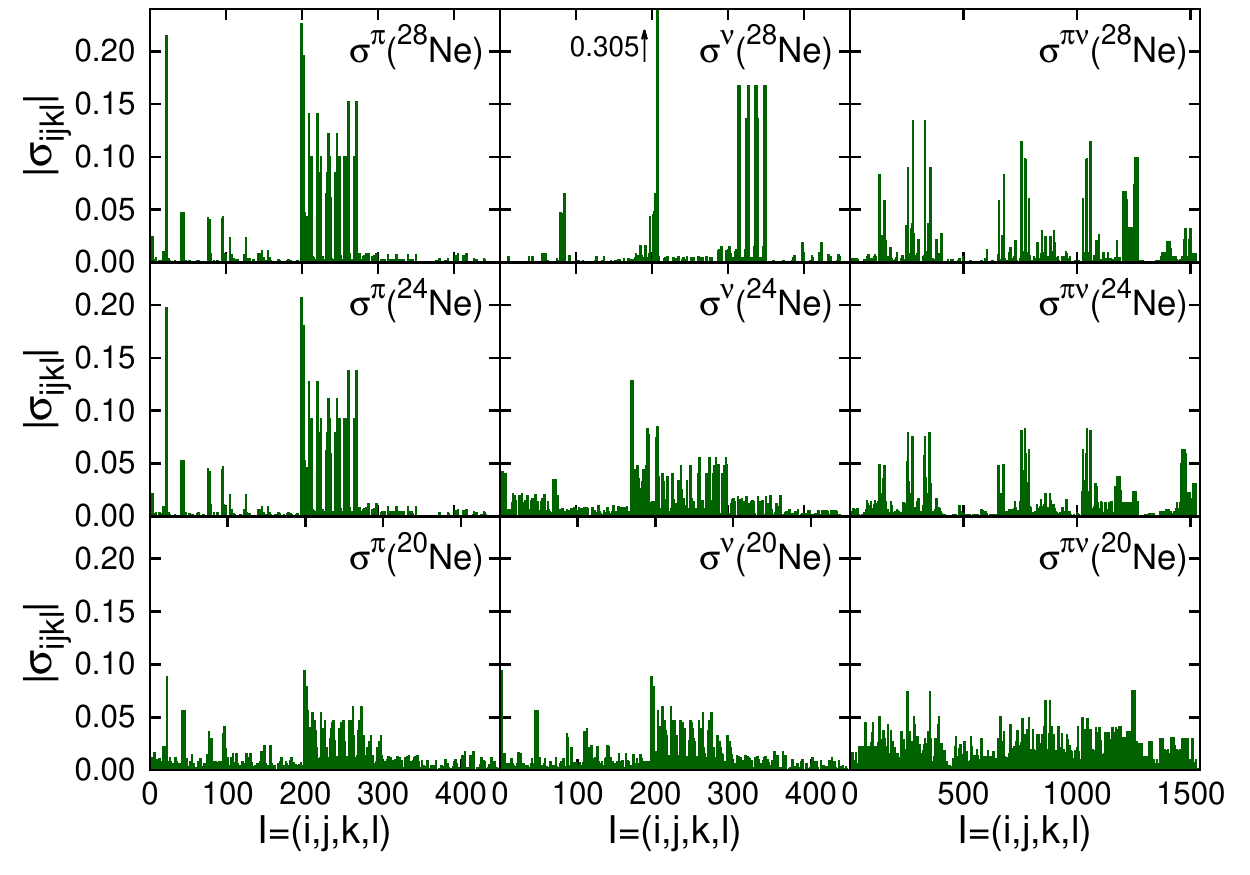}
 \caption{(Color Online) Proton correlations $\sigma^\pi$ (left), neutron correlations $\sigma^\nu$ (center) and proton-neutron correlations $\sigma^{\pi\nu}$ (right), for $^{28}$Ne, $^{24}$Ne and $^{20}$Ne. They are calculated when full consistency between orbitals and correlations is reached (level 3 of the method).}
 \label{f:sigma1}
 \end{figure}

\begin{figure}
 \centering
 \includegraphics[width=\columnwidth] {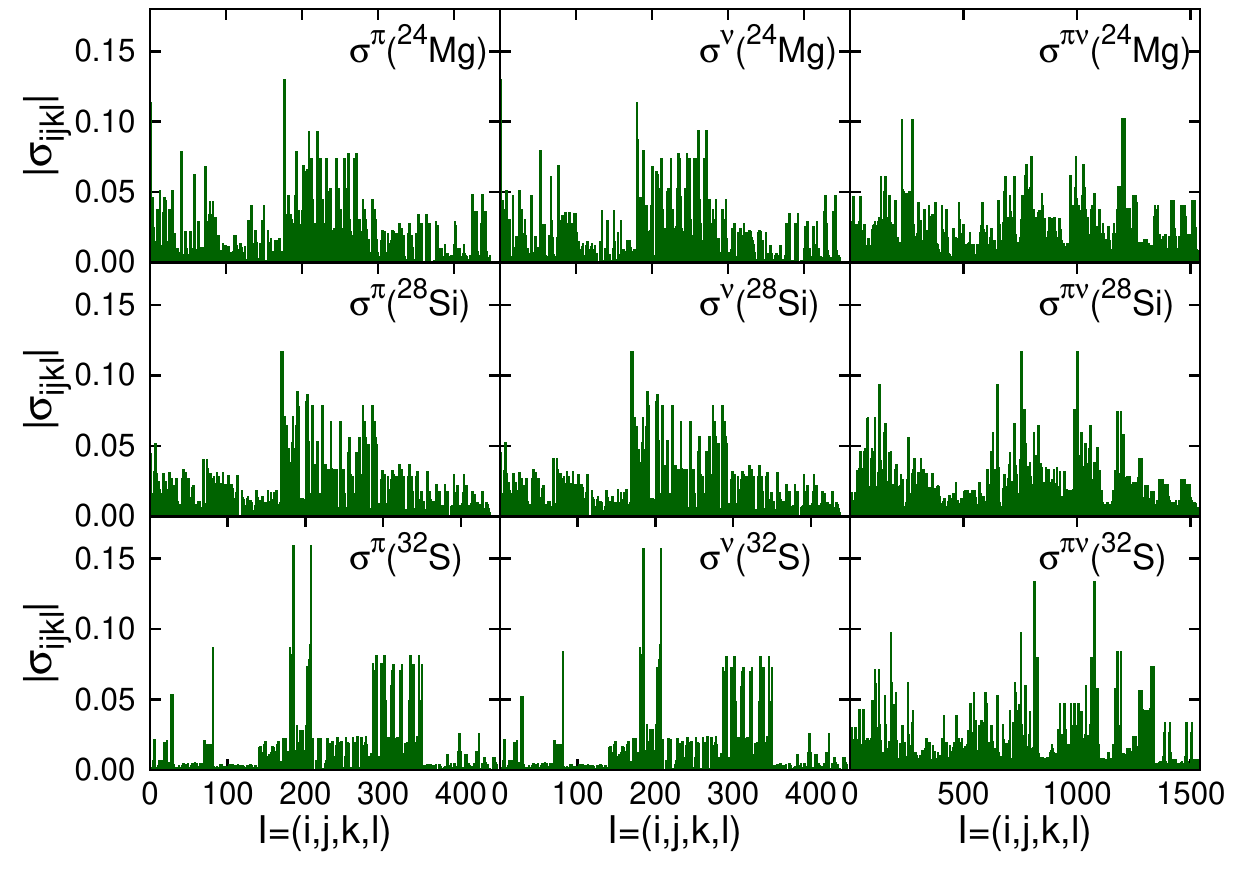}
 \caption{(Color Online) Proton correlations $\sigma^\pi$ (left), neutron correlations $\sigma^\nu$ (center) and proton-neutron correlations $\sigma^{\pi\nu}$ (right), for $^{24}$Mg, $^{28}$Si and $^{32}$S. They are calculated when full consistency between orbitals and correlations is reached (level 3 of the method).}
 \label{f:sigma2}
 \end{figure}

As MPMH explicitly preserves spherical symmetry, the information about the deformation of a nucleus should be contained to some extent in the correlation matrix $\sigma$. 
To illustrate this, we show in Fig. (\ref{f:sigma1}) the calculated correlations for three Neon isotopes obtained when full self-consistency of orbitals and correlations is reached (level 3 of the method). The linear index $I$ represents a quadruplet of single-particle states $(i,j,k,l)$.
Correlations between neutrons are drastically modified with the neutron number. Proton correlations appear to slowly increase when $N$ decreases, as $\sigma_\pi$ appears more fragmented. This behavior is likely to be caused by the proton-neutron interaction. Indeed, we note the importance of correlations between both isospin which are generally enhanced in nuclei with equal numbers of protons and neutrons, such as $^{20}$Ne, since the two types of nucleons occupy the same orbitals and highly overlap spatially. This effect is also illustrated in Fig. (\ref{f:sigma2}) where we display the correlation content of the three other $N=Z$ notable nuclei. We also note the strength of pure neutron and proton correlations in $^{28}$Si and $^{24}$Mg, compared to other nuclei under study. \\ \\

Using these correlation matrices to calculate the source term $G[\sigma]$ of the orbital equation (\ref{e:eq2_Gogny}), we obtain the values shown in Table \ref{source_sd}. Since $G[\sigma]$ couples single-particle states in the valence space to orbitals in the rest of the basis ---characterized by the same angular momentum $j$ and parity $\pi$--- we obtain a total of 10 couplings for each isospin.

\begin{table}
\begin{subtable}[]{1 \columnwidth}
\centering
 \begin{tabular}{ccccccccccccc}
 \hline
($j_k,j_l$)                        &&    $^{28}$Ne    &&     $^{24}$Ne                    &&     $^{20}$Ne                   && $^{24}$Mg &&      $^{28}$Si   && $^{32}$S   \\
\hline
\hline
$(0d_{\frac{5}{2}},1d_{\frac{5}{2}})$         && 0.0361                &&  0.00325                               && 0.0926                              & &   0.238         && 0.161   && 0.0590  \\ 
\hline
$(0d_{\frac{5}{2}},2d_{\frac{5}{2}})$         && 0.226                 && 0.422                                  && 0.576                                 &&   0.539          && 0.387   && 0.156  \\
\hline
$(0d_{\frac{5}{2}},3d_{\frac{5}{2}})$         && 0.297                  && 0.300                                  && 0.431                                &&  0.591           && 0.567   && 0.243\\
\hline
\hline
$(1s,0s)$                                                           && {\bf 0.257}          && {\bf 0.334 }            && {\bf 1.40  }                      && {\bf 2.02 }           && {\bf 2.08}    && {\bf 1.22}\\
\hline
$(1s,2s)$                                                           && 0.0561                 && 0.0705                                && 0.433                                 && 0.580            && 0.694    && 0.543\\
\hline
$(1s,3s)$                                                          && 0.0262                 && 0.0203                                && 0.152                                 && 0.0586            && 0.0245   &&  0.131\\
\hline
$(1s,4s)$                                                          && 0.0444                  && 0.0495                                 && 0.298                                 && 0.431            && 0.519   && 0.259\\
\hline
\hline
$(0d_{\frac{3}{2}},1d_{\frac{3}{2}})$       && 0.0264                  && 0.0470                                 && 0.191                                  && 0.256          && 0.274   && 0.214\\
\hline
$(0d_{\frac{3}{2}},2d_{\frac{3}{2}})$       && 0.0136                  && 0.0190                                 && 0.263                                  && 0.211           && 0.163  && 0.296\\
\hline
$(0d_{\frac{3}{2}},3d_{\frac{3}{2}})$       && 0.0271                  && 0.0702                                 && 0.311                                  && 0.395           && 0.411   && 0.528\\
\hline   
\end{tabular} 
\subcaption{Protons couplings $|G^\pi_{j_k,j_l}[\sigma]|$ (in MeV). \\ $\;$}
\end{subtable} 

\begin{subtable}[]{1 \columnwidth}
\centering
 \begin{tabular}{ccccccccccccc}
 \hline
($j_k,j_l$)                         &&    $^{28}$Ne    &&     $^{24}$Ne                    &&     $^{20}$Ne                   && $^{24}$Mg &&      $^{28}$Si && $^{32}$S   \\
\hline
\hline
$(0d_{\frac{5}{2}},1d_{\frac{5}{2}})$         && 0.0317                  &&  0.0977                               && 0.0368                              & &   0.159         && 0.0918      &&0.0251\\
\hline
$(0d_{\frac{5}{2}},2d_{\frac{5}{2}})$         && 0.0869               && 0.172                                  && 0.600                             &&   0.578          && 0.422        && 0.172\\ 
\hline
$(0d_{\frac{5}{2}},3d_{\frac{5}{2}})$         && 0.0152                  && 0.303                                 && 0.420                              &&  0.579           && 0.553      && 0.233\\
\hline
\hline
$(1s,0s)$                                                           && {\bf 0.289}         && {\bf 1.43   }          && {\bf 1.39    }                            && {\bf 2.02}            && {\bf 2.07}   && {\bf 1.20}\\
\hline
$(1s,2s)$                                                           && 0.122                  && 0.448                                   && 0.393                                && 0.531           && 0.619    && 0.478\\
\hline
$(1s,3s)$                                                          && 0.0505                  && 0.00147                                && 0.182                                 && 0.104            && 0.0876  &&  0.0706\\
\hline
$(1s,4s)$                                                          && 0.0450                  && 0.335                                 && 0.292                                && 0.429           && 0.513 &&  0.254\\
\hline
\hline
$(0d_{\frac{3}{2}},1d_{\frac{3}{2}})$       && 0.284                  && 0.154                                 && 0.168                                 && 0.226           && 0.237     && 0.151\\
\hline
$(0d_{\frac{3}{2}},2d_{\frac{3}{2}})$       && 0.0395               && 0.102                                 && 0.274                                 && 0.230           && 0.189  && 0.338\\
\hline
$(0d_{\frac{3}{2}},3d_{\frac{3}{2}})$       && 0.261                && 0.245                                 && 0.304                                 && 0.389           && 0.406   && 0.520\\
\hline   
\end{tabular} 
\subcaption{Neutrons couplings $|G^{\nu}_{j_k,j_l}[\sigma]|$ (in MeV).}
\end{subtable}
\caption{Proton (top) and neutron (bottom) source terms $|G^{\tau}_{j_k,j_l}[\sigma]|$ ($\tau=\pi,\nu$) (in MeV) between the sub-shells $j_k$ of the valence space and other sub-shells $j_l$ outside of the model space.}
\label{source_sd}
\end{table}

\noindent First we note that some values of the source term are not negligible. In particular, we observe a systematic high value of the coupling between the $1s$ and the $0s$ shells (shown in bold) compared to other couplings. They are $> 1$ MeV in the nuclei described as the most deformed by mean-field calculations, and reach $\sim 2$ MeV in $^{24}$Mg and $^{28}$Si. Dynamical correlations related to the source term therefore seem to act toward a strong mixing of these shells. The $1s$ and $2s$ shells, as well as some $d_{5/2}$ and $d_{3/2}$ sub-shells, also appear importantly coupled.
\\ \\
Regarding the Neon isotopic chain, the proton source term generally increases as the neutron number $N$ decreases, in accordance with the correlation matrices from Fig. (\ref{f:sigma1}). The behavior of the neutron source term, is less clear. For instance, these couplings appear globally quite strong in $^{24}$Ne compared to the proton ones.
\\ \\ 
Finally, let us look more carefully at the evolution of the couplings $G_{kl}[\sigma]$ with the single-particle energy difference $\Delta \varepsilon = |\varepsilon_k - \varepsilon_l|$. In an intuitive way using perturbative arguments, one would expect the values of $G[\sigma]$ to decrease as $\Delta \varepsilon$ increases. However, this behavior is not clear from the calculated values. We remind that these calculations are realized using a spherical mean-field. Thus if correlations associated to deformation are strong, important couplings to high energy orbitals can appear. 

\subsubsection{Correlation energies} 
Table \ref{Ecorr_sd} displays the correlation energy $E_{corr} = E_{HF} - E_0$ of the selected benchmark nuclei, defined as the difference between the energy $E_0$ of the correlated ground state and the energy $E_{HF}$ of the spherical Hartree-Fock ground state. The results are shown at the three stages \ref{level1}, \ref{level2} and \ref{level3} of implementation of the method, described at the beginning of the section. As expected from the values of $\sigma$ and $G[\sigma]$, the correlation energy of the Neon isotopes increases considerably as the neutron number decreases. At the non self-consistent stage (column 2), among the presented nuclei, $^{24}$Mg appears as the most correlated one.
Introducing the correlated density in the interaction and the rearrangement terms that account for medium effects (column 3), leads to a gain of correlation energy that is less than 1 MeV in all nuclei.
On top of that, the renormalization of single-particle states has a considerable effect (column 4). The most significant one appears in $^{28}$Si, for which optimizing the orbitals brings an additional 1.83 MeV. $E_{corr}$ is increased by 1.76, 1.22, 1.18 and 0.98 MeV in $^{20}$Ne, $^{22}$Ne, $^{32}$S and $^{24}$Mg respectively. The effect is weaker in the other nuclei under study.

\begin{table}
 \centering
\begin{tabular}{cccc}
\hline     
                     &  Level 1:                                  &    Level 2:                                  & Level 3:  \\
                     & Eq. (\ref{e:eq1_Gogny}) with        &   Full                                          &  Full \\
                     & $\rho = \rho_{HF}$, $\sigma=0. \; $         &  Eq. (\ref{e:eq1_Gogny}).  $\;$  & Eqs. (\ref{e:eq1_Gogny})$\&$ (\ref{e:eq2_Gogny}). \\
\hline  
\hline
 $^{28}$Ne &    1.15      &   1.28 &   1.58                      \\           
\hline
$^{26}$Ne &    0.41     &  0.88   &  1.55                      \\    
\hline
$^{24}$Ne &   5.75      &6.23     & 6.98              \\        
\hline
 $^{22}$Ne & 10.48     &10.90   &     12.12       \\       
\hline
$^{20}$Ne &   10.93    &11.54       & 13.30      \\     
\hline
\hline
$^{24}$Mg & 14.24          &     15.06                   &   16.04                   \\
\hline
$^{28}$Si &   5.89           & 6.25                           &   8.08                    \\
\hline
$^{32}$S &   3.37           & 4.58                           &   5.76                    \\
\hline
\end{tabular}
\caption{Correlation energy $E_{corr}=E_{HF}-E_{0}$  for the Neon isotopes and other benchmark nuclei, in MeV.}
\label{Ecorr_sd}
\end{table}

\subsubsection{Composition of the ground-state wave function} \label{sec:wf_compo}
In order to obtain a more precise description of the correlations incorporated by MPMH in the ground state, it is necessary to analyze the composition of the wave function in terms of the different configurations. We show in Table \ref{t:gs_compo} the main components of the wave function obtained again at the three levels of implementation of the MPMH method. When the orbitals are not modified (levels \ref{level1} and \ref{level2}), we show the weights of the most important configurations built on pure Hartree-Fock single-particle states. Conversely, when full self-consistency is applied (level \ref{level3}), the many-body Slater determinants are constructed on optimized orbitals.

\begin{table*}
 \centering

\begin{tabular}{cccccc}
\hline     
                    &                                    &  Level 1:                                  &    Level 2:                                  & Level 3:                                                                                      &Level 3:\\
   Nucleus    &   Configuration           & Eq. (\ref{e:eq1_Gogny}) with        &   Full                                          &  Full 										    &  Full \\
                     &                                   & $\rho = \rho_{HF}$, $\sigma=0. \; $         &  Eq. (\ref{e:eq1_Gogny}).  $\;$  & Eqs. (\ref{e:eq1_Gogny})$\&$ (\ref{e:eq2_Gogny}). & Eqs. (\ref{e:eq1_Gogny})$\&$ (\ref{e:eq2_Gogny}). \\
                     &                                   & (HF orbitals)                                        & (HF orbitals)                                    & (SC orbitals)                         					 & (HF orbitals) \\
\hline  
\hline
\multirow{3}{*}{$^{28}$Ne} &   {\bf 0p-0h }                                                       & {\bf 86.24 }  & {\bf 84.11}       &  {\bf  83.63  }               &  \\
\cline{2-6}
                                                   &(1p-1h)$_\pi$ ($0d_{5/2}\rightarrow 1s$)            &  3.49    & 3.19                  &   2.85                   & \\
\cline{2-6}
                                                   &(1p-1h)$_\nu$ ($1s\rightarrow0d_{3/2}$)           & 3.26       &  4.04             &   4.50                  &  \\
\hline
\hline
\multirow{4}{*}{$^{26}$Ne} &      {\bf 0p-0h }                                     & {\bf 77.11 }   & {\bf 70.88}       &   {\bf 69.62   }               &   {\bf 61.50 } \\
\cline{2-6}
                                                   &       (1p-1h)$_\nu$ ($1s\rightarrow0d_{3/2}$)               &   6.02     & 7.28      &   7.59                &    \\
\cline{2-6}
&(1p-1h)$_\nu$ ($0d_{5/2}\rightarrow 0d_{3/2}$)  &  4.45    &4.87       &   5.20                   & \\
\cline{2-6}
&(2p-2h)$_{\pi\nu}$ ($1s^\nu \otimes 0d_{5/2}^\pi \rightarrow 0d_{3/2}^\nu \otimes 1s^\pi$)  &  2.27  & 2.75         &   2.38                  &  \\
\hline
\hline
\multirow{4}{*}{$^{24}$Ne} &      {\bf 0p-0h }                                     & {\bf 56.51 }   & {\bf 53.45}       &   {\bf 49.41   }              &  \\
\cline{2-6}
                                                   &       (1p-1h)$_\nu$ ($0d_{5/2}\rightarrow1s$)               &   17.81     & 17.48      &   17.81                  &  \\
\cline{2-6}
                                                  &(1p-1h)$_\nu$ ($0d_{5/2}\rightarrow 0d_{3/2}$)  &  5.60    & 6.27       &   6.34                   & \\
\cline{2-6}
                                                   &       (2p-2h)$_\nu$ ($0d_{5/2}\rightarrow1s$)               &   6.17     & 6.56      &   7.54                  &  \\
\hline
\hline
\multirow{8}{*}{$^{20}$Ne} &{\bf 0p-0h }                                                                     & {\bf 45.36} & {\bf 43.05}           &   {\bf 33.05   }             &   \\
\cline{2-6}
&(2p-2h)$_{\pi\nu}$ ($0d_{5/2}^\pi \otimes 0d_{5/2}^\nu \rightarrow 1s^\pi \otimes 1s^\nu$)  &  8.15     & 6.80      &   8.86                &    \\
\cline{2-6}
&(1p-1h)$_\pi$ ($0d_{5/2}\rightarrow0d_{3/2}$)               &   6.91    &8.26       &   8.65                &   \\
\cline{2-6}
&(1p-1h)$_\nu$ ($0d_{5/2}\rightarrow0d_{3/2}$)               &   6.94    & 8.30       &   8.58                &   \\
\cline{2-6}
&(1p-1h)$_\pi$ ($0d_{5/2}\rightarrow 1s$)  &  5.29     &4.44      &   5.08                 &   \\
\cline{2-6}
&(1p-1h)$_\nu$ ($0d_{5/2}\rightarrow 1s$)  &  5.40    &4.50       &   5.13                 &   \\
\cline{2-6}
&(2p-2h)$_\pi$ ($0d_{5/2} \otimes 0d_{5/2}  \rightarrow 1s \otimes 1s $)  &  2.32  & 1.89         &   2.46                &    \\
\cline{2-6}
&(2p-2h)$_\nu$ ($0d_{5/2} \otimes 0d_{5/2} \rightarrow 1s \otimes 1s$)  &  2.44     &1.95      &   2.52                 &   \\
\hline
\hline
\multirow{8}{*}{$^{24}$Mg} &{\bf 0p-0h }                                                  & {\bf 34.63}    & {\bf 32.45}       &   {\bf 23.82 }                &  \\
\cline{2-6}
&(1p-1h)$_\nu$ ($0d_{5/2}\rightarrow1s$)               &   8.31       & 7.13    &   6.49               &   \\
\cline{2-6}
&(1p-1h)$_\pi$ ($0d_{5/2}\rightarrow1s$)               &   8.08      & 6.98     &   6.37               &   \\
\cline{2-6}
&(2p-2h)$_{\pi\nu}$ ($0d_{5/2}^\pi \otimes 0d_{5/2}^\nu \rightarrow 1s^\pi \otimes 1s^\nu$)  &  5.30  & 4.32         &   5.16               &     \\
\cline{2-6}
&(1p-1h)$_\nu$ ($0d_{5/2}\rightarrow0d_{3/2}$)               &   4.43     & 4.83      &   3.94               &   \\
\cline{2-6}
&(1p-1h)$_\pi$ ($0d_{5/2}\rightarrow0d_{3/2}$)               &   4.37     & 4.83      &   3.96                &  \\
\cline{2-6}
&(2p-2h)$_\nu$ ($0d_{5/2} \otimes 0d_{5/2}  \rightarrow 1s \otimes 1s $)  &  2.24   & 1.83        &   2.26               &     \\
\cline{2-6}
&(2p-2h)$_\pi$ ($0d_{5/2} \otimes 0d_{5/2} \rightarrow 1s \otimes 1s$)  &  2.12     & 1.76      &   2.17                 &   \\
\hline
\hline
\multirow{4}{*}{$^{28}$Si} &{\bf 0p-0h   }                                                         & {\bf 26.02}     & {\bf 38.68}      &   {\bf 17.80 }               &    {\bf 16.99} \\
\cline{2-6}
&(2p-2h)$_{\pi\nu}$ ($0d_{5/2}^\pi \otimes 0d_{5/2}^\nu \rightarrow 1s^\pi \otimes 1s^\nu$)  &  12.36     & 8.11      &   8.98                &    \\
\cline{2-6}
&(2p-2h)$_\nu$ ($0d_{5/2} \otimes 0d_{5/2}  \rightarrow 1s \otimes 1s $)  &  5.03     & 3.28      &   3.66                 &   \\
\cline{2-6}
&(2p-2h)$_\pi$ ($0d_{5/2} \otimes 0d_{5/2} \rightarrow 1s \otimes 1s$)  &  4.87     & 3.17      &   3.54                  &  \\
\hline
\hline
\multirow{4}{*}{$^{32}$S} &{\bf 0p-0h }                                                                     & {\bf 60.30}     & {\bf 47.23}      &   {\bf 26.20}            &     {\bf 24.26}   \\
\cline{2-6}
&(2p-2h)$_{\pi\nu}$   ($1s^\pi \otimes 1s^\nu \rightarrow 0d_{3/2}^\pi \otimes 0d_{3/2}^\nu$)  &  8.36      & 9.31     &   11.20                &    \\
\cline{2-6}
&(2p-2h)$_\nu$ ($  1s \otimes 1s \rightarrow 0d_{3/2} \otimes 0d_{3/2}$)  &  3.80      & 4.38     &   5.47                &    \\
\cline{2-6}
&(2p-2h)$_\pi$ ($ 1s \otimes 1s \rightarrow  0d_{3/2} \otimes 0d_{3/2}$)  &  4.11    & 4.80       &   5.87               &     \\
\hline
\end{tabular}
\caption{Main components of the ground-state of different nuclei, expressed in percents ($\%$).}
\label{t:gs_compo}
\end{table*}

\noindent At the non self-consistent level 1 (column 3), the Hartree-Fock 0p-0h state always appears as the major component, and absorbs most of the wave function in weakly correlated nuclei ($> 86 \%$ in $^{28}$Ne). The rest of the weight is distributed among many other configurations, mostly of 1p-1h and 2p-2h types. As already stated, configurations involving excitations of both protons and neutrons are more important in $N=Z$ nuclei, where their interaction is favored. For instance, the second main component in $^{28}$Si is a $(2p-2h)_{\pi\nu}=(1p-1h)_\pi \otimes (1p-1h)_\nu$ excitation, with a weight $> 12 \%$ while the Hartree-Fock states only embodies $\sim 26 \%$ of the wave function.
The medium effects simulated by the rearrangement terms (level 2, column 4) fragment the wave function by decreasing the 0p-0h component in most cases. Only $^{28}$Si makes exception with a Hartree-Fock component that increases from $\sim 26$ to $\sim 39 \%$. 
When full self-consistency is reached (level 3, column 5) the composition of the ground state wave function is again notably modified. The weight of the 0p-0h reference state is the most affected. In the Neon chain it decreases by a few percents in the heavier isotopes while the reduction is more important in the lighter ones. In particular the 0p-0h component is lowered from $\sim 43$ to $\sim 33 \%$ in $^{20}$Ne.  The wave function of $^{24}$Mg already appeared fragmented before self-consistency was introduced with a 0p-0h Hartree-Fock component of $\sim 35 \%$.  Still, self-consistency leads to an additional loss of $\sim 11 \%$ of this weight. In $^{28}$Si, the rise of the 0p-0h component due to the rearrangement terms is now counterbalanced by the transformation of the 
single-particle states, which brings it back down to only $\sim 18 \%$. Finally, the most striking effect is seen on  $^{32}$S, for which the reference state component decreases from $\sim60\%$ to $\sim 45 \%$ with rearrangement terms and to only $\sim 26 \%$ after orbital optimization. Looking at the other components, we note that this systematic reduction of the 0p-0h state in all the nuclei under study is not 
transferred to another particular configuration: the missing weight seems to be rather equally distributed over many components. 
\\
\\
Finally, it is always informative to analyze the evolution of the pure Hartree-Fock component, that is, the weight of the 0p-0h component built on \textit{non-optimized 
 Hartree-Fock orbitals} at the three stages 1, 2 and 3 of the MPMH method. To obtain this quantity after reaching self-consistency, we follow the procedure described in Ref. \cite{Robin}. As it is not trivial, we can only apply this procedure to nuclei with at least one closed sub-shell. We show the results in the sixth column of Table \ref{t:gs_compo}. 
Comparing them to the values shown in the fifth column, we note that the weight of the optimized reference state $\ket{0p-0h}$ is systematically slightly higher than the weight of the Hartree-Fock state $\ket{HF}$, illustrating the fact that this new reference state incorporates a higher physical content and minimizes the effect of correlations. This phenomenon is however in competition with the tendency to fragment the wave function and the evolution of the single-particle spectrum. Indeed, if gaps around the Fermi level are reduced, certain excitations may become more favorable and their weight might increase.

\subsubsection{Single-particle energies} 

\begin{figure}
\centering
\subcaptionbox{Proton SPE difference in $^{20}$Ne.}%
[\linewidth]{\includegraphics[width=0.8\columnwidth] {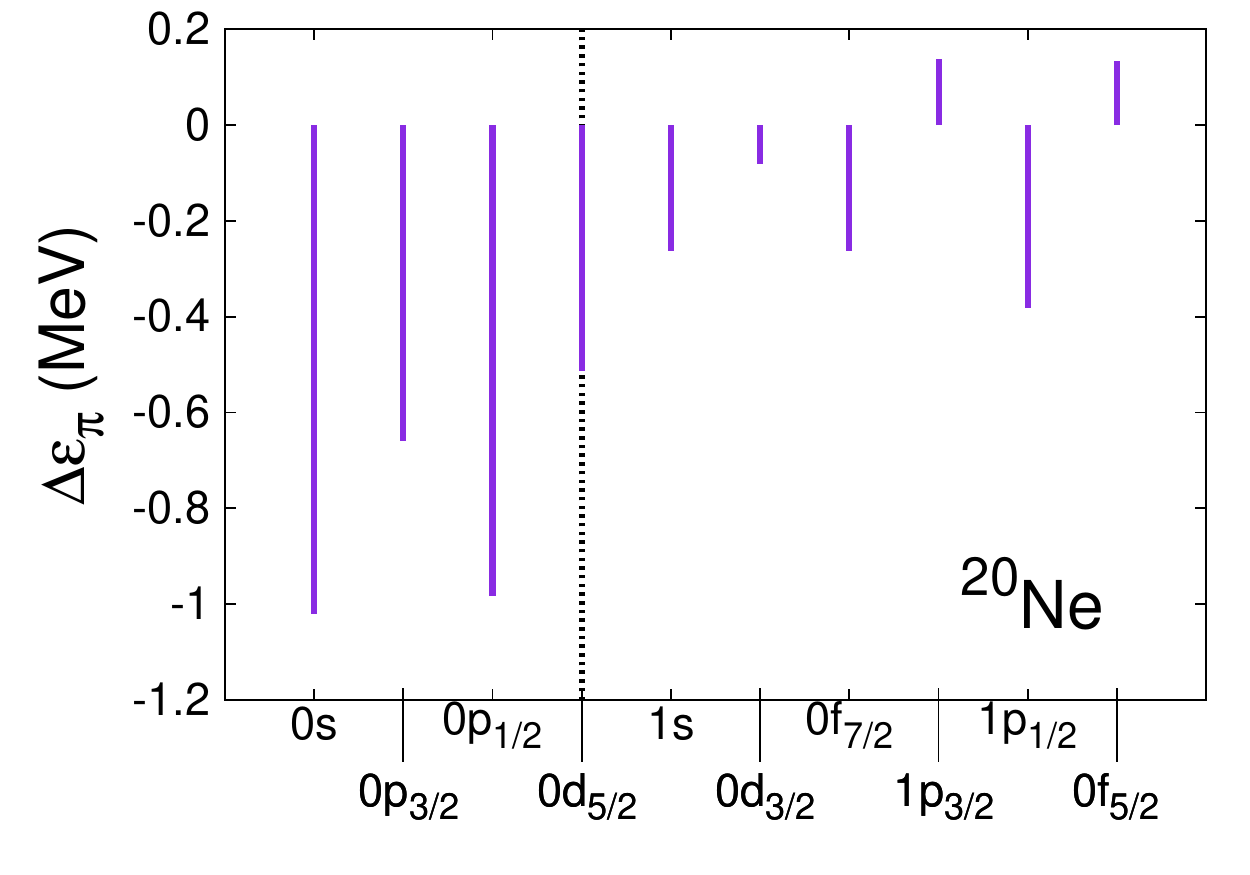}} \hfill %
\subcaptionbox{Neutron SPE difference in $^{20}$Ne.}%
[\linewidth]{\includegraphics[width=0.8\columnwidth] {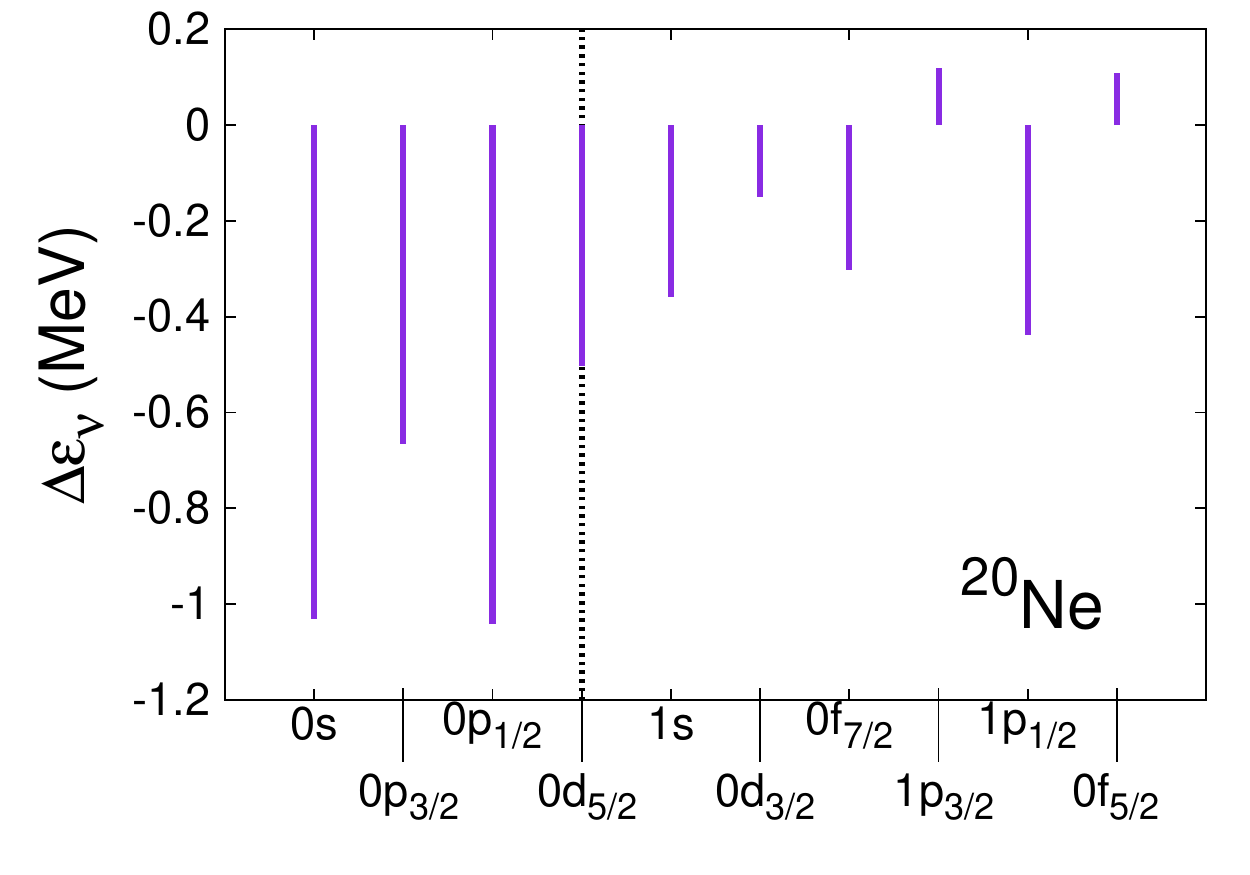}} \hfill %
\subcaptionbox{Proton SPE difference in $^{28}$Ne.}%
[\linewidth]{\includegraphics[width=0.8\columnwidth] {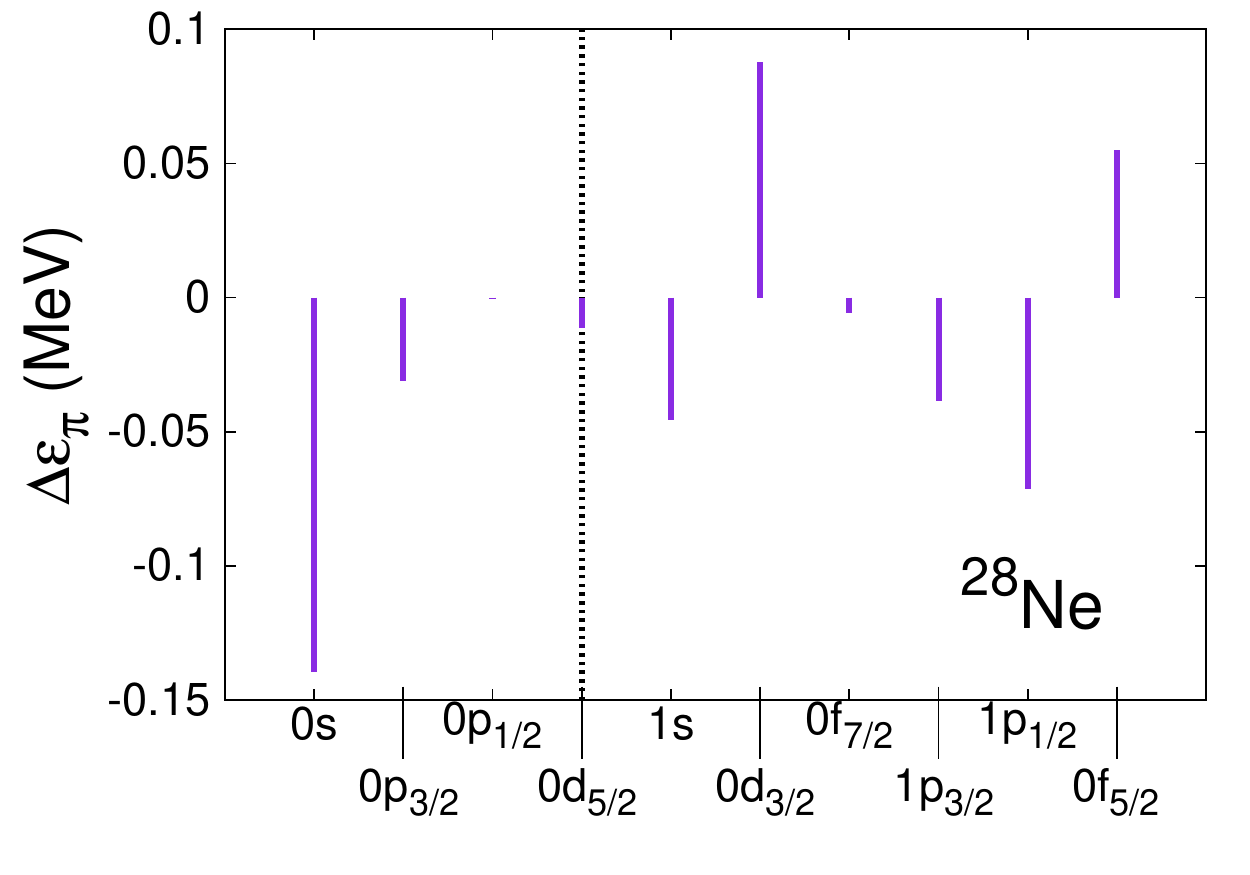}} \hfill %
\subcaptionbox{Neutron SPE difference in $^{28}$Ne.}%
[\linewidth]{\includegraphics[width=0.8\columnwidth] {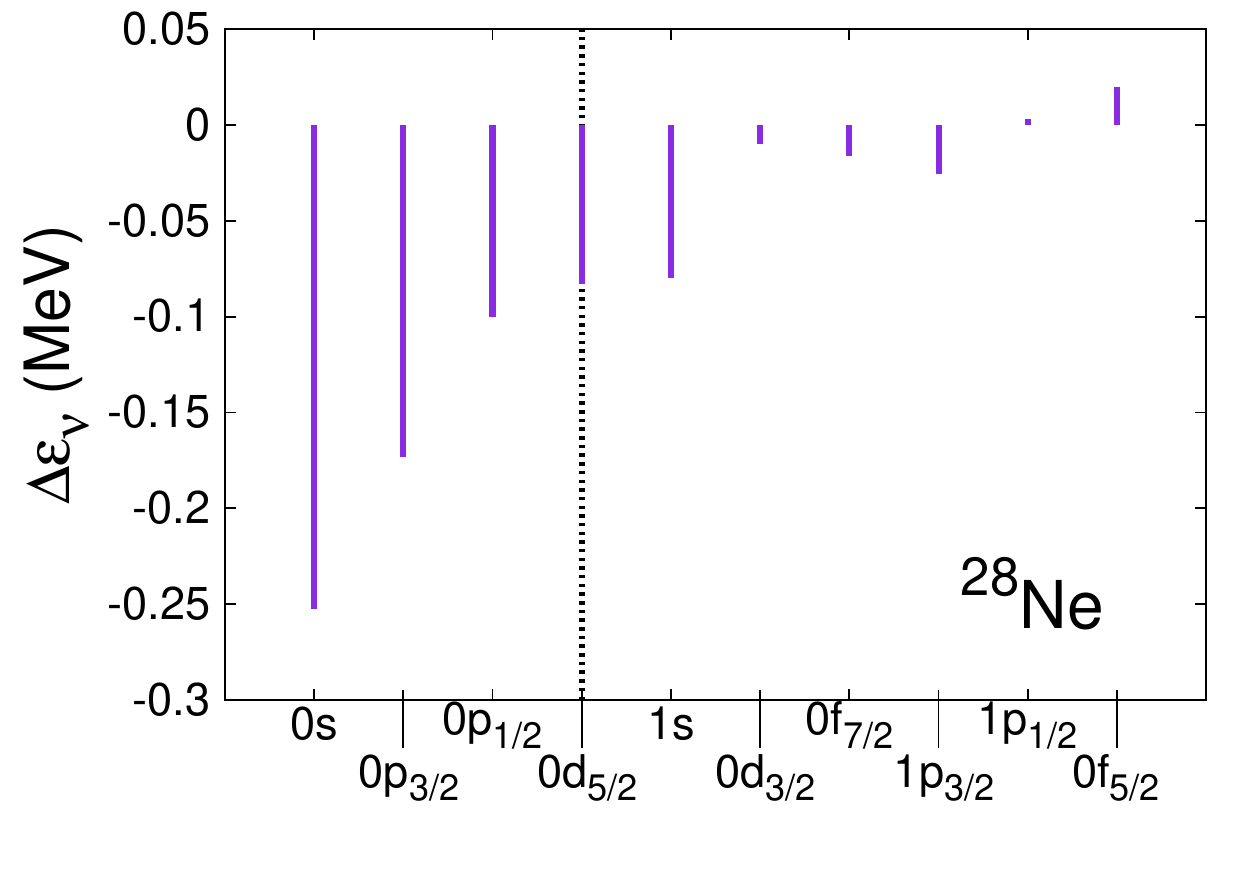}} \hfill %
\caption{(Color Online) Differences $\Delta \varepsilon = \varepsilon_{HF} - \varepsilon$ between Hartree-Fock and self-consistent single-particle energies, expressed in MeV. The Fermi level is marked by a dashed line.}
\label{SPE_sd}
\end{figure}

We present here the modification of single-particle energies (SPE) when the mean-field is constructed consistently with the correlations present in the system.
We show in Fig. (\ref{SPE_sd}) the difference between Hartree-Fock SPE $\varepsilon_{HF}$ and optimized SPE taken as eigenvalues $\varepsilon$ of the general mean-field of Eq. (\ref{e:mean-field})                                                                   
for the lightest and heaviest Neon isotopes under study. The proton and neutron spectra appear very similar for the $N=Z$ nucleus $^{20}$Ne. They are globally more compressed than the Hartree-Fock ones  (by $\sim 1$ MeV), and in particular the gaps under and above the Fermi level are decreased. The deepest shells $0s$ and $0p$ undergo the biggest modification and are shifted up by $>600$ keV ($\sim 1$ MeV for the $0s$ and the $0p_{1/2}$). The change is less important in $^{28}$Ne where the biggest shifts are of order $\sim 250$ keV. Although a smooth compression of the neutron spectrum is observed, the behavior of the proton one is less systematic and seems to indicate an important influence of the proton-neutron interaction. The gap at the Fermi level is actually slightly increased in this case (by $\sim 40$ keV). We emphasize that these results strongly depend on the properties of the effective interaction. In particular, particle-vibration coupling effects have only been very scarcely tested with the D1S Gogny force used here.

\subsubsection{Single-particle orbitals}
\label{sec:orbitals}
\begin{figure*}
\centering
\subcaptionbox{$^{20}$Ne \label{20Ne_orbitals}}%
[.49\linewidth]{\includegraphics[width=.99\columnwidth] {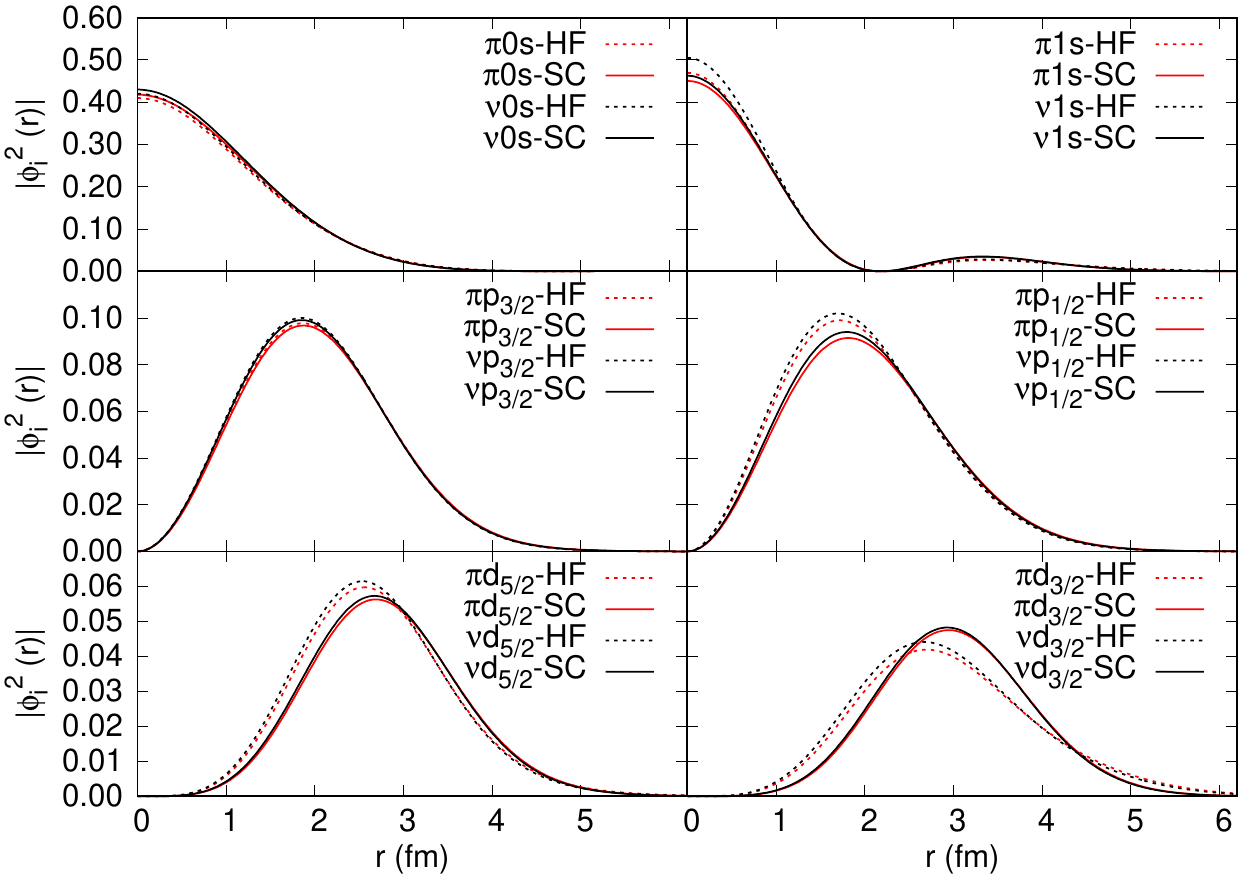}} \hfill %
\subcaptionbox{$^{28}$Ne \label{28Ne_orbitals}}%
[.49\linewidth]{\includegraphics[width=.99\columnwidth] {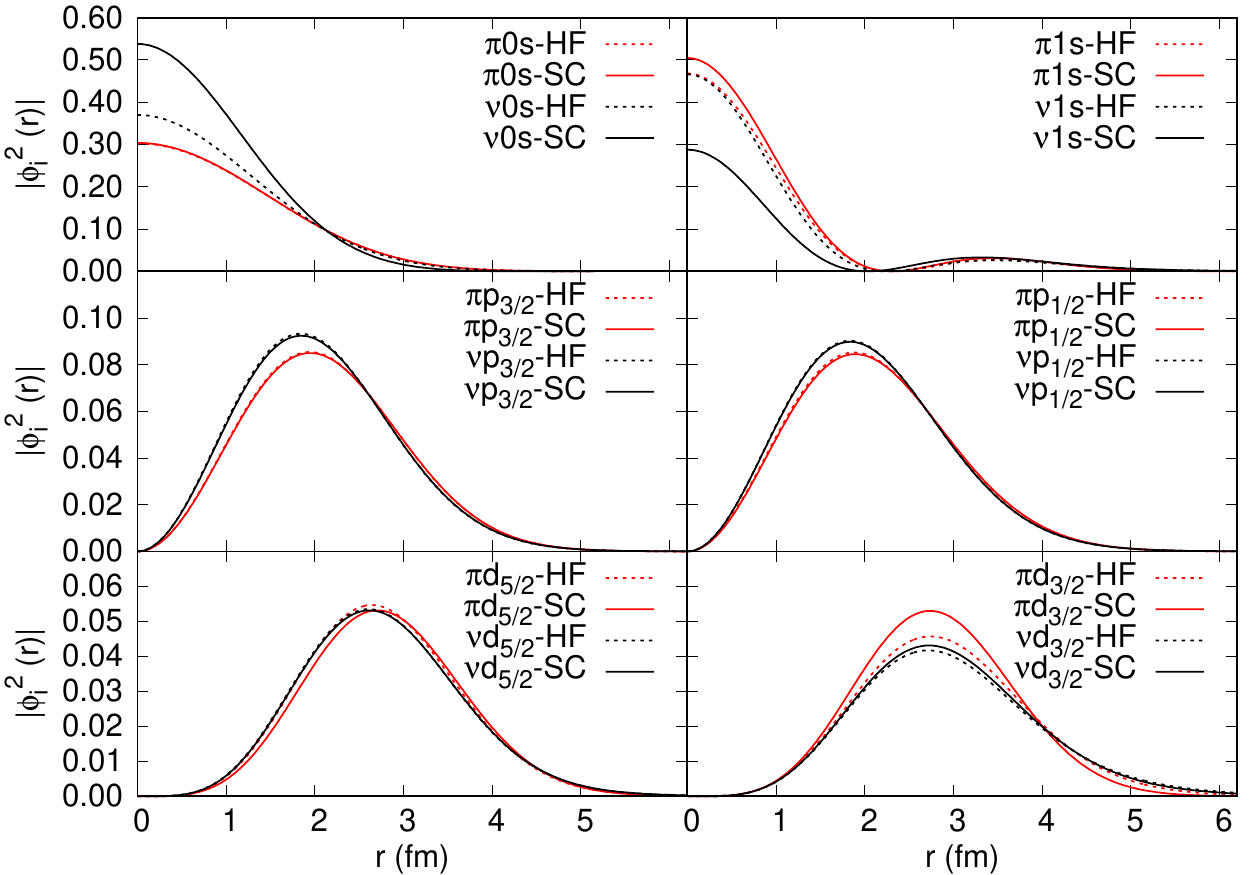}} \hfill %
\subcaptionbox{$^{24}$Mg \label{24Mg_orbitals}}%
[.49\linewidth]{\includegraphics[width=.99\columnwidth] {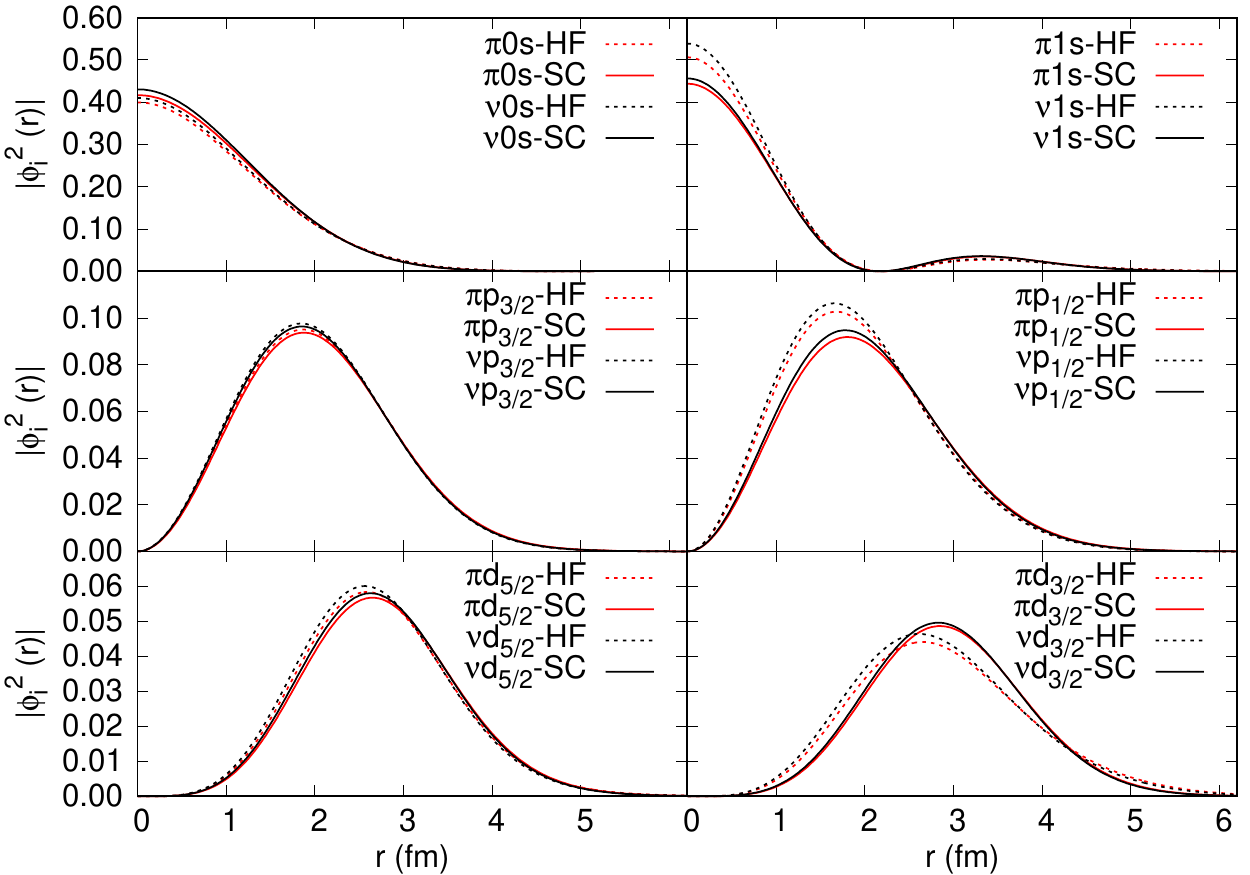}} \hfill %
\subcaptionbox{$^{28}$Si \label{28Si_orbitals}}%
[.49\linewidth]{\includegraphics[width=.99\columnwidth] {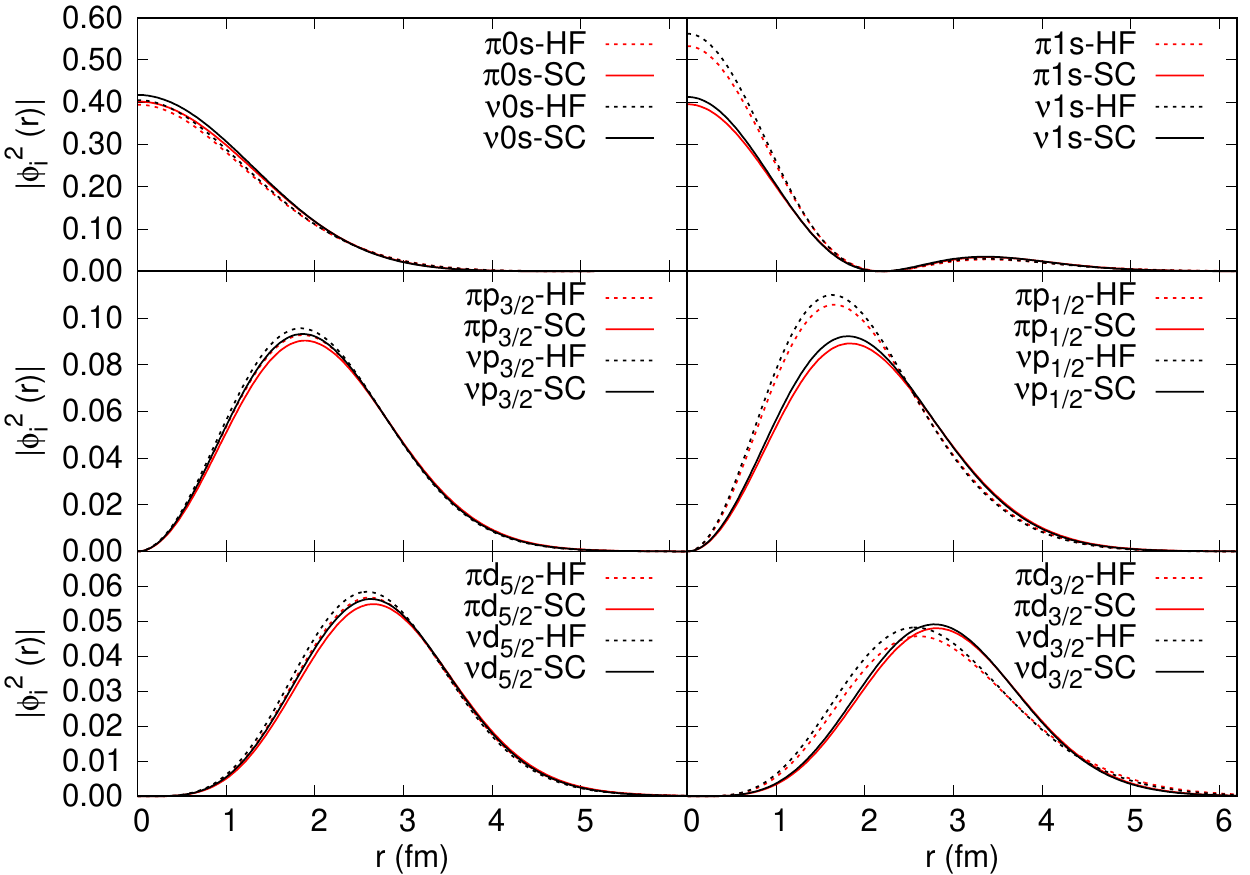}} \hfill %
\subcaptionbox{$^{32}$S \label{32S_orbitals}}%
[.49\linewidth]{\includegraphics[width=.99\columnwidth] {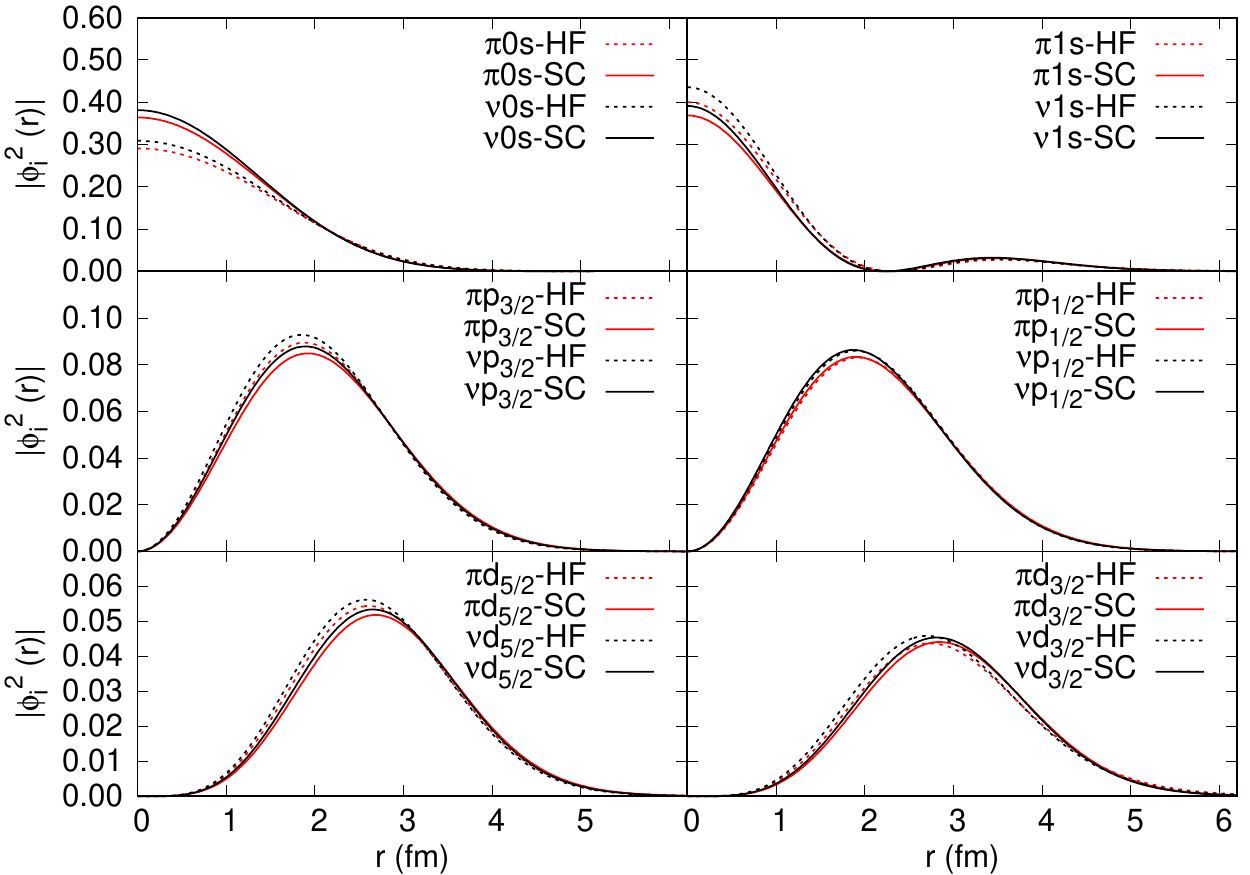}} \hfill %
\caption{(Color Online) Squared modulus of the radial part of the single-particle orbitals. The proton (neutron) self-consistent (SC) orbitals are shown with full red (black) lines while the pure Hartree-Fock (HF) orbitals are shown with dashed lines.}
\label{fig:orbitals} 
\end{figure*}

We now examine the modification of the single-particle orbitals due to the correlations. We show in Fig. \ref{fig:orbitals} the squared modulus of the radial part of the core and valence single-particle orbitals $0s, 0p_{3/2}, 0p_{1/2},0d_{5/2},1s,0d_{3/2}$ for both isospin. The proton $\pi$ (neutron $\nu$) self-consistent (SC) orbitals are shown with full red (black) lines while the pure Hartree-Fock (HF) orbitals are shown with dashed lines. 
Globally, although small, the modification of the radial orbitals due to the correlations of the many-body wave function is noticeable both in the deeply bound orbitals of the core and in the valence-space ones.
Depending on the nucleus, the $s$, $p$ and $d$ single-particle states are more or less sensitive to the correlations. When the change compared to the HF states is appreciable, the extrema of the $p$ and $d$ wave functions are globally shifted towards larger distance $r$. The $0d_{5/2}$, $0p_{3/2}$ and $0p_{1/2}$ are rather smoothed by the self-consistency compared to the HF result, and their surface component becomes larger. On the contrary, the $0d_{3/2}$ orbital acquires a stronger volume component as the peak around $3$ fm increases and the extension towards the surface decreases. The biggest changes in the $p$ and $d$ states are observed in $^{20}$Ne, $^{28}$Si and $^{24}$Mg. In $^{28}$Ne the peak of the $\pi 0d_{3/2}$ is enhanced but not shifted. The magnitude of the $s$ orbitals can be strongly modified. Systematically, the maximum of the $0s$ shell at $r=0$ is increased by the correlations of the system while the $1s$ is decreased (except for the $\pi 1s$ in $^{28}$Ne). The strongest modifications for the neutron $s$ orbitals appear in $^{28}$Ne and $^{28}$Si.  

\subsection{Systematic description of even-even $sd$-shell nuclei}
In this part we study systematically the properties of the ground states of even-even $sd$-shell nuclei with $10 \leqslant (Z,N) \leqslant 18$. Observables such as charge radii and energies are calculated and compared to experiment at the three levels of implementation of the MPMH method.

\subsubsection{Binding and two-nucleon separation energies} 
\begin{figure}
\centering
\includegraphics[width=\columnwidth] {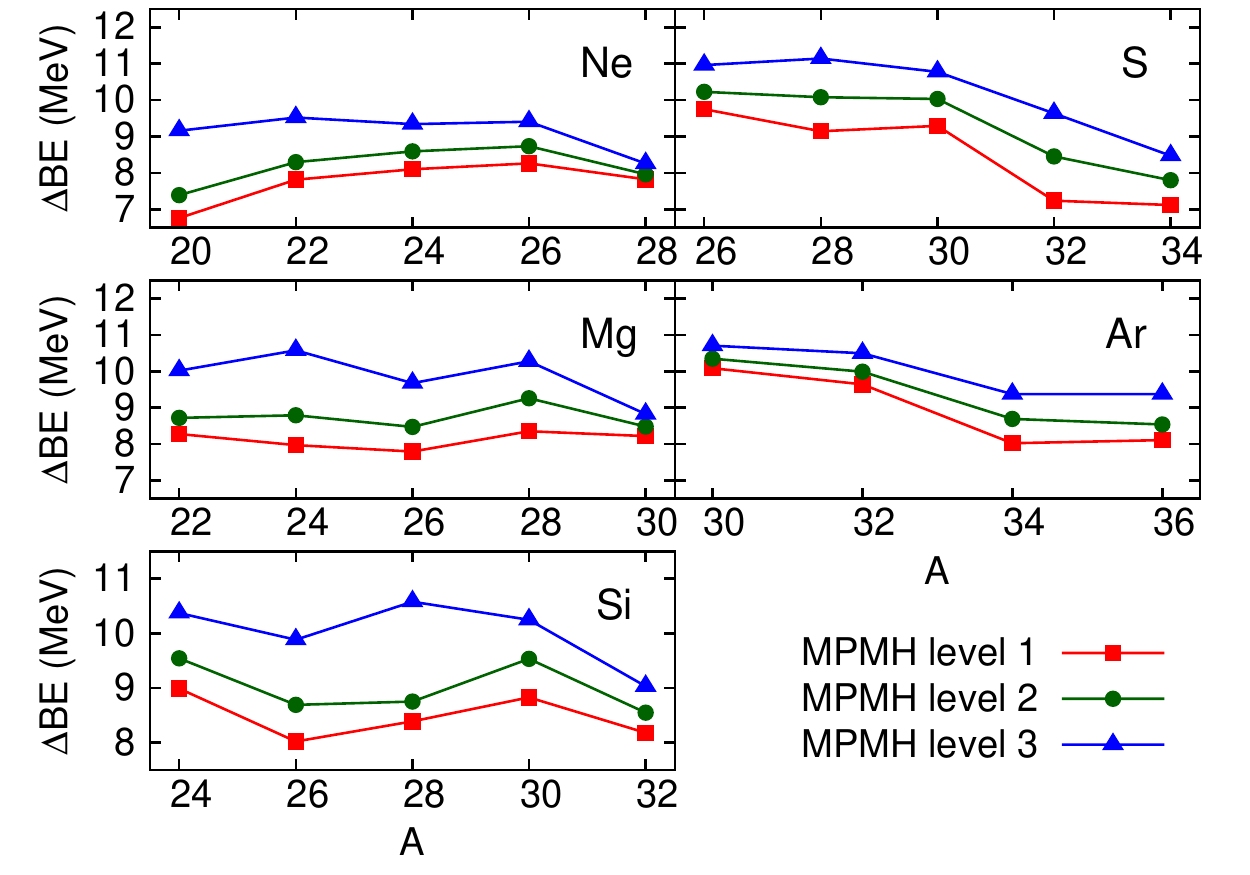}
\caption{(Color Online) Difference between theoretical and experimental binding energies (in MeV) at the different levels of implementation of the MPMH method. Experimental data are taken from \cite{Wang}.}
\label{f:BE}
\end{figure}

We plotted in Fig.~\ref{f:BE} the difference $\Delta BE$ between experimental \cite{Wang} and theoretical binding energies $BE(N,Z)=\braket{\Psi_0^{(N,Z)} | K + V^{2N}_{D1S}[\rho] | \Psi_0^{(N,Z)}}$  for the different isotopic chains. At the non-self consistent level~1 (red squares), an average difference to experiment $\langle \Delta BE \rangle \sim 8.34$ MeV is found when averaging over the $sd$-shell nuclei. 
This global shift is related to the use of the D1S Gogny interaction, which was fitted at the mean-field level, only leaving room for reasonable extensions aiming at the treatment of long range correlations, and thus includes already in a phenomenological way some of the correlations that are explicitly treated by the MPMH method \cite{LeBloas}.
Accordingly, $\langle \Delta BE \rangle$ increases when introducing self-consistency: the use of the correlated density in the D1S interaction and resulting rearrangement terms (green circles) leads to an average difference of $8.91$ MeV, and the additional orbital optimization (blue triangles) gives $\langle \Delta BE \rangle = 9.84$ MeV. On the other hand, the standard deviation $\sigma_{dev}(BE)$ slightly improves when going from no to full self-consistency as it varies from $0.82$ to $0.79$ MeV. We note that this small deviation compared to experiment leads to a relatively good description of the two-nucleon separation energies $S_{2n}$ and $S_{2p}$. Statistically, when averaging the difference to experiment \cite{Wang} over the $sd$-shell nuclei we find: $\langle \Delta S_{2n} \rangle = 641$ keV and $\langle \Delta S_{2p} \rangle = 577$ keV, when the full MPMH method is applied. The standard deviations to data are $\sigma(S_{2n})=453$ keV and $\sigma(S_{2p})=332$ keV.

\subsubsection{Charge radii} 
\begin{figure}
\centering
\includegraphics[width=\columnwidth]  {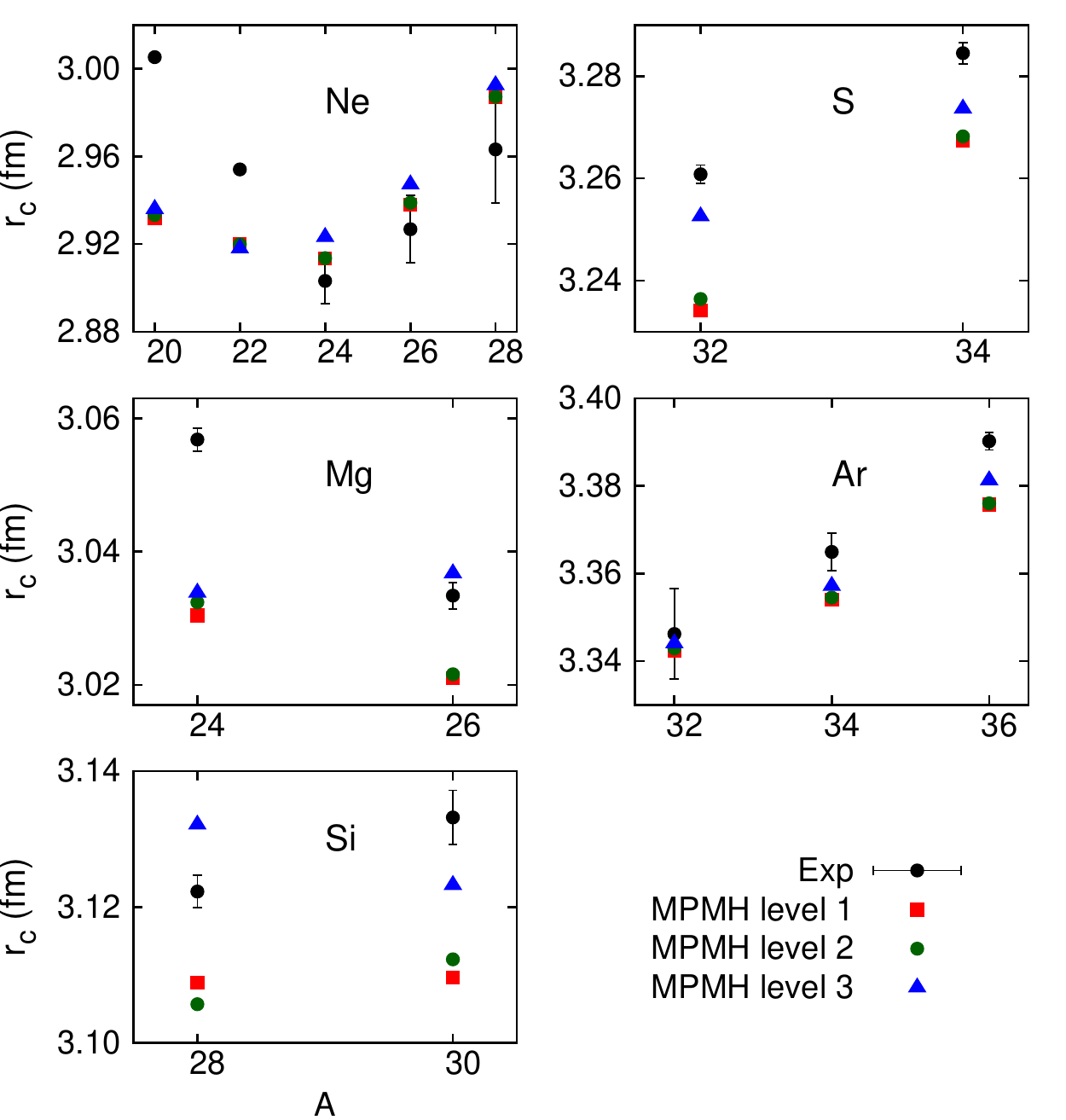}
\caption{(Color Online) Comparison between experimental and MPMH charge radii (in fm). Experimental values are taken from \cite{Angeli}.}
 \label{f:charge_radii} 
\end{figure}

\begin{figure*}
\centering
\subcaptionbox{$^{20}$Ne}%
[.49\linewidth]{\includegraphics[width=\columnwidth] {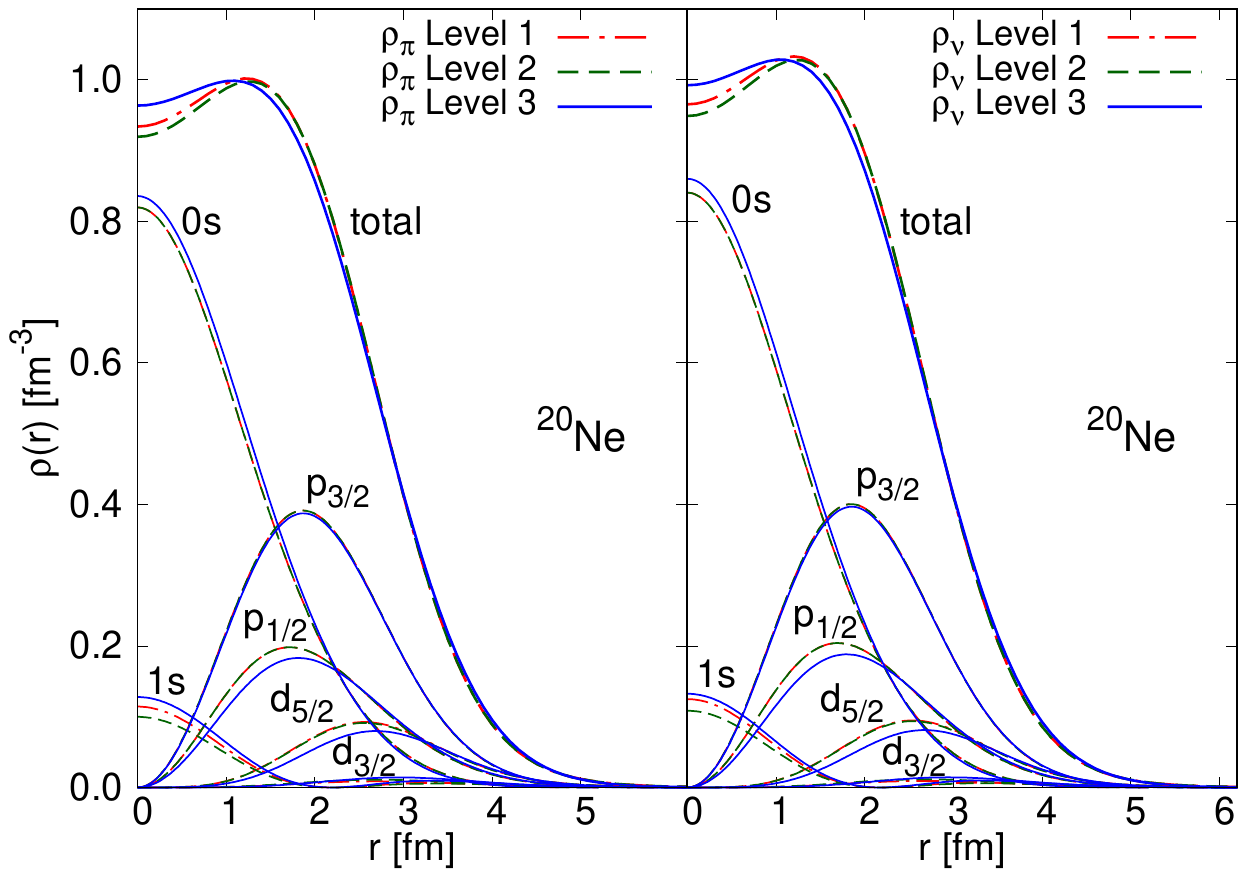}} \hfill %
\subcaptionbox{$^{28}$Ne}%
[.49\linewidth]{\includegraphics[width=\columnwidth] {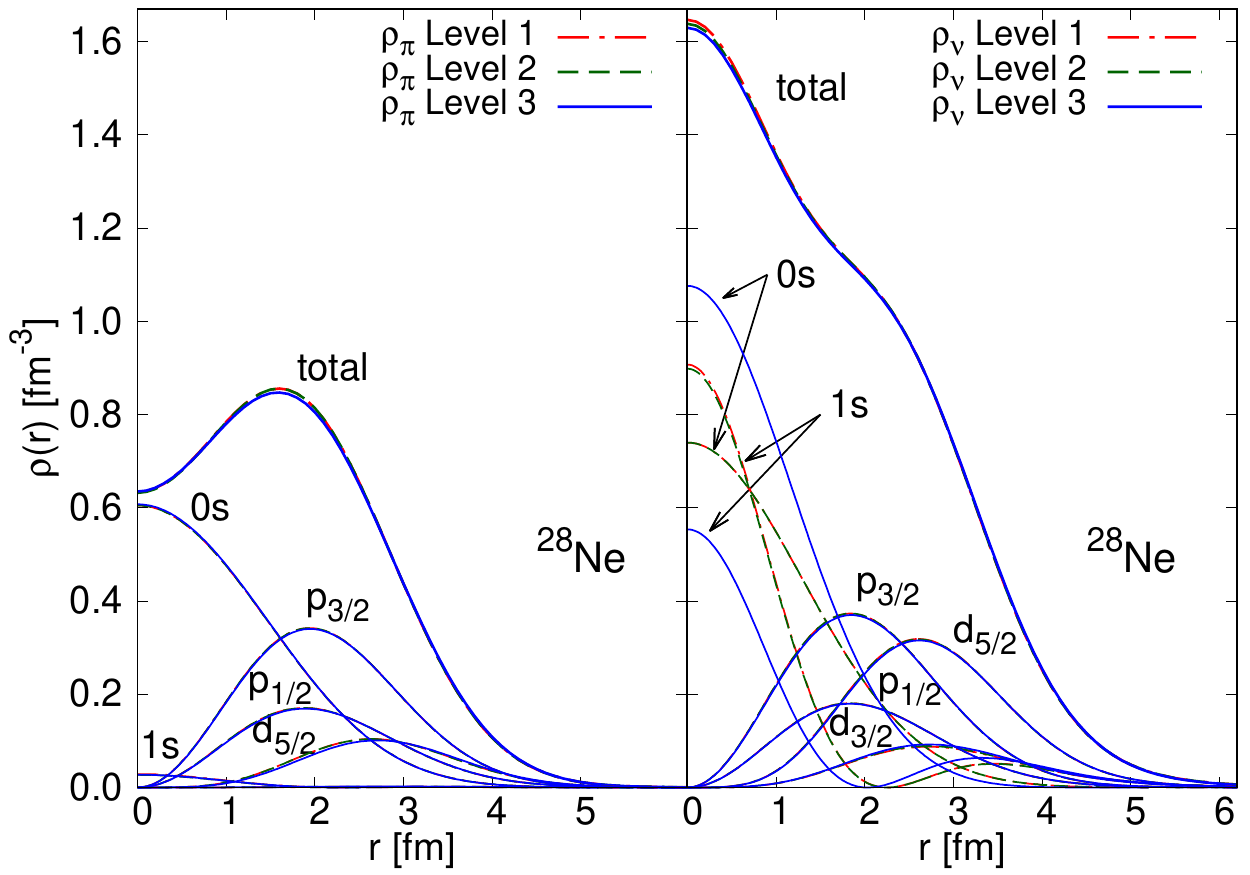}} \hfill %
\subcaptionbox{$^{24}$Mg}%
[.49\linewidth]{\includegraphics[width=\columnwidth] {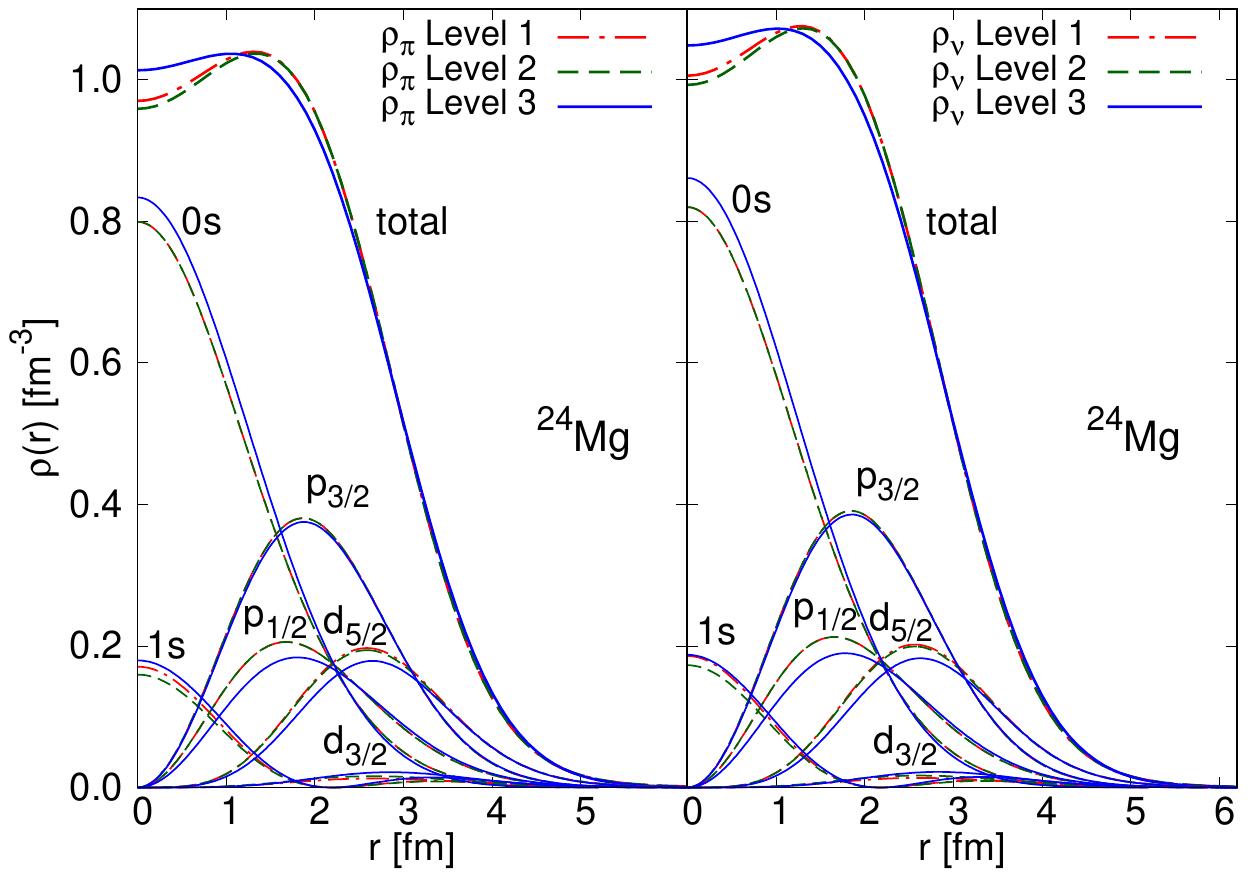}} \hfill %
\subcaptionbox{$^{26}$Mg}%
[.49\linewidth]{\includegraphics[width=\columnwidth] {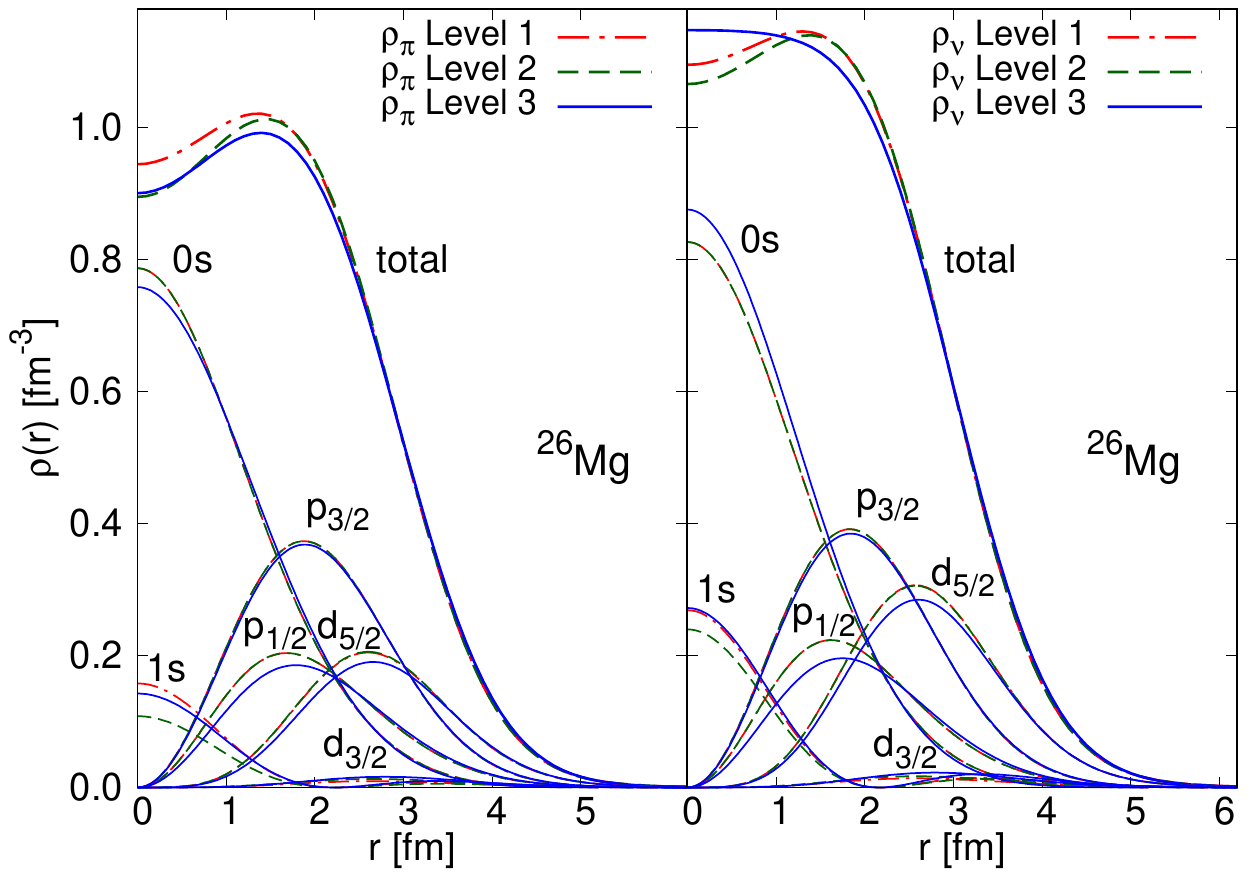}} \hfill %
\subcaptionbox{$^{28}$Si}%
[.49\linewidth]{\includegraphics[width=\columnwidth] {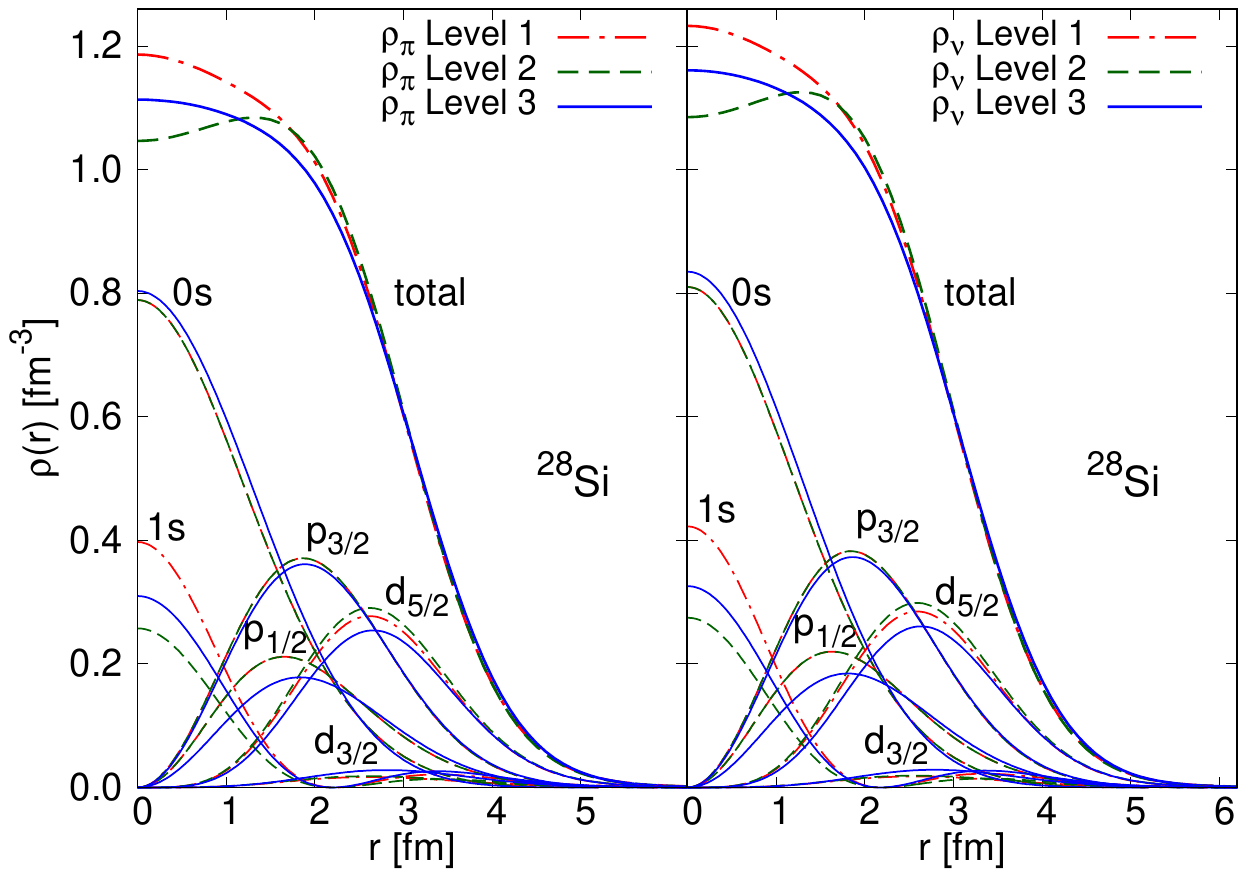}} \hfill %
\subcaptionbox{$^{32}$S}%
[.49\linewidth]{\includegraphics[width=\columnwidth] {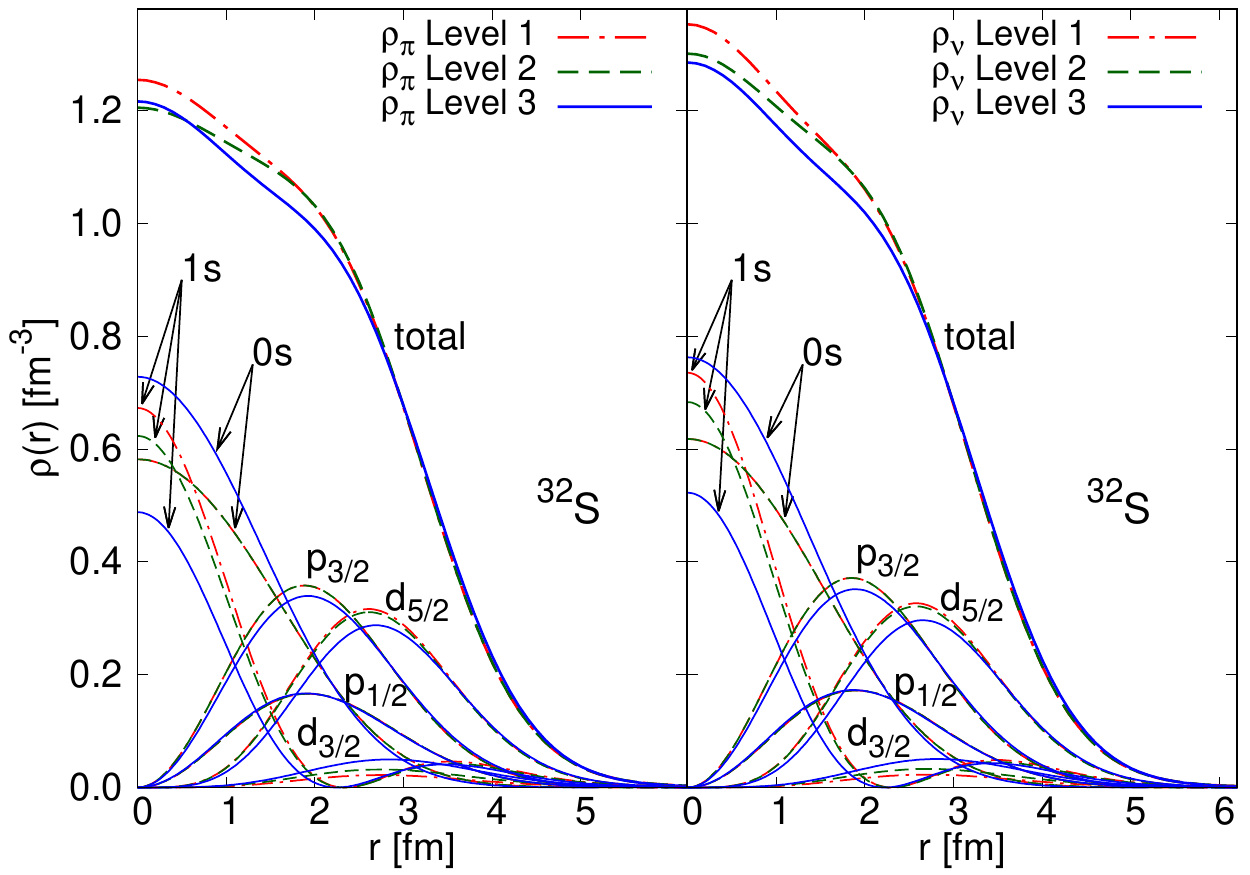}} \hfill %
\caption{(Color Online) Radial proton density $\rho_\pi (r)$ (left) and neutron density $\rho_\nu(r)$ (right) for $^{20}$Ne, $^{28}$Ne, $^{24}$Mg, $^{26}$Mg, $^{28}$Si and $^{32}$S, and contribution of each orbital, in fm$^{-3}$.}
\label{fig:radial_density} 
\end{figure*}

Charge radii are measurable quantities very sensitive to the correlation content of nuclei and related to nuclear deformation. The root mean-square charge radius $r_c$ is expressed as,
\begin{equation}
 r_c = \sqrt{r_p^2 + \frac{3}{2}(B^2 - b^2) - 0.1161 \frac{N}{Z}} \; ,
\end{equation} 
where $r_p$ denotes the proton root mean square radius, 
\begin{equation}
 r_p = \sqrt{\frac{\int d^3 r \rho_{\pi}(r) r^2}{Z}} \; ,
\end{equation} 
with $\rho_{\pi}(r)$ the proton radial density in the ground state
\begin{eqnarray}
\rho_{\pi}(r) &=& \braket{\Psi| \hat{a}^\dagger_\pi (r) \hat{a}_\pi(r) |\Psi } \nonumber \\
                   &=& \sum_{{i_\pi}{j_\pi}} \phi_{i_\pi}^* (r) \phi_{j_\pi}(r) \braket{\Psi| \hat{a}^\dagger_{i_\pi} \hat{a}_{j_\pi} |\Psi } \nonumber \\
                   &=& \sum_{{i_\pi}} |\phi_{i_\pi} (r)|^2 n_{i_\pi}\; ,
\end{eqnarray} 
where we used the fact that the density matrix is diagonal in the basis used for the configuration mixing due to the small valence space, and $n_{i_\pi}$ denotes the proton occupation numbers.
The charge radius $r_c$ is corrected by $\frac{3}{2}(B^2 - b^2)$ where $B=0.7144$ fm results from the proton form factor, and $b$ is a center of mass correction.
Finally $0.1161 \frac{N}{Z}$ denotes a correction due to neutron electromagnetic properties.
\\
In Fig. \ref{f:charge_radii} we compare the theoretical charge radii to available experimental data \cite{Angeli}. To understand these results we also show in Fig. \ref{fig:radial_density} the radial densities of a few nuclei (the proton densities are on the left panels). We also show on these figures the contribution $n_i |\phi_i (r)|^2 $ of each orbital $i$.
\\ 
At the non self-consistent stage (red squares in Fig. \ref{f:charge_radii}), charge radii are either lying in the experimental error bars or underestimated, leaving room for unaccounted correlations.
The worst discrepancy is encountered in the most deformed nuclei, exhibiting important collectivity. This behavior can be anticipated since the configuration mixing has been 
restricted to the $sd$-shell, and therefore "surface" orbitals with a larger spatial extension such as the $0f_{7/2}$ are not populated.
The introduction of the correlated density in the interaction and of the rearrangement terms (green circles) slightly improves the theoretical values. 
The only exception is $^{28}$Si. Looking at Fig. \ref{fig:radial_density} we note that, in most nuclei, the rearrangement terms do not induce a change in the profile of the proton density, but only lead to a slight decrease of the proton density at the center, counterbalanced by a slightly larger extension of the density towards the surface. In $^{28}$Si, the change is stronger as we observe a modification of the density shape with a strong depopulation at the center which is accompanied by an increase around $r\simeq 2-3$ fm. We also note a slight shrinking of the density at high distance, leading to a smaller predicted radius.
\\
At the fully self-consistent level (blue triangles), when the orbitals are optimized according to the correlations, the charge radii are (almost) systematically increased.
The radii of the Argon isotopes, rather poorly correlated, are all improved. The radii of Sulfur nuclei are drastically increased and in better accordance with experiment. Let us remind the important fragmentation introduced in the ground state of $^{32}$S via the orbital renormalization. An important effect is also seen on the Silicon and Magnesium isotopes, although it appears too important in $^{26}$Mg and $^{28}$Si, leading to an 
overestimation of the radii and the wrong trend along the isotopic chains. Of course one should remember that the Gogny interaction used to perform the calculation is not adapted to approaches such as the MPMH configuration mixing method, which introduces explicitly all types of correlations that are already included phenomenologically in the force parameters. Moreover the use of the correlated density in the interaction may lead to uncontrolled over-counting effects. Concerning the Neon nuclei, the radii for $^{28-26-24}$Ne, barely inside the error bars, are slightly overestimated when full self-consistency is considered. Very little effect is seen on the lighter and more correlated isotopes $^{20-22}$Ne, whose radii remain largely undervalued. Globally we see from Fig. \ref{fig:radial_density} that in the nuclei that show the largest increase of $r_c$, such as $^{32}$S, $^{28}$Si and $^{26}$Mg, the density 
is significantly reduced at $r=0$ by the optimization of orbitals, while the shape of the density profile is kept similar to the one obtained when no self-consistency is applied (level 1 of MPMH). In the nuclei where the improvement is small, such as $^{24}$Mg or the Neon isotopes, the density is on the other hand increased at the center, and thus extends to smaller distances.
Finally, let us note that the good experimental trend is obtained for the Ne, S and Ar isotopes.

\section{Description of low-lying excited states and applications to reaction calculations} \label{sec:excited_states}
In this section we study the low-lying excited states, and put to test the MPMH ground and excited wave functions by using them as input for reaction calculations.
As explained in Ref. \cite{Robin}, the excited states are obtained in the following way:
\\
At the non self-consistent stage (level 1), they are calculated by extracting several eigenstates of the Hamiltonian matrix $H[\rho_{HF}]$ with the Lanczos algorithm.
\\
When the first equation is solved iteratively with the rearrangement terms (level 2), we first iterate the diagonalization of $\mathpzc{H}[\rho_{gs},\sigma_{gs}] = H[\rho_{gs}] + R[\rho_{gs},\sigma_{gs}]$, where $\rho_{gs}$ and $\sigma_{gs}$ refer to the densities of the correlated ground-state. Once this procedure has converged, we extract several eigenvalues of $\mathpzc{H}[\rho_{gs},\sigma_{gs}]$  in order to obtain the excited states.
\\
Similarly, when full self-consistency is applied, we first perform the doubly iterative procedure described in Ref.~\cite{Robin} for the ground state only. In this way, we obtain single-particle orbitals that are consistent with the ground state correlations. Again, once this doubly-iterative procedure has converged, we extract several eigenvalues of $\mathpzc{H}[\rho_{gs},\sigma_{gs}]$  in order to obtain the excited states.

\subsection{Excitation energies}

\begin{figure}
\centering
\includegraphics[width=\columnwidth] {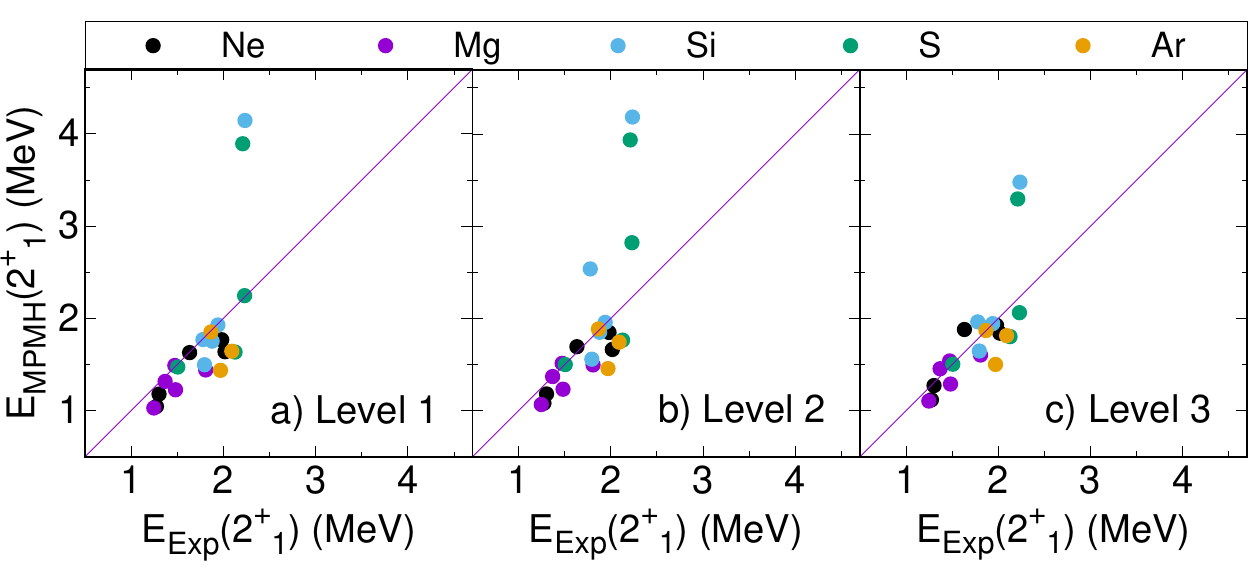}
\caption{(Color Online) Theoretical excitation energies of the $2_1^+$ states compared to experiment. Experimental data are taken from \cite{nndc}. Results are expressed in MeV.}
\label{f:Exc_E}
\end{figure}

\begin{figure*}
\centering
\subcaptionbox{\label{Ne-E}}%
{\includegraphics[width=\columnwidth] {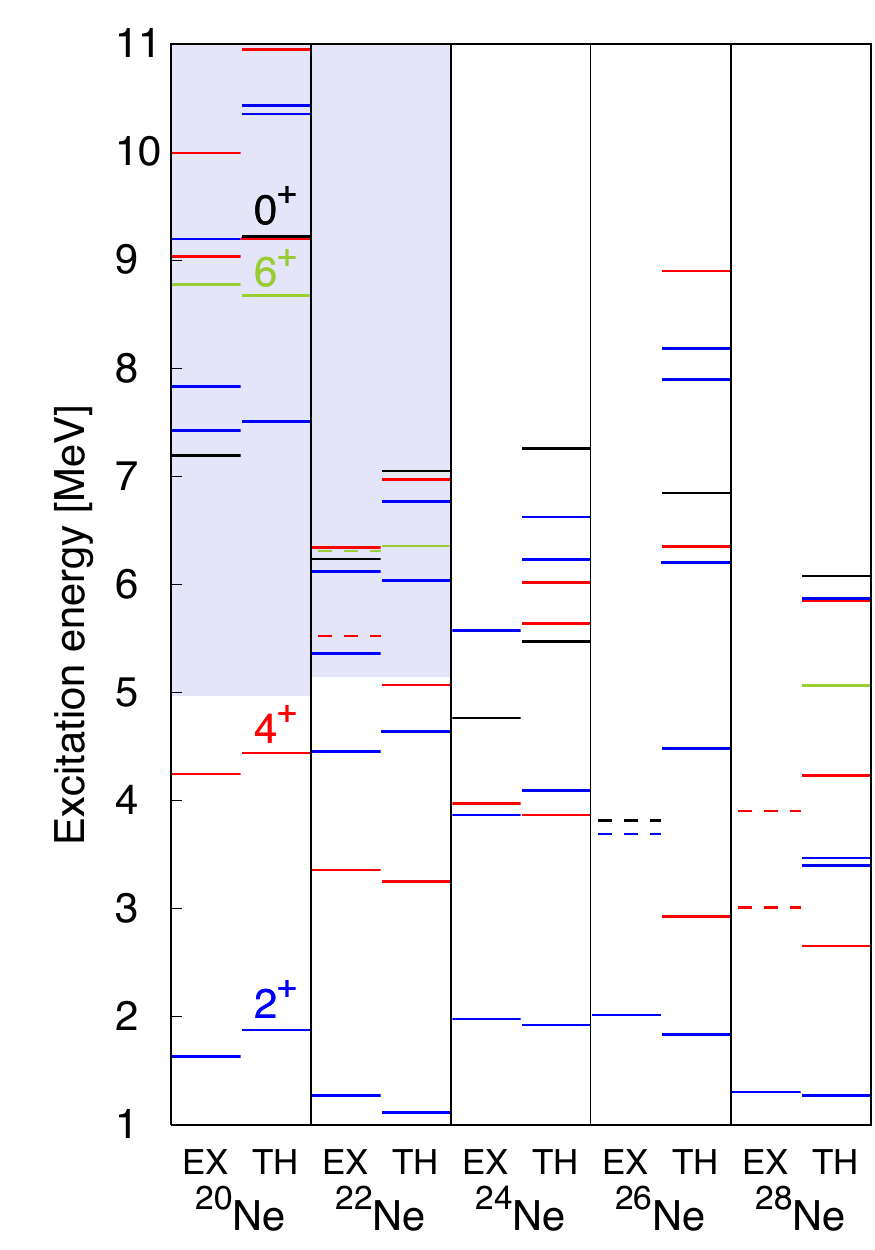}} \hfill %
\subcaptionbox{\label{Mg-E}}%
{\includegraphics[width=\columnwidth] {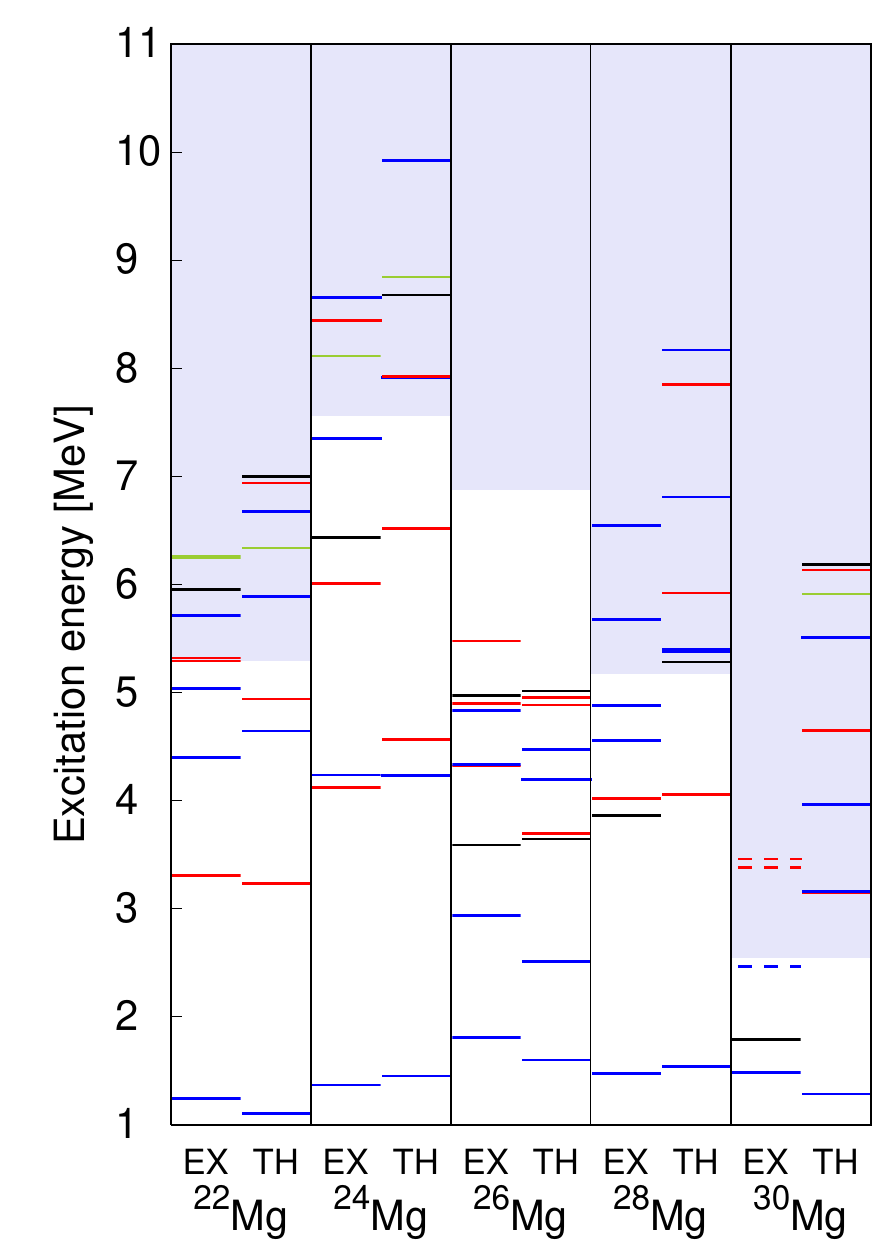}} \hfill %
\caption{(Color Online) Low-energy spectra of Neon and Magnesium isotopes. For each nucleus the theoretical spectrum is on the right while the corresponding experimental states are shown on the left. 
$0^+,2^+,4^+$ and $6^+$ states are shown with black, blue, red and green lines respectively.
The experimental plain lines show states with spin and parity that have been surely assigned, while the dashed lines denote states for which the spin and/or parity have been assigned "based on weak arguments" \cite{nndc}}
 \label{fig:energy1} 
\end{figure*}

\begin{figure*}
\centering
\subcaptionbox{\label{Si-E}}%
{\includegraphics[width=0.4\textwidth,height=10cm] {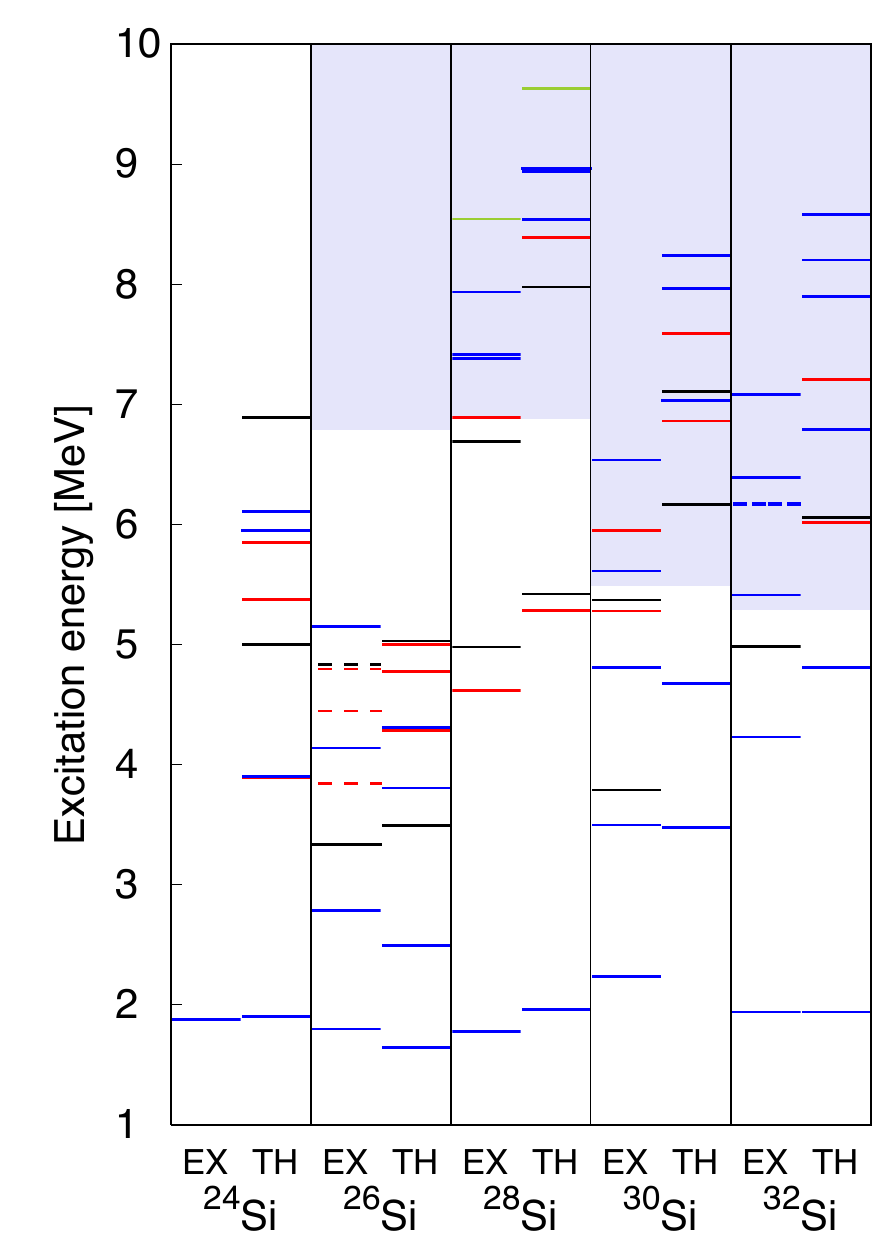}} \hfill %
\subcaptionbox{\label{S-E}}%
{\includegraphics[width=0.32\textwidth,height=10cm] {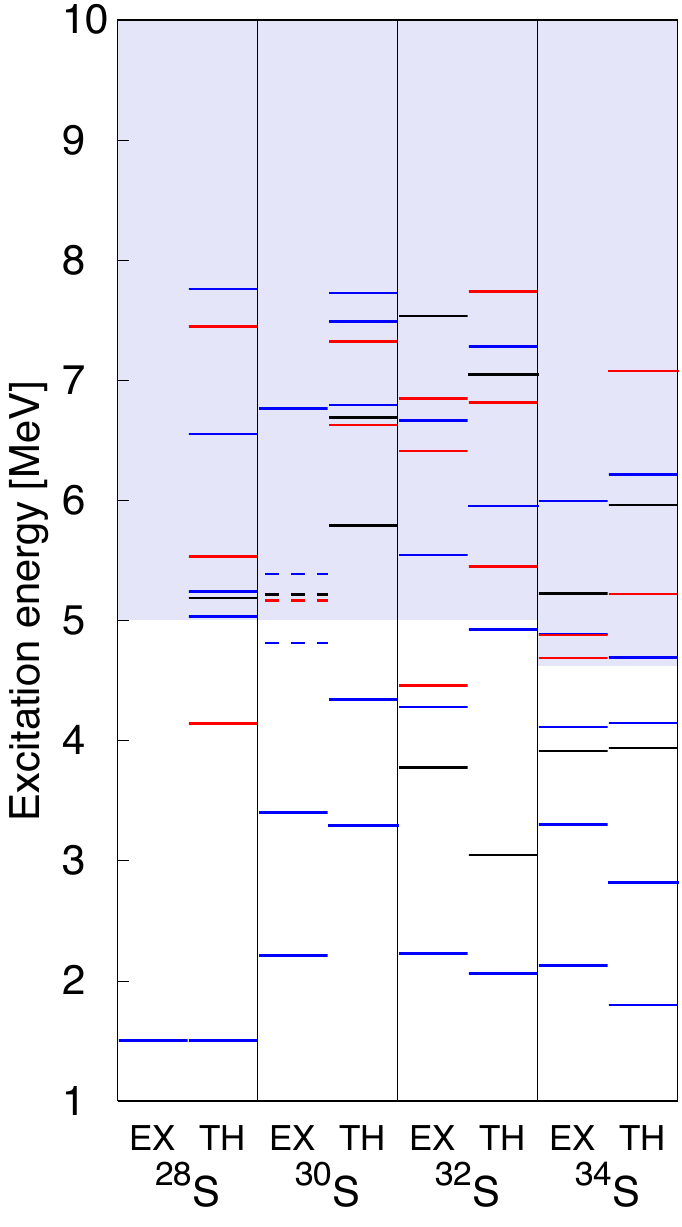}} \hfill %
\subcaptionbox{\label{Ar-E}}%
{\includegraphics[width=0.24\textwidth,height=10cm] {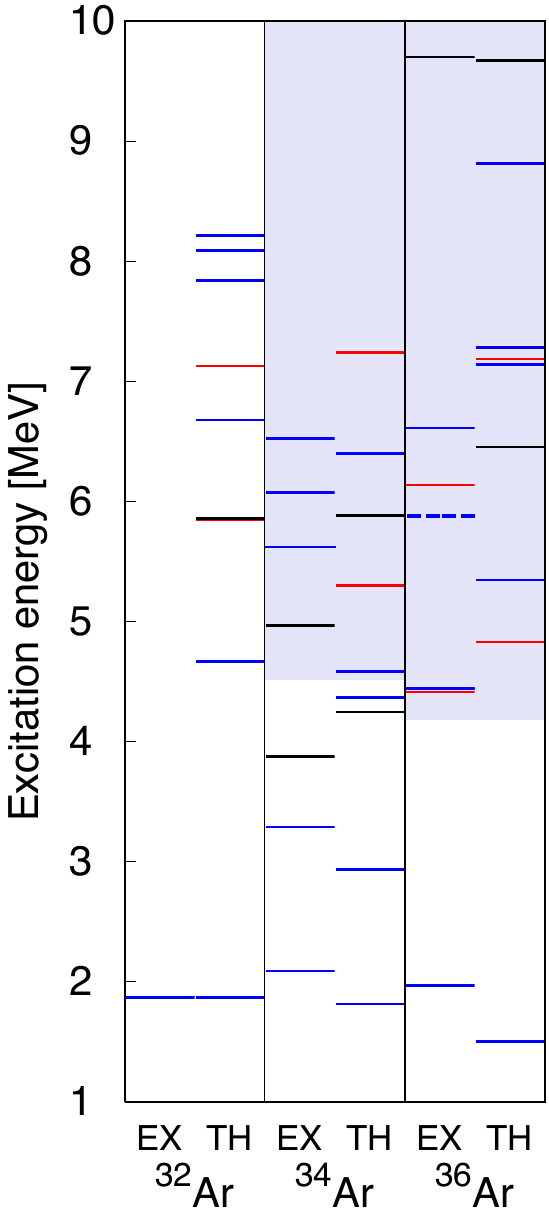}} \hfill %
\caption{(Color Online) Same as Fig. \ref{fig:energy1} for the Silicon, Sulfur and Argon chains.}
 \label{fig:energy2} 
\end{figure*}

\noindent  We plotted in Fig. \ref{f:Exc_E} the theoretical excitation energies of the first excited $J^\pi=2^+$ state of $sd$-shell nuclei as a function of experimental ones \cite{nndc}. 
We have excluded from this study $^{26}$S, $^{28}$Ar and $^{30}$Ar, as these nuclei are predicted with negative two-proton separation energies $S_{2p}$, in accordance with experiment \cite{Wang}.
For the majority of nuclei, a good agreement with experiment is found at the levels 1 and 2 of implementation of MPMH, although we note a small systematic underestimation of the experimental values. Discrepancies are seen for a few Silicon and Sulfur nuclei. In particular, the energy of the first $2^+$ state is overestimated by $\sim 600$ and $\sim 800$ keV in  $^{32}$S and  $^{28}$Si. More importantly, it is overestimated by 1.7 and 1.9 MeV in the two mirror nuclei  $^{30}$S and $^{30}$Si, respectively. We found that this shift also appears in the next excited states (see Figs. \ref{fig:energy1} and \ref{fig:energy2}). 
This behavior has in fact already been noticed and investigated in a previous study \cite{Pillet2}, where it was found that this systematic shift is due to uncontrolled proton-neutron matrix elements of the Gogny interaction in the $T=0$ channel.
However the optimization of orbitals modifies the single-particle spectra and in particular the sizes of the gaps around the Fermi level. Solving the variational equation (\ref{e:eq2_Gogny}) is thus expected to modify the values of the matrix elements occurring in the Hamiltonian matrix to diagonalize, and hence to have an impact on the low-lying spectroscopy. Indeed, we observe on the right panel of Fig. \ref{f:Exc_E} a downward shift of $\sim 600$ keV in the spectra of $^{30}$Si and $^{30}$S. This effect is very encouraging but still insufficient to reach the experimental values. The rest of the discrepancy 
may be attributed to the aforementioned uncontrolled $T=0$ matrix elements of the D1S Gogny interaction.
For the rest of the $sd$-shell nuclei, including $^{32}$S and $^{28}$Si, the theoretical energies are very close to the experimental values.
Overall, the average difference to experimental results is decreased from 418 to 286 keV when optimizing the orbitals, and the standard deviation is lowered from 680 to 494 keV. 
Excluding the peculiar cases $^{30}$S and $^{30}$Si, the average difference to experiment is decreased from 227 to 149 keV, while the standard deviation decreases from 214 to 121 keV. 
Globally, the results concerning the $2^+_1$  excitations energies are in very good agreement with experiment.
\\
\\
In order to evaluate further the capability of the MPMH approach to describe the low-energy spectroscopy, we show in Figs. \ref{fig:energy1} and \ref{fig:energy2} the low-energy spectra of the same $sd$-shell nuclei. From a theoretical point of view, the description of the spectra of these light nuclei is a great challenge as important variations can be seen in nuclei differing by only one nucleon.
The spectra shown here contain the first nine excited states with natural and positive parity calculated by the complete MPMH method (level 3). These include $2_1^+,2_2^+,2_3^+,2_4^+,2_5^+$ shown with blue lines, $4_1^+,4_2^+,4_3^+$ shown with red lines, as well as $0_2^+,0_3^+$ with black lines and $6_1^+$ with green lines. They are compared to the corresponding experimental states taken from \cite{nndc}. The plain lines show states with spin and parity that have been surely assigned, while the dashed lines denote states for which the spin and/or parity have been assigned "based on weak arguments" \cite{nndc}. As the configuration mixing is restricted to the $sd$-shell which contains positive-parity states only, we show by a blue area the energy region above the first experimental negative-parity state where our predictions could be potentially polluted by negative-parity orbitals.
\\
Concerning the Neon isotopic chain, the $2_1^+$ and $4_1^+$ states are in good agreement with the data (experimentally, $^{26}$Ne has a state at $\sim 3.5$ MeV with spin and parity potentially equal to $3^+$ or $4^+$, which is not shown here). In $^{22}$Ne and $^{24}$Ne the position of the $2_2^+$ state is also well predicted by MPMH. More discrepancies are found for higher $2^+$ and $4^+$ states. The position of the $6^+_1$ state, measured in $^{20}$Ne and $^{22}$Ne is in very good agreement with experiment. However, the position of the $0_2^+$ is found systematically overestimated by $0.7$ to $3$ MeV.
Similar conclusions are obtained for the Magnesium chain, although the $0_2^+$ and $0_3^+$ states in $^{26}$Mg are found in excellent agreement with the data. In $^{30}$Mg the $0_2^+$ state is however overestimated by $4.4$ MeV.
The low-lying spectrum of $^{26}$Si is rather well reproduced despite some inversions of states above $3.5$ MeV which are close in energy. The worst discrepancy is found for the $2^+_4$ state which is underestimated by $\sim 840$ keV. In $^{28}$Si, the sequence of state is perfectly reproduced. Although the $2^+_1$ state is in accordance with the data, we note an upward shift of the theoretical states which appears to grow with energy. A similar situation is seen in $^{32}$Si, although several states are unassigned experimentally which makes the comparison more difficult. In $^{30}$Si and $^{30}$S, as already noted, the entire spectrum appears to be shifted by more than 1 MeV. In $^{32}$S, the $2_2^+,2_3^+,2_4^+$ and $4_1^+,4_2^+,4_3^+$ states tend to be a bit overestimated by the theory (by $\sim 400$ keV to $\sim 1$ MeV), while the position of the $0^+_2$ and $0^+_3$ states are underestimated by $\sim 500-700$ keV. A rather good agreement is obtained for the spectrum of $^{34}$S in the region below the first experimental negative-parity state. This is also true for the Argon isotopes. Several states with unassigned spin and parity are however present at higher energy, which makes the comparison to data more difficult.
\\
\\

\begin{table}
\centering
\begin{tabular}{cccc}
State & EXP & MPMH & Ref. \cite{Rodriguez2010} \\
\hline
\hline
$2^+_1$ & 1.369 & 1.453 & 1.202 \\
$2^+_2$ & 4.238 & 4.230 & 5.616 \\
$2^+_3$ & 7.349 & 7.914 & 12.686 \\
$4^+_1$ & 4.123 & 4.564 & 3.875 \\
$4^+_2$ & 6.011 & 6.518 & 7.990 \\
$4^+_3$ & 8.439 & 7.923 & 14.363 \\
$6^+_1$ & 8.114 & 8.843 & 8.256  \\
$0^+_2$ & 6.432 & 8.676 & 11.265 \\
\end{tabular}
\caption{Comparison between the low-energy states of $^{24}$Mg calculated by the MPMH method, the GCM method of Ref. \cite{Rodriguez2010} and the experimental data \cite{nndc}. The energies are in MeV.}
\label{T:GCM}
\end{table}

Recently developed beyond-mean-field methods based on symmetry-breaking and projection techniques using the same Gogny interaction \cite{Rodriguez2007,Rodriguez2010,Egido2016} can be an interesting point of comparison. For instance, Ref. \cite{Rodriguez2010} studied the nucleus $^{24}$Mg within a GCM approach including particle-number and angular-momentum projection of a triaxially-deformed wave function. We compare our results to the states predicted in that study in Table \ref{T:GCM}. While the GCM method predicts the position of yrast states ($2^+_1$,  $4^+_1$ and $6^+_1$) with a similar accuracy as MPMH, the position of other excited states appears to be overestimated. On the other hand, the GCM approach is very powerful for the description of nuclear collectivity and deformation, which cannot be reproduced in the present application of the MPMH method, due to the small valence space used here (see the discussion and Fig.~\ref{f:BE2} of transition probabilities in the next section).

\subsection{Transition densities and inelastic scattering on discrete states}
In this last section, we aim to test the structure description provided by the MPMH approach by using it as ingredient for reaction calculations. 
In this study, we are interested in inelastic scattering of electrons and protons from $sd$-shell nuclei, when the target nucleus is excited from its ground state to a low-lying excited state.

\subsubsection{Inelastic electron scattering}
\begin{figure*}
\centering
\subcaptionbox{Transition density in $^{28}$Si. \label{28Si_rho}}%
[.49\linewidth]{\includegraphics[width=\columnwidth] {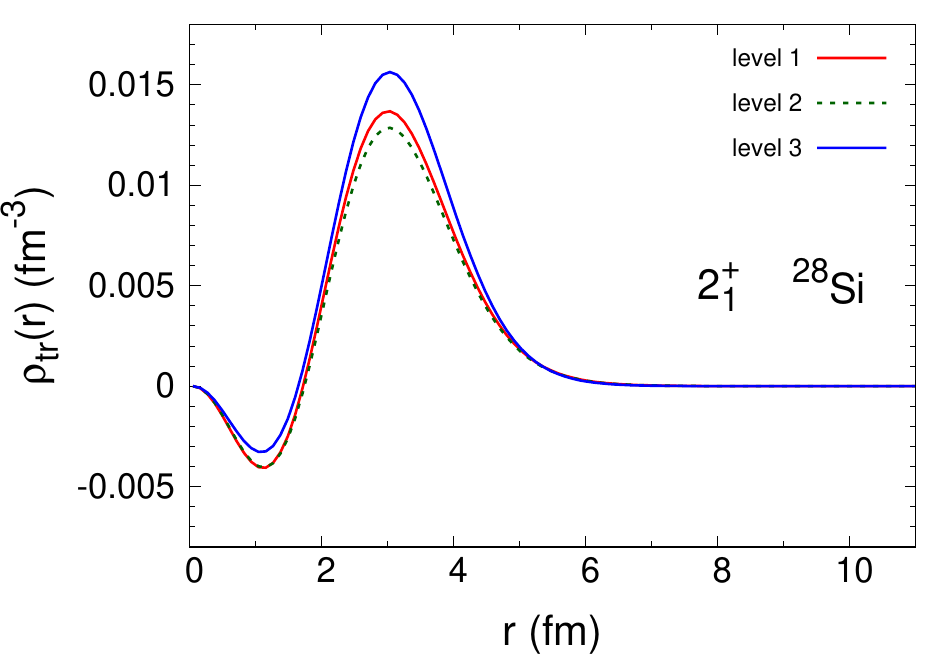}} \hfill %
\subcaptionbox{Form factor in $^{28}$Si \label{28Si_fql}}%
[.49\linewidth]{\includegraphics[width=\columnwidth] {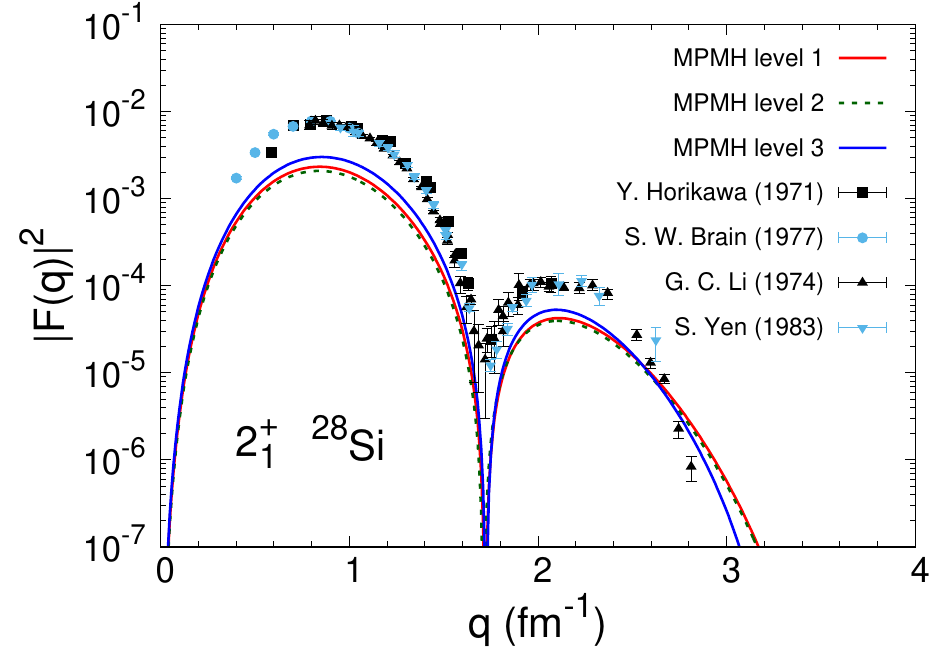}} \hfill %
\subcaptionbox{Transition density in $^{32}$S \label{32S_rho}}%
[.49\linewidth]{\includegraphics[width=\columnwidth] {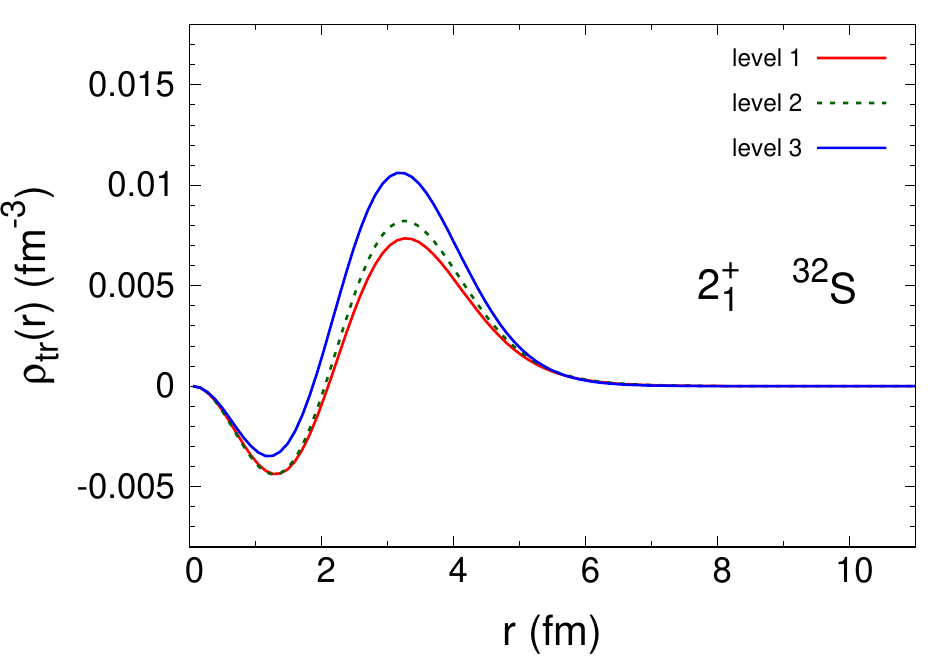}} \hfill %
\subcaptionbox{Form factor in $^{32}$S \label{32S_fql}}%
[.49\linewidth]{\includegraphics[width=\columnwidth] {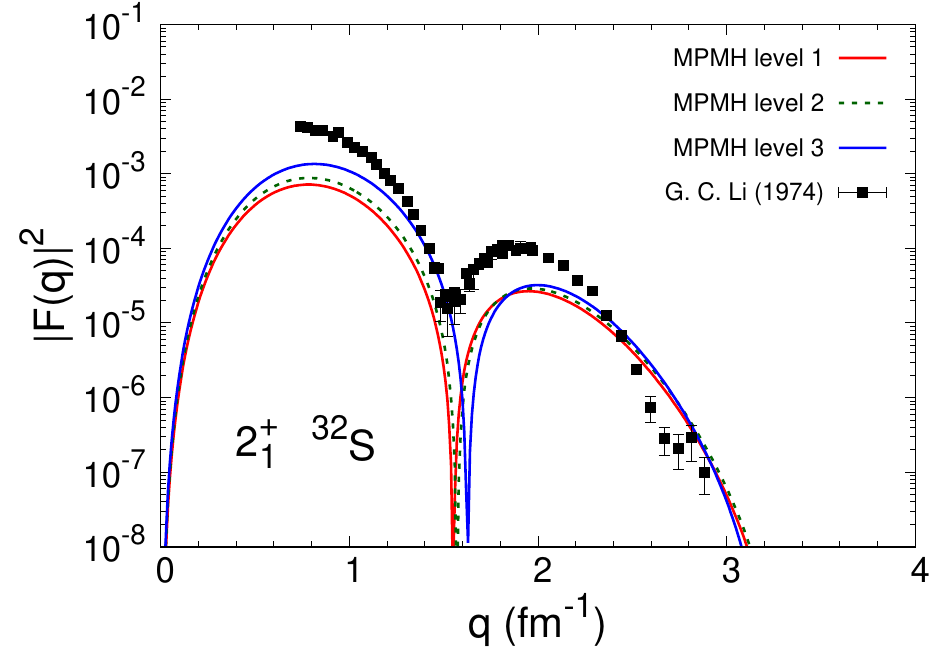}} \hfill %
\subcaptionbox{Transition density in $^{20}$Ne \label{20Ne_rho}}%
[.49\linewidth]{\includegraphics[width=\columnwidth] {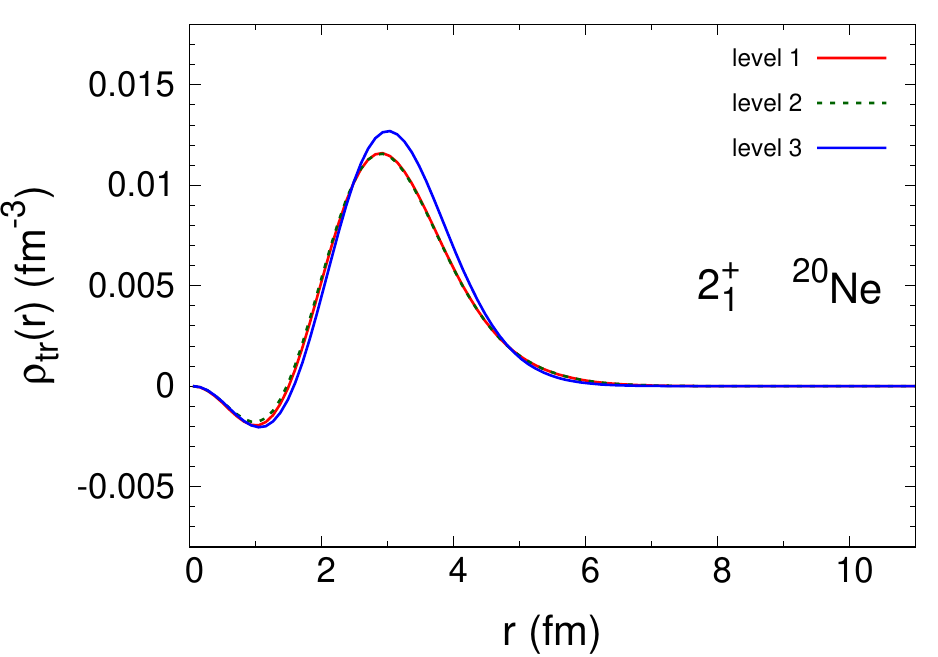}} \hfill %
\subcaptionbox{Form factor in $^{20}$Ne \label{20Ne_fql}}%
[.49\linewidth]{\includegraphics[width=\columnwidth] {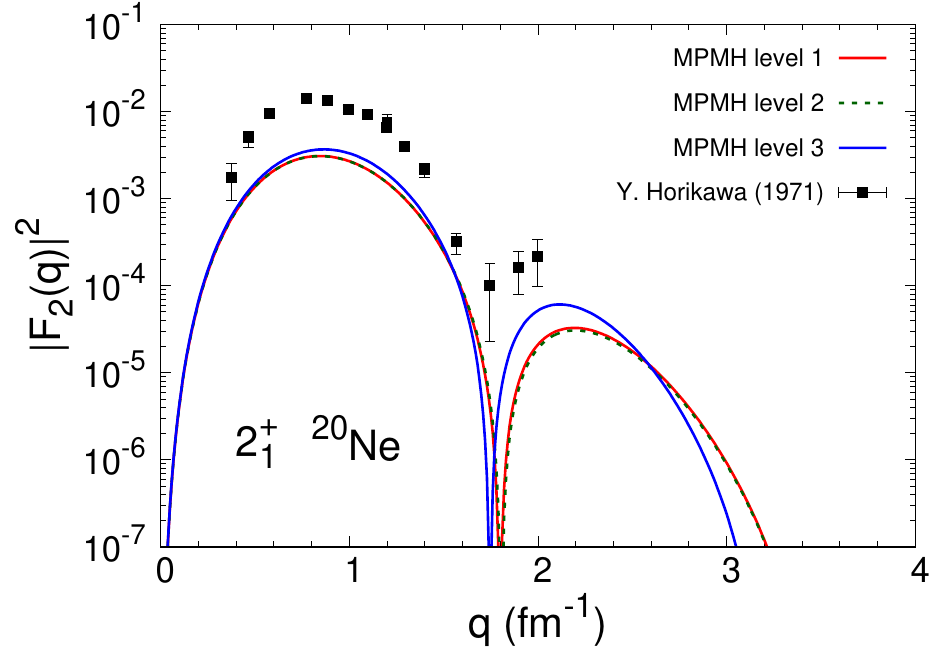}} \hfill %
\caption{(Color Online) Charge transition densities and corresponding form factors for $^{28}$Si, $^{32}$S and $^{20}$Ne. The experimental data are taken from \cite{Horikawa71,Brain77,Li74,Yen83}.}
 \label{fig:form_factor} 
\end{figure*}

Using transition densities calculated in the framework of the MPMH method, we calculated form factors for electron scattering on nuclei of the $sd$-shell. Here we consider the Plane Wave Born Approximation (PWBA), which treats the electron as a plane wave and the electromagnetic interaction as an exchange of a single virtual photon.  We consider only the longitudinal part of the form factors due to the Coulomb interaction between the electron and the nucleus, \textit{i.e.} we neglect their transverse part which arises from the interaction with the electromagnetic currents of the target nuclei. In this case, the form factors read
\begin{equation}
|F_{C,\lambda=J}(q)|^2 = \frac{4\pi}{Z^2}\sqrt{\frac{2J_f+1}{2J_i+1}} \int_0^\infty \rho_{tr}(r) j_\lambda(qr) r^2 dr \; ,
\label{e:form_f}
\end{equation} 
where  $j_\lambda(qr)$ are spherical Bessel functions. The radial charge transition density between initial and final states $\rho_{tr}(r)=\braket{\Psi_f |\hat \rho_{ch}(r)| \Psi_i}$ constitutes the input calculated within the MPMH method. The correction to the charge density due to the finite size of the nucleons is obtained by folding the proton (resp. neutron) density with the distribution of the proton (resp. neutron) which is normalized to unity (resp. zero). 
\\
\\
We show in Fig.~\ref{fig:form_factor} the results obtained for the transitions $0_1^+ \rightarrow 2_1^+$ for $^{28}$Si, $^{32}$S and $^{20}$Ne. The theoretical curves are compared to experiment \cite{Horikawa71,Brain77,Li74,Yen83}. To interpret the results we also show the MPMH transition densities used in the calculations. \\

\begin{figure*}
\centering
\subcaptionbox{\label{20Ne_rho_sdfp}}%
[.49\linewidth]{\includegraphics[width=\columnwidth] {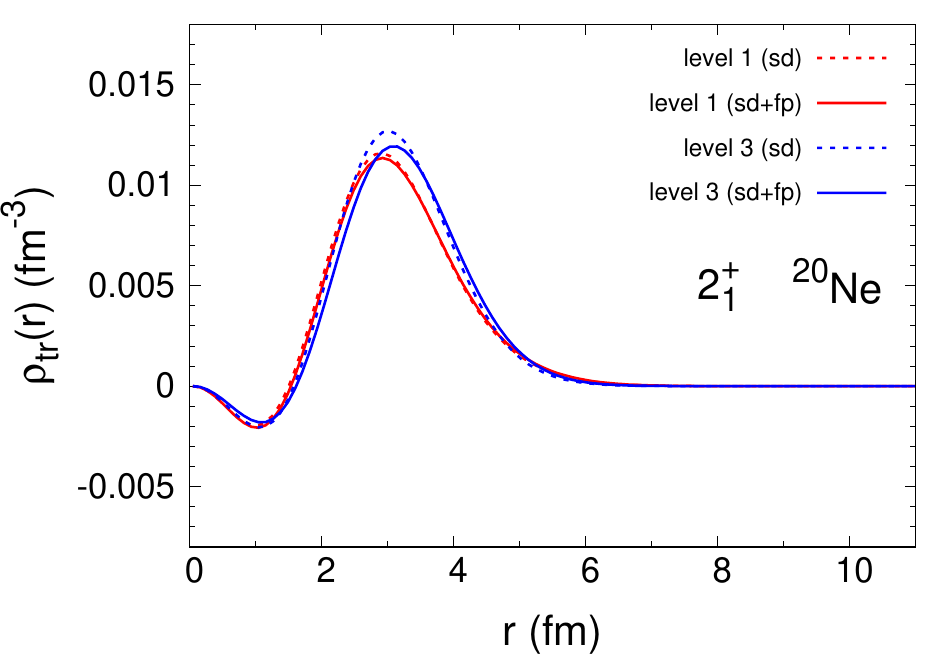}} \hfill %
\subcaptionbox{\label{20Ne_fql_sdfp}}%
[.49\linewidth]{\includegraphics[width=\columnwidth] {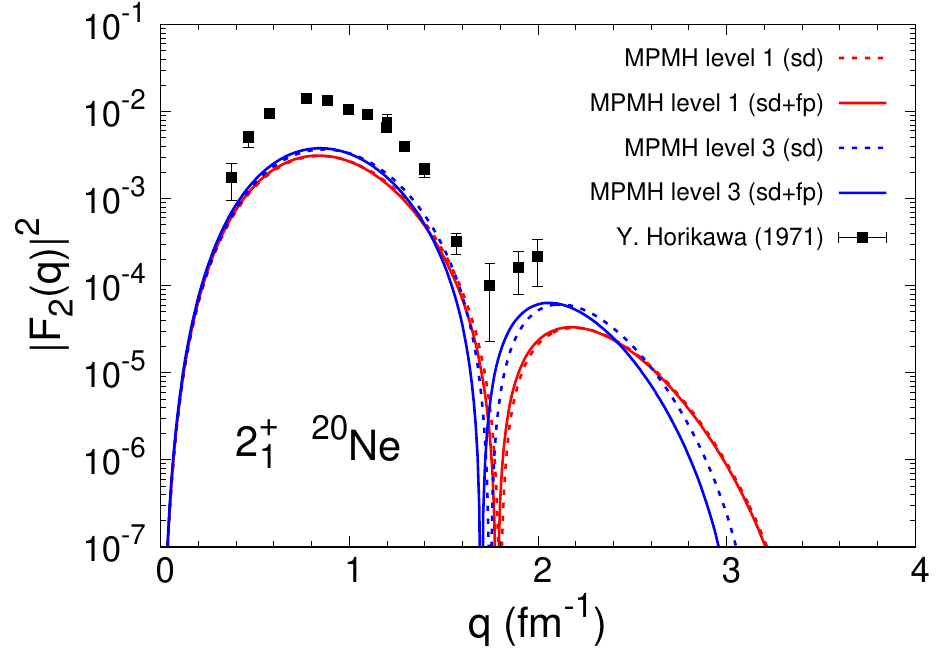}} \hfill %
\caption{(Color Online) Transition densities and corresponding form factors for $^{20}$Ne obtained with two valence spaces: $sd$ shell and $sd+fp$ shells.}
 \label{fig:form_factor_sdfp} 
\end{figure*}

\noindent At the non self-consistent level 1, the trends of the experimental form factor are rather well reproduced by the theoretical results at low momenta, while the tail appears too spread toward high momenta.  We also observe a clear lack of magnitude of the calculated form factors which underestimate the experimental data by a factor $\sim 4$ to $\sim 5$ depending on the nucleus. 
When introducing the correlated density in the interaction while keeping the Hartree-Fock orbitals frozen (level 2), the results are not very much modified. We observe a slight increase and decrease of the global magnitude in $^{32}$S and $^{28}$Si, respectively, while there is no change in $^{20}$Ne.
\\
\\
As already stated in \cite{Robin}, this systematic underestimate of the electric quadrupole properties is a well-known behavior which is due to the restriction of the configuration mixing to the $sd$-shell, unable to fully account for the quadrupole collectivity, as it only generates $0\hbar\Omega$ excitations. In traditional Shell-Model studies, the missing $2\hbar\Omega$ transitions that can be generated by the electric quadrupole operator are usually implicitly taken into account via the introduction of effective charges, that are often fitted to data. 
In this work we aim to quantify the effect of the optimization of orbitals on transition probabilities, with no use of effective charges.
Indeed, as discussed in \cite{Robin}, the source term $G[\sigma]$ of the orbital equation (\ref{e:eq2_Gogny}) introduces a coupling between the initial blocks of single-particle states: core, valence space and initially empty pure HF states. Thus, solving Eq.~(\ref{e:eq2_Gogny}) implicitly generates np-nh excitations (as products of 1p-1h excitations) spanning on the entire initial HF single-particle basis, on top of the Slater determinants of the $sd$-shell. Part of the neglected Hilbert space is thus implicitly accounted for. For instance, the most important effect is encountered in $^{32}$S for which an analysis of the final wave function shows that the self-consistent ground state contains more than $7 \%$ of Slater determinants built outside of the initial HF $sd$-shell.
\\
The missing $2\hbar \Omega$ configurations can result either from 1p-1h excitations between shells differing by $\Delta N_{shell}=2$, or from 2p-2h excitations between shells with $\Delta N_{shell}=1$.
\\
 The former type of excitations are mostly excitations from the $sd$ to the $sdg$ shell or from the $0s$ to the $sd$ shell. As seen from Table \ref{source_sd} the $0s$ and $sdg$ shells are largely influenced by the source term $G[\sigma]$. In particular, the coupling $sd-0s$ is always very strong. Moreover, 1p-1h excitations being one-body excitations, they can be generated through the optimization of orbitals. For these two reasons, the effect of the orbital equation should be maximal in this case, and should allow to partly account for this type of excitations. Less importantly, 1p-1h excitations from the $p$ to the $fp$ shells can also come into play. The source term does not couple to such negative parity states. The orbital equation will however allow to mix the single-particle orbitals from $p$ sub-shells with same angular momentum $j$ through $\left[ \hat{\mathpzc{h}}[\rho,\sigma] , \hat{\rho} \right] = 0$. Again we remind that we are reasoning here in terms of the initial non-optimized orbitals. In practice these are taken as spherical Hartree-Fock states. Since the Hartree-Fock field already incorporates much physical information (compared to \textit{e.g.} pure harmonic oscillator potential), the mixing between orbitals remains usually rather small, as illustrated in section \ref{sec:orbitals}.
\\
On the other hand, 2p-2h excitations between shells with $\Delta N_{shell}=1$ remain unaccounted for. Indeed, because symmetries are explicitly conserved in the MPMH approach, single-particle states from $sd$ cannot mix with orbitals from $p$ or $fp$, and therefore 2p-2h excitations such as $sd \rightarrow fp$ or $p \rightarrow sd$ cannot be generated by the transformation of orbitals.
\\
\\
When looking at the form factors obtained when full self-consistency is applied (level 3), we note that the optimization of orbitals globally improves the results. The effect in  $^{28}$Si and $^{32}$S are quite noticeable, as the magnitude of the theoretical form factors is increased and the global factor needed to reach experiment is now reduced to a value of $\sim 2.5$ for both of these nuclei. This is caused by an important increase of the transition density at the surface. Indeed the peak around 3 fm varies from a value of $\sim 0.013$ to $\sim 0.016$ fm$^{-3}$ in $^{28}$Si, and from 0.0082 to 0.011 fm$^{-3}$ in $^{32}$S.  
Moreover, we also observe a slight extension of the transition density toward larger distance $r$, which induces a shrinking of the tail of the form factor toward lower momenta $q$ and leads to a better agreement with experiment, in particular in $^{28}$Si. This is in accordance with the results shown in section \ref{sec:orbitals}, where it was found that the self-consistent orbitals tend to have a larger spatial extension than the pure HF ones.
In $^{32}$S however we note a displacement of the minimum of the form factor toward higher momenta as self-consistency is included, which is in disagreement with the data.
In $^{20}$Ne, on the contrary we note a displacement of the minimum of $|F_{C,2}|^2$ toward smaller momenta due to the orbital transformation which leads to a better agreement with experiment. A strong shrinking at the tail also appears, however there is no data at high momenta $q$. The magnitude of the form factor is only slightly increased by self-consistency effects. The global factor needed to reach experiment varies from a value of $\sim 4.5$ at the levels 1 and 2 of implementation of MPMH to a value of $\sim 4$ when the correlations and orbitals are fully consistent (level 3).
\\
\\
To end this study, we also investigate the effect of the explicit inclusion of the $fp$ shell in the configuration mixing. We note that in principle, when using more than one shell as the valence space, one would need an exact treatment of the center of mass motion. However here we only aim to provide an illustration of the effect caused by the introduction of explicit $2\hbar\Omega$ configurations.
We show in Fig. \ref{fig:form_factor_sdfp} the form factors calculated for the transition $0_1^+ \rightarrow 2_1^+$ in $^{20}$Ne when the $sd+fp$ valence space is considered, without and with full self-consistency (levels 1 and 3 of MPMH). We compare them to the ones previously obtained with the $sd$ valence-space only. With no consistency (red curves), little change is observed when the $fp$ shell is included. Self-consistency effects are however increased. The shift of the minimum toward smaller momenta $q$ is found larger, and is accompanied by a global compression of the form factor, in particular at the tail. Magnitude is also gained and the discrepancy with experiment is reduced to a factor of $\sim 3.5$.\\ \\

\begin{figure}[h!]
\centering
\includegraphics[width=\columnwidth] {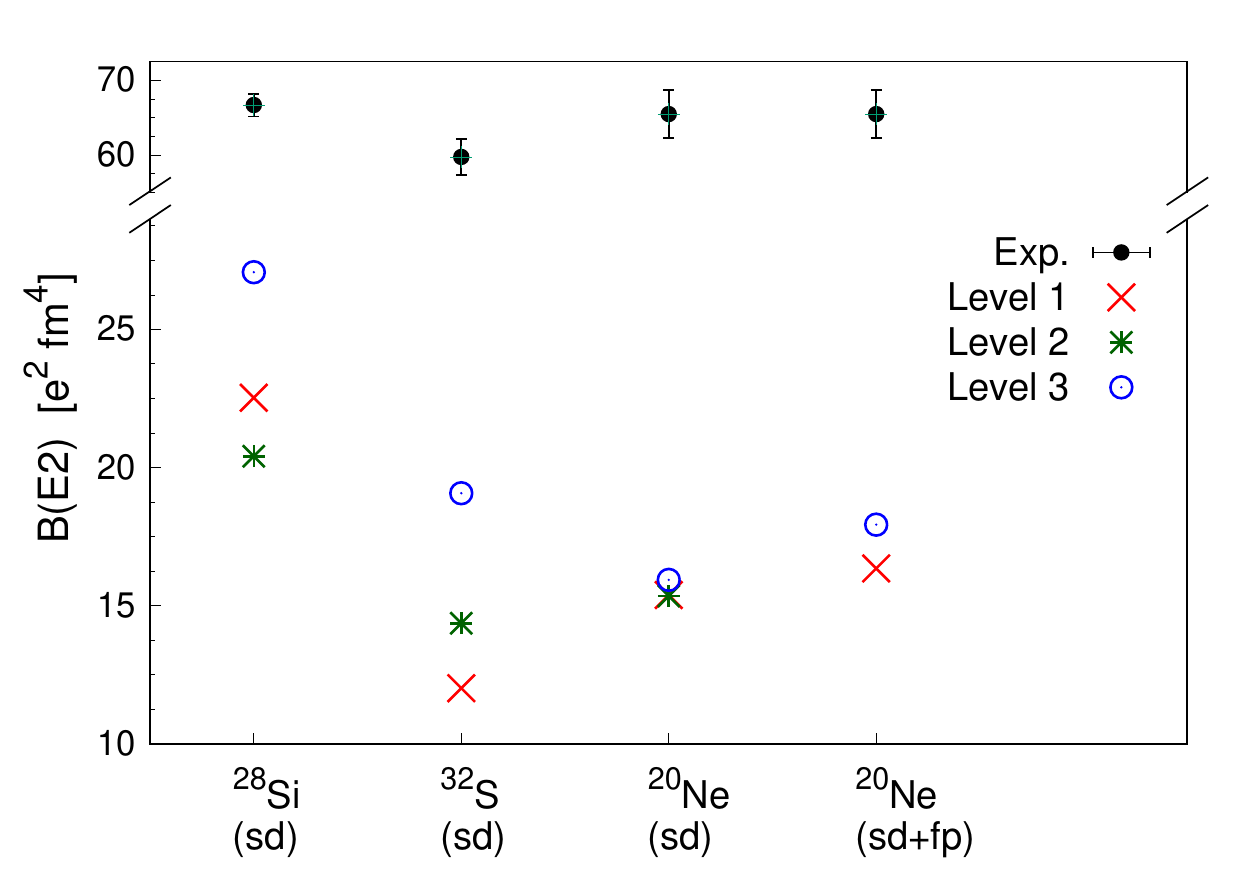}
\caption{(Color Online) Comparison of theoretical and experimental quadrupole transition probabilities $B(E2;2^+_1 \rightarrow 0^+_1)$ expressed in e$^2$.fm$^4$. The experimental data (in black) are taken from \cite{nndc}. Results without and with full self-consistency are shown in red and blue respectively. Results obtained with introducing rearrangement terms in the first variational equation are shown in green. Theoretical results for $^{28}$Si, $^{32}$S and $^{20}$Ne obtained using the $sd$-shell as a valence space are shown on the left of the figure. For comparison, the results obtained for $^{20}$Ne with an $sd+fp$ valence space, without and with full self-consistency, are displayed on the right.}
\label{f:BE2}
\end{figure}

\noindent Finally, for information, we note that, as expected, the results concerning the form factors are consistent with $B(E2)$ calculations that we show in Fig.~\ref{f:BE2} .

\subsubsection{Inelastic proton scattering}
\begin{figure}
\includegraphics[width=\columnwidth] {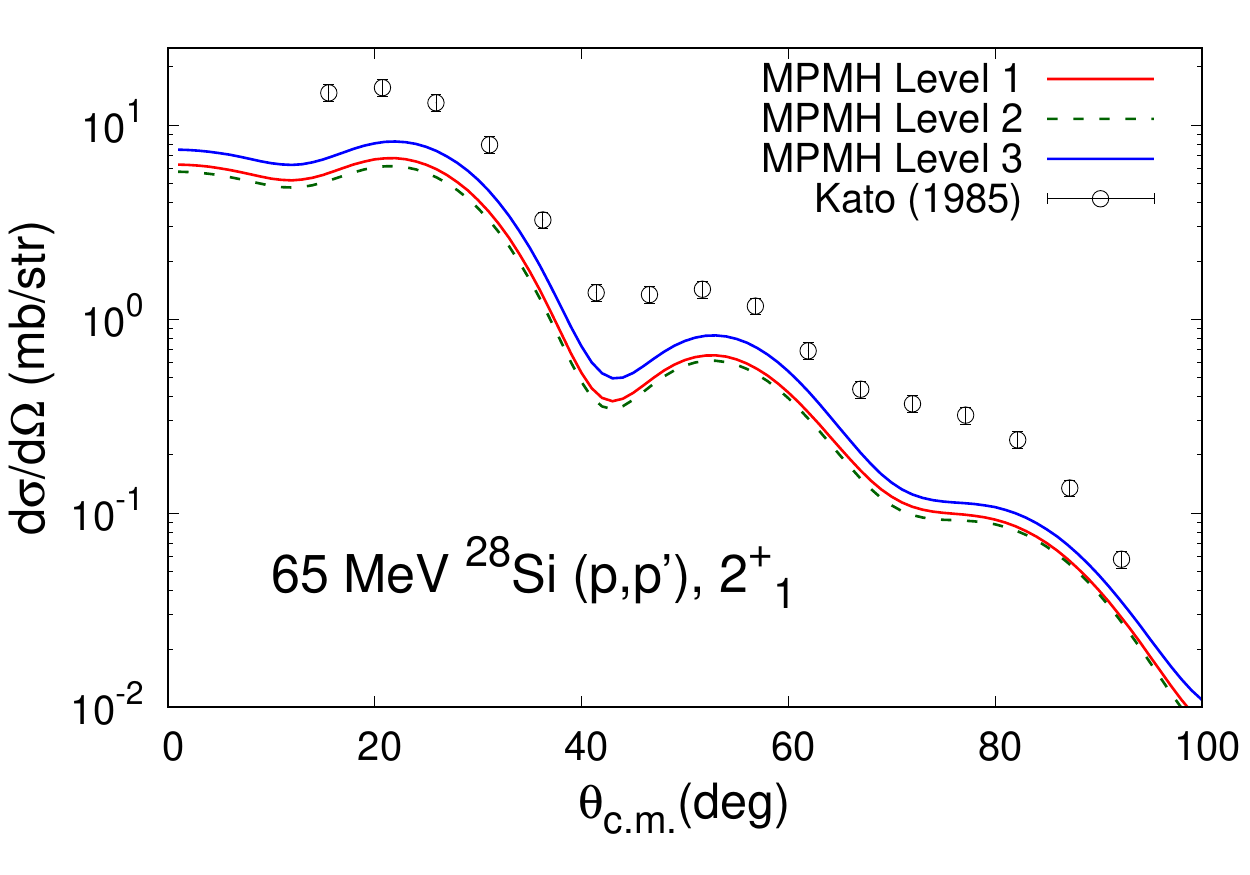}
\caption{(Color Online) Inelastic proton scattering cross section for the transition $0_1^+ \rightarrow 2_1^+$ in $^{28}$Si.}
\label{fig:28Si_pp}
\end{figure}

While charge, thus proton, density distributions can be precisely inferred from electron scattering measurements, the scattering of hadrons provides insight on the distribution of both neutrons and protons. Besides, for nucleon incident energies around 50~MeV, the strength of the proton-neutron interaction has been shown to be approximately three times larger than the strength of proton-proton or neutron-neutron interaction  \cite{Bern81}. Proton inelastic scattering is therefore very sensitive to the neutron collectivity in transitions. We propose to illustrate these aspects through the study of inelastic proton scattering off $^{28}$Si that leads to the excitation of the yrast $2^+$ state, for an incident energy of 65~MeV.
\\
In this energy range, nucleon direct inelastic scattering off spherical and near spherical target is
well described within the Distorted Wave Born Approximation (DWBA) \cite{Satchler}, as the elastic
channel remains strongly dominant over one particular inelastic channel.
Below, we remind the main steps of this approach. We omit the various projectile and target intrinsic spin dependencies
 for simplicity, as these details can be found in \cite{Satchler}.
In the DWBA framework, the transition amplitude expressing an excitation of the nucleus from its ground state $\ket{\Psi_0}$ to an excited state $\ket{\Psi_n}$ reads
\begin{equation}
 T_{fi} \simeq \braket{\chi_f^-({\bf k_f}) \Psi_n| \hat V_{eff}  |\chi_{i}^+({\bf k_i}) \Psi_0} \; .
\label{eq:Tfi}
\end{equation}
The distorted waves $\chi_{i}^+({\bf k_i})$ and $\chi_f^-({\bf k_f})$ are solutions of the equations
\begin{eqnarray}
\Bigl( E_{i} - T_0 + \braket{\Psi_0|\hat{V}_{eff}|\Psi_0} \Bigr)  \chi_{i}^+({\bf k_i}) &=& 0 \; , \\
\Bigl( E_{f} - T_0 + \braket{\Psi_0|\hat{V}^\dagger_{eff}|\Psi_0} \Bigr)  \chi_f^-({\bf k_f}) &=& 0 \; ,
\end{eqnarray}
where $T_0$ is the kinetic energy operator for the relative motion in the center of mass system, and $E_{i/f}$ is the initial/final kinetic energy, that is $E_{f}=E_{i}-E^*$ ($E^*$ is the excitation energy of the state $\ket{\Psi_n}$). The subscript $+$ (-) refers to the boundary condition of a plane wave plus an outgoing scattered wave (plane wave plus incoming scattering wave).
If we express the two-body effective interaction $V_{eff}$ in the second quantification framework, the transition and optical potentials, $\hat{U}_{n0} \equiv \braket{\Psi_n|\hat{V}_{eff}|\Psi_0}$ and $\hat{U}_{00} \equiv \braket{\Psi_0|\hat{V}_{eff}|\Psi_0}$ respectively, read
\begin{eqnarray}
 \hat{U}_{nm}  &=&  \frac{1}{2} \sum_{i j \; k k'} \braket{k' j |\hat{V}_{eff}| \widetilde{k i}}    \braket{\Psi_n | a^{\dagger}_j a_i   |\Psi_{m}}  a^\dagger_{k'} a_k \; , \nonumber \\
\end{eqnarray}
where $(k,k')$ denote states of the projectile nucleon, while $(i,j)$ are states for nucleons in the target.
In this work, the effective interaction $V_{eff}$ between the projectile and the target nucleons is taken as the Melbourne  G-matrix, which is a solution of the Brueckner-Bethe-Goldstone equation with the Bonn-B bare interaction~\cite{Amos00}. The one-body ground-state density matrix elements  $\braket{\Psi_0| a^{\dagger}_j a_i   |\Psi_{0}}$, as well as the one-body transition density matrix elements $\rho^{m\rightarrow n}_{ij} = \braket{\Psi_n| a^{\dagger}_j a_i   |\Psi_{m}}$, are calculated within the MPMH approach as detailed in the appendix.
\\
This reaction approach based on the Melbourne G-matrix interaction was previously successfully used
to describe elastic and direct inelastic nucleon scattering, for various closed-shell or near closed-shell nuclei and various excitations \cite{Amos00}.
Calculations based on the Melbourne G-matrix and  nuclear structure information calculated within the Random
Phase Approximation formalism implemented with the D1S Gogny force, also provided a very good account of elastic and direct inelastic scattering data for various doubly closed-shell nuclei \cite{Dupuis06,Dupuis08}.
We show in Fig.~\ref{fig:28Si_pp} the theoretical cross sections for inelastic proton scattering on a $^{28}$Si target.
Inelastic cross section calculated considering three various levels of implementation of the MPMH method are compared
to experiment \cite{Kato1985}.
Qualitatively, the results are in accordance with the previous electron scattering study. In particular, we note that the
rearrangement terms do not produce a noticeable effect, while the optimization of orbitals globally improves
the cross section by a factor $\sim 1.25$.


\section{Conclusion and perspectives} \label{sec:conclusion}
In this paper, the full formalism of the variational multiparticle-multihole configuration mixing method (MPMH) was systematically applied to the study of even-even $sd$-shell nuclei.
Using the D1S Gogny force, we have investigated many properties of ground and first excited states. 
We have also put an emphasis on the effect of the modification of single-particle states due to the correlations of the system, which we found can modify both the valence states and the deeper orbitals of the core.
We have analyzed in more detail the correlation content of the ground state of some notable nuclei, and found that the optimization of the orbitals leads to a fragmentation of the wave function, which can exceed $20\%$ in a few cases. In most cases the change of orbitals also led to a depletion of the spatial density at the center, which was counterbalanced by a larger extension towards the surface, and therefore led to an increase of the charge radii, in accordance with experiment.
While the binding energies were found to be overestimated by several MeVs -- as expected from the Gogny interaction -- very good results were found concerning the excitation energies of the first excited states. 
Finally, in order to test the nuclear structure description provided by the present approach, we have used the transition densities calculated within the MPMH framework as input for the calculation of (e,e') form factors and (p,p') cross sections. The experimental trends were found to be well reproduced in most cases, in particular when using optimized orbitals. An underestimation of the magnitude, reflecting a lack of collectivity in the nuclear wave functions, was observed, although the use of optimized of single-particle states led to a small improvement. 
\\
This first systematic study has shed light on the potential and mandatory extensions of the MPMH approach. The determination of the single-particle orbitals consistently with the correlations of the system is a desirable and satisfying feature of this framework, which also leads to better results. However in its present status, the MPMH method shows some insufficiencies. 
The use of the D1S Gogny interaction, fitted at the mean-field level, inevitably leads to double-counting when used in an approach such as MPMH which explicitly treats many-body correlations. To avoid this problem, we envisage for the future to implement better-suited interactions.
Additionally, currently available computational power imposes limits on the complexity and quantity of nucleon configurations in the model space, making very difficult to properly describe the collectivity of the system.
This issue may be alleviated using the Feshbach formalism \cite{Feshbach} to renormalize the Hamiltonian within the space of Slater determinants that are explicitly introduced in the many-body wave function. The resulting effective Hamiltonian would implicitly account for the missing space that is left untreated by the configuration mixing built on optimal orbitals. Such an extension would make it possible to improve the description of the collectivity and to tackle the description of mid- to heavy-mass nuclei, without introducing phenomenological effective charges.


\appendix

\section{Transition density matrix}

The structure ingredient provided by the MPMH approach for the reaction calculations is the one-body transition density.
The matrix elements of the transition density matrix between states $\ket{\Pn}$ and $\ket{\Pm}$ are defined by
\begin{equation}
\rhoij = \elmx{i}{\hat{\rho}^{n\rightarrow m}}{j}
= \elmx{\Pm}{\aC{j}\aA{i}}{\Pn}\:,
\end{equation}
where $i$ and $j$ are single-particle states with angular momentum $j$, projection $m$ and other quantum number $\zeta$: 
\begin{equation}
\ket{j} = \aC{j}\ket{0} = \ket{\zeta_j j_j m_j}\:.
\end{equation}
Within the MPMH approach, nuclear states are explicitly characterized by a good projection $K$  of the angular momentum on the quantization axis in the laboratory frame and a good parity $\pi$. As calculations are performed at the spherical point, and configuration mixings are done in spaces preserving rotational symmetry, the nuclear states are also characterized by a good total angular momentum $J$.
Here we detail the calculation of the matrix element of $\hat \rho$ between two MPMH states $\ket{\Pn^{JK\pi}}$ and $\ket{\Pm^{J^\prime K^\prime\pi^\prime}}$, which are linear combinations of the same Slater determinants:
\begin{eqnarray}
\ket{\Pn^{JK\pi}} &=& \sum_\alpha A_\alpha^{(n)} \ket{\Phi_\alpha^{K\pi}}\; , \\
\ket{\Pm^{J^\prime K^\prime \pi^\prime}} &=& \sum_\alpha A_\alpha^{(m)} \ket{\Phi_\alpha^{K^\prime\pi^\prime}} \; ,\label{defMPMH}
\end{eqnarray}
in the natural single-particle basis, used to build the many-body configurations. \\ \\

\noindent Denoting by $\bar{a}$, $\bar{b}$... the states belonging to the core and by $a$, $b$... the active orbitals, we can decompose the transition density matrix into parts:

\begin{itemize}
\item The transition density between two orbitals of the core matrix is diagonal and reads
\begin{equation}
\rho_{\bar{a}\bar{b}}^{n\rightarrow m} = \delta_{\bar{a}\bar{b}}\,\delta_{nm}\:.
\end{equation}

\item For pure active orbitals, the transition density can be decomposed into fractional diagonal elements and a non-diagonal part as
\begin{equation}
\rho_{ab}^{n \rightarrow m} = \delta_{ab}\,v_a^{nm} + (1-\delta_{ab})\,w_{ab}^{nm}\:.
\end{equation}

\item Finally, matrix elements between core and active orbitals cancel:
\begin{equation}
\rho_{\bar{a} a}^{n\rightarrow m} = \rho_{a \bar{a}}^{n\rightarrow m} = 0
\qquad\forall \bar{a},\,a,\,n,\,m\:.
\end{equation}
\end{itemize}

\multieq{
\rhoij \equiv
\begin{pmatrix}
\begin{array}{ccc|}
\delta_{nm} & & 0 \\
 & \ddots & \\
0 & & \delta_{nm}
\end{array}
&
\begin{array}{ccc}
 \phantom{\delta_{nm}} & \phantom{\ddots} & \\
 & 0 & \\
 & & \phantom{\delta_{nm}}
\end{array}\\
\hline
\begin{array}{ccc|}
 & & \\
 \phantom{\delta_{nm}} & \phantom{\ddots} & \\
 & 0 & \\
 & & \phantom{\delta_{nm}}\\
 & &
\end{array}
&
\begin{array}{ccc}
 & & \\
 v_1^{nm} & & (w^{nm})_{i<j}\\
 & \ddots & \\
 (w^{nm})_{i>j} & & v_N^{nm}\\
 & &
\end{array}
\end{pmatrix}\\
\qquad core\ orbitals \qquad\qquad active\ orbitals\nonumber
}

$\;$\\ \\
Finally, we calculate the transition density matrix coupled to a good angular momentum $\lambda$
\begin{equation}
\rhobab = \elmx{\Pm^{J^\prime K^\prime}}{[\aC{i}\otimes\aT{j}]^\lambda_\mu}
{\Pn^{JK}}\:,\label{defsig}
\end{equation}
where the tilde operator $\aT{j}$ is defined as
$(-)^{j_j-m_j}\aA{j}$ to transform under rotation as a spherical
tensor of rank $j_j$. 
This object can be deduced from the transition density $\rhoij$ by
\multieqal{
\rhobab &= \elmx{\Pm^{J^\prime K^\prime}}{[\aC{i}\otimes\aT{j}]^\lambda_\mu}
{\Pn^{JK}}\nonumber\\
&= \sum_{m_j m_i} \left( j_i\,m_i\,\,j_j-\!m_j |
\lambda \mu \right)\,
(-)^{j_j-m_j} \nonumber \\ 
& \hspace{1.5cm} \times \elmx{\Pm^{J^\prime K^\prime}}{\aC{i}\aA{j}}{\Pn^{JK}}\:.
\label{sig1}
}

We now consider the case of transitions between the ground state of
an even-even nucleus ($J=K=0$) and one of its excited states also characterized
by $K=0$. The transition density will be coupled to $(\lambda,\mu) =
(J,0)$. In this case, the contribution of time-reversed single-particle
states can be simply related to the one of the ``positive'' states.
\multieqal{
\rhobabz &= \elmx{\Pn^{J0}}{[\aC{i}\otimes\aT{j}]^J_0}
{\Psi_{gs}^{00}}\nonumber\\[6pt]
&= \sum_{0<|m|\leqslant\min(j_j,j_i)} \left( j_i\,m\,\,j_j-\!m | J 0
\right)\,(-)^{j_j-m}\nonumber \\
& \hspace{1.5cm} \times \elmx{\Pn^{J0}}{\aC{i}\aA{j}}{\Psi_{gs}^{00}} 
\label{RHOJGS}\\[6pt]
&= \sum_{m=1/2}^{\min(j_j,j_i)} \left( j_i\,m\,\,j_j-\!m | J 0
\right)\,(-)^{j_j-m}\,\elmx{\Pn^{J0}}{\aC{i}\aA{j}}{\Psi_{gs}^{00}}
\nonumber\\
&+\sum_{m=1/2}^{\min(j_j,j_i)} \left( j_i-\!m\,\,j_j\,m | J 0
\right)\,(-)^{j_j+m}\,\elmx{\Pn^{J0}}{\aC{-i}\aA{-j}}{\Psi_{gs}^{00}}\:.
\label{sig2}
}

In the previous equations $\ket{-j}$ denotes the state having all the same quantum numbers
as $\ket{j}$ with $m_j$ changed into $-m_j$:
\begin{equation}
\ket{-j} = \aC{-j}\ket{0} = \ket{\zeta_j j_j -\!m_j}\:.
\end{equation}
The state $\ket{-j}$ corresponds to the time-reversal conjugate state
of $\ket{j}$ up to a phase $(-)^{l_j + j_j - m_j}$,
\begin{equation}
\ket{-j} = (-)^{l_j + j_j - m_j} \ketbar{j}
= (-)^{l_j + j_j - m_j} \ketbar{\zeta_j j_j m_j}\:.
\end{equation}
In addition, the two Clebsch-Gordan coefficients in (\ref{sig2}) are related by
\begin{equation}
\left( j_i-\!m\,\,j_j\,m | J 0 \right) = (-)^{j_j+j_i+J}
\left( j_i\,m\,\,j_j-\!m | J 0 \right)\:.
\end{equation}
The excited nuclear state is either time-reversal symmetric or
time-reversal antisymmetric according to the parity of its total angular
momentum $J$,
\begin{equation}
\hat{T}\ket{\Pn^{J0}} = (-)^J \ket{\Pn^{J0}}\:.
\end{equation}
Therefore, the last matrix element in (\ref{sig2}) satisfies,
\multieqal{
\elmx{\Pn^{J0}}{\aC{-i}\aA{-j}}{\Psi_{gs}^{00}} &=
(-)^{l_j + j_j-m_j}(-)^{l_i + j_i-m_i} \nonumber \\
& \hspace{1.5cm} \times \elmx{\Pn^{J0}}{\aC{\overline{i}}\aA{\overline{j}}}{\Psi_{gs}^{00}}\\
&=(-)^{l_j + j_j-m_j}(-)^{l_i + j_i-m_i}(-)^J \nonumber \\
& \hspace{1.5cm} \times \elmx{\Pn^{J0}}{\aC{i}\aA{j}}{\Psi_{gs}^{00}}\:.
}
Finally, the matrix elements of $\rhobabz$ become
\multieqal{
\rhobabz &= \sum_{m=1/2}^{\min(j_j,j_i)} \left( j_i\,m\,\,j_j-\!m | J 0
\right)\,(-)^{j_j-m}\,\elmx{\Pn^{J0}}{\aC{i}\aA{j}}{\Psi_{gs}^{00}}
\nonumber\\[6pt]
&+\sum_{m=1/2}^{\min(j_j,j_i)} \left( j_i\,m\,\,j_j-\!m | J 0
\right)\,(-)^{j_j-m}\,(-)^{2j_j}\,(-)^{2j_i} \nonumber \\
& \hspace{1.5cm} \times (-)^{2J}\,(-)^{l_j+l_i}\,
\elmx{\Pn^{J0}}{\aC{i}\aA{j}}{\Psi_{gs}^{00}} \nonumber \\[6pt]
&= \sum_{m=1/2}^{\min(j_j,j_i)} \left( j_i\,m\,\,j_j-\!m | J 0 
\right)\,(-)^{j_j-m}\,\left(1+(-)^{l_j+l_i}\right)\nonumber \\
& \hspace{1.5cm} \times  \elmx{\Pn^{J0}}{\aC{i}\aA{j}}{\Psi_{gs}^{00}}\:.
\label{sigfin}
}

\end{document}